\documentclass[12pt]{article} \usepackage{latexsym} \textwidth 15cm
\begin{document}
\title{{\bf CLASSICAL AND QUANTUM DECAY OF OSCILLATONS:
\\ OSCILLATING SELF-GRAVITATING REAL SCALAR FIELD SOLITONS}
\thanks{Alberta-Thy-10-03, gr-qc/0310006}}
\author{
Don N. Page
\thanks{Internet address:
don@phys.ualberta.ca}
\\
Institute for Theoretical Physics\\
Department of Physics, University of Alberta\\
Edmonton, Alberta, Canada T6G 2J1
}
\date{(2003 Oct. 2)}
\maketitle
\large

\begin{abstract}
\baselineskip 18 pt

The oscillating gravitational field of an oscillaton of finite mass
$M$ causes it to lose energy by emitting classical scalar field waves,
but at a rate that is non-perturbatively tiny for small $\mu \equiv
GMm/\hbar c$, where $m$ is the scalar field mass: $dM/dt \approx -
3\,797\,437.776\,333\,015 (c^3/G) \mu^{-2}
e^{-39.433\,795\,197\,160\,163/\mu}[1+O(\mu)].$ Oscillatons also decay
by the quantum process of the annihilation of scalarons into
gravitons, which is only perturbatively small in $\mu$, giving by
itself $dM/dt \approx - 0.008\,513\,223\,934\,732\,692 (m^2 c^2/\hbar)
\mu^{5} [1+O(\mu^2)]$.  Thus the quantum decay is faster than the
classical one for $\mu \,\stackrel{<}{\sim}\, 39.4338/[\ln{(\hbar
c/Gm^2)} + 7 \ln{(1/\mu)} +19.9160]$.  The time for an oscillaton to
decay away completely into free scalarons and gravitons is $t_{\rm
decay} \sim 2\hbar^6 c^3/G^5 m^{11} \sim 10^{324} {\rm yr} (1\, {\rm
meV}/m c^2)^{11}$.  Oscillatons of more than one real scalar field of
the same mass generically asymptotically approach a static-geometry
$U(1)$ boson star configuration with $\mu = \mu_0$, at the rate
$d(GM/c^3)/dt \approx [(C/\mu^4)e^{-\alpha/\mu}+Q(m/m_{\rm
Pl})^2\mu^3] (\mu^2-\mu_0^2)$, with $\mu_0$ depending on the
magnitudes and relative phases of the oscillating fields, and with the
same constants $C$, $\alpha$, and $Q$ given numerically above for the
single-field case that is equivalent to $\mu_0=0$.

\end{abstract}
\normalsize
\baselineskip 18 pt
\newpage

\section{Introduction}

Seidel and Suen \cite{SS, SS2} have found numerically that there exist
non-singular oscillating self-gravitating solitonic configurations of
a real scalar field, which they called oscillating soliton stars.
These have also been studied by several other authors \cite{I, U-L,
AGMNULW, ULMB, HC, ABGMNUL, GUL} and are now generally called {\it
oscillatons}.  In the simplest case, which is what I shall consider
here, they arise from the Einstein-Klein-Gordon (EKG) equations for
gravity plus one or more minimally coupled massive real scalar fields.

The previous numerical evidence suggested that although these
oscillatons are oscillating, they appeared to be periodic and stable
\cite{SS, SS2, U-L, AGMNULW, ULMB, HC, ABGMNUL, GUL}, so that
classically, at least, an isolated oscillaton might be expected to
last forever.  However, here I shall show that oscillatons of finite
mass actually decay classically as the oscillating gravitational field
leads to the emission of scalar waves.  The decay rate is calculated
for the case of low-mass classical oscillatons and is found to be
nonperturbatively tiny (nonanalytic in the oscillaton mass at zero
mass), given by Eq. (\ref{eq:113}) below, which may be why it has not
been clearly seen numerically.

Seidel and Suen \cite{SS} did recognize that their numerical results
were consistent with quasiperiodic oscillations analogous to the orbit
of two black holes that spiral inward while emitting gravitational
waves.  My results are also similar to this analogy, with the
oscillatons classically emitting scalar waves instead of gravitational
waves, except that here the classical decay rate goes to zero faster
than any power law of the appropriate small parameter (here $\mu$,
which is the mass $M$ of the oscillaton, multiplied by the scalar
field mass $m$, and divided by the square of the Planck mass $m_{\rm
Pl}$) as $\mu$ is taken to zero.

In addition, oscillatons decay quantum mechanically by the
annihilation of scalarons into gravitons, at a rate that is also
calculated here for low-mass oscillatons, given by Eq. (\ref{eq:129})
below.  Although this mass-loss rate is also small, it is perturbative
and goes as the fifth power of the oscillaton mass, so for
sufficiently small oscillaton mass, this quantum decay dominates over
the classical mass-loss rate.  The time for an oscillaton with an
initially large number of scalarons to decay away completely into free
scalarons and gravitons then goes as the inverse 11th power of the
scalaron mass and hence is very large if the scalaron mass is much
less than the Planck mass.

Although my numerical results for the classical and quantum decay
rates are for a single real scalar field in a spherical configuration
(and for any number of such fields of the same mass oscillating in the
same mode except for possible phase shifts \cite{HC}, which do give a
nontrivial effect), I shall start by giving the formalism for the
classical decay rates for an arbitrary nearly-Newtonian configuration
of an arbitrary number of massive scalar fields, and then do a
detailed numerical analysis of the single-field nearly-Newtonian and
nearly-periodic spherical case for both the classical and quantum
decays.  Then I shall return to a discussion of the classical and
quantum decay rates for multi-field oscillatons.

Throughout this paper I shall assume that the mass of each scalar field
is much less than the Planck mass, which is a necessary (though not
sufficient) requirement for doing a classical analysis and is also
necessary for the validity of various equations I shall use for the
quantum decay of oscillatons.

\section{Notation and Units}

Consider the case in which there are $n$ real scalar fields
$\Phi_{IJ}$, each with mass $m_I$, minimally coupled to Einstein
gravity, and with no other self-interactions.  The index $I$ labels
the different mass values, and the index $J$ labels the different
scalar fields that have the same mass.

In the classical analysis, I shall often use units in which $c=1$
(though sometimes for results I shall insert the appropriate power of
$c$ in order to be able to evaluate quantities in conventional units),
but I shall not set $\hbar$ or $G$ equal to unity.  However, to avoid
having $\hbar$'s in most of my equations, I shall let the masses $m_I$
have units of inverse time, which is indeed what they would have (at
least if $c=1$) in the classical Klein-Gordon equation that each
scalar field obeys,
 \begin{equation}
 (\Box-m_I^2)\Phi_{IJ}=0.
 \label{eq:0}
 \end{equation}

I.e., a free zero-spatial-momentum real scalar solution in flat
Minkowski spacetime in orthonormal Minkowski coordinates, with the
appropriate zero of time, would have the time dependence $\cos{(m_I
t)}$.  In terms of the conventional scalaron particle masses, which I
shall hereafter denote with the starred subscript, $m_{*I}$, one has
$m_I = m_{*I}c^2/\hbar$.  Perhaps it is more perspicuous to write this
relation as
 \begin{equation}
 m_{*I}c^2 = \hbar m_I,
 \label{eq:1}
 \end{equation}
so that in terms of the classical quantity $m_I$ (the natural
frequency of the scalar field, in radians per second), the energy
$m_{*I}c^2$ of a one-particle quantum excitation of the scalar field
is indeed $\hbar$ times the frequency of the excitation.

That is, I am taking the view that it is the natural frequencies $m_I$
that are the classical parameters of the scalar fields, and that the
masses $m_{*I}$ of the scalaron particles are quantum properties that
will not show up in a classical analysis or in the classical decay of
the oscillatons (though they will when one considers the quantum
annihilation of scalarons into gravitons).

Analogously, to avoid factors of Newton's gravitational constant $G$
in most of my equations, when I consider the gravitational mass $M$ of
a scalar field configuration or oscillaton, it is convenient to
include $G$ in it (or actually $G/c^3$ if one uses units in which the
speed of light, $c$, is not unity), so that my $M$ has units of time
and is thus half the gravitational (Schwarzschild) radius of the
configuration divided by the speed of light.  Therefore, if I let
$M_{*}$ be the mass in conventional mass units (e.g., grams or
kilograms), what I shall use is
 \begin{equation}
 M \equiv {GM_{*}\over c^3}.
 \label{eq:2}
 \end{equation}

With these conventions, I can avoid using $\hbar$ and $G$ in most of
my intermediate equations, even without using units in which those
quantities are set equal to unity.

For example, the simplest classical spherical oscillatons of a single
real scalar field (with no nodes) are characterized (up to the overall
scale, into which the natural frequency $m$ enters) by the single
dimensionless parameter
 \begin{equation}
 \mu \equiv Mm = {GM_{*}m_{*} \over \hbar c}
 \approx 7.483\,138\,84\times 10^9 \left({M_{*}\over M_{\odot}}\right)
    \left({m_{*}c^2 \over {\rm eV}}\right),
 \label{eq:3}
 \end{equation}
where $M_{\odot} \approx 1.989\times 10^{33}\, g$ is the mass of the sun
in conventional units.  (In temporal units, the mass of the sun is
$4.925\,490\,95 \times 10^{-6}\, {\rm s}$, almost 5 microseconds, known
to much higher accuracy than in conventional units, since the
gravitational effect of the sun, proportional to $GM_{\odot}$, is known
much more accurately than $G$ is in conventional units.)  By
dimensional analysis, one can then easily see that if there is a
classical decay of such an oscillaton, one must have (since my $M$ has
the dimension of time)
 \begin{equation}
 {dM \over dt} = - f(\mu),
 \label{eq:4}
 \end{equation}
a function purely of the only dimensionless parameter of the
oscillaton, $\mu$.  (In Eq. (\ref{eq:113}) below I shall give this
function for $\mu \ll 1$, finding that it is nonanalytic at $\mu=0$.)

The price of this simplicity in the units is that one must get used to
the mass of the oscillaton having the dimension of time, which is the
inverse of the dimension of frequency that is the classical `mass' of the
scalar field in its classical Klein-Gordon equation.

We can get a further simplification by using an appropriate
redefinition of the scalar fields $\Phi_{IJ}$.  Since the square of
the time derivative of a scalar field has the dimension of energy
density, the square of a scalar field has the dimension of mass
divided by length, which is the same as the dimension of $c^2/G$.
Thus (considering also the $8\pi$ in Einstein's equations) it is
convenient to define the dimensionless scalar field values
 \begin{equation}
 \phi_{IJ} \equiv \sqrt{8\pi G/c^2}\,\Phi_{IJ}.
 \label{eq:5}
 \end{equation}

Then by Einstein's equations, the Ricci tensor generated by the
stress-energy tensor of the scalar fields is
 \begin{equation}
 R_{\alpha\beta} = \sum_{IJ} [\phi_{IJ,\alpha}\phi_{IJ,\beta} +
 {1\over 2} g_{\alpha\beta} m{_I}^2 \phi_{IJ}^2].
 \label{eq:6}
 \end{equation}
This will have the dimension of inverse time squared if the
coordinates have the dimension of time and if the metric components
$g_{\alpha\beta}$ are dimensionless.

Following the examples of \cite{HBG, GUL}, it is also convenient to
combine each rescaled dimensionless real scalar field $\phi_{IJ}$ and
its time derivative $\dot{\phi}_{IJ} \equiv \partial
\phi_{IJ}/\partial t$ into a single dimensionless complex quantity,
 \begin{equation}
 \Psi_{IJ}\equiv{1\over 2}e^{i m_I t}(\phi_{IJ}+{i\over m_I}\dot{\phi}_{IJ}),
 \label{eq:7}
 \end{equation}
so
 \begin{equation}
 \phi_{IJ} = \Psi_{IJ}e^{-i m_I t} + \bar{\Psi}_{IJ} e^{i m_I t}
 \label{eq:8}
 \end{equation}
and
 \begin{equation}
 \dot{\phi}_{IJ} = -im_I\Psi_{IJ}e^{-i m_I t} +im_I\bar{\Psi}_{IJ} e^{i m_I t}.
 \label{eq:9}
 \end{equation}

In terms of the complex $\Psi_{IJ}$ and its complex conjugate
$\bar{\Psi}_{IJ}$, the time-time, time-space, and space-space
components of the Ricci tensor are (using 0 to denote the time
coordinate $t=x^0$ and lower-case Latin letters to
denote spatial coordinates $x^i$)
 \begin{equation}
 R_{00} = \sum_{IJ} \{m_I^2 [(2+g_{00})\Psi_{IJ}\bar{\Psi}_{IJ}
     -(1-{1\over 2}g_{00})(\Psi_{IJ}^2 e^{-2i m_I t}
                           +\bar{\Psi}_{IJ}^2 e^{-2i m_I t})]\},
 \label{eq:10}
 \end{equation}
 \begin{eqnarray}
 R_{0i} &=& \sum_{IJ} \{m_I^2 g_{0i}[\Psi_{IJ}\bar{\Psi}_{IJ}
                           +{1\over 2}(\Psi_{IJ}^2 e^{-2i m_I t}
                           +\bar{\Psi}_{IJ}^2 e^{2i m_I t})]
 \nonumber \\
   &+&im_I[\bar{\Psi}_{IJ}\Psi_{IJ,i}-\Psi_{IJ}\bar{\Psi}_{IJ,i}
                    -\Psi_{IJ}\Psi_{IJ,i}e^{-2i m_I t}
                    +\bar{\Psi}_{IJ}\bar{\Psi}_{IJ,i}e^{2i m_I t}]\},
 \label{eq:11}
 \end{eqnarray}
 \begin{eqnarray}
 R_{ij} &=& \sum_{IJ} \{m_I^2 g_{ij}[\Psi_{IJ}\bar{\Psi}_{IJ}
                           +{1\over 2}(\Psi_{IJ}^2 e^{-2i m_I t}
                           +\bar{\Psi}_{IJ}^2 e^{2i m_I t})]
 \nonumber \\
   &+&\Psi_{IJ,i}\bar{\Psi}_{IJ,j}+\bar{\Psi}_{IJ,i}\Psi_{IJ,j}
                    +\Psi_{IJ,i}\Psi_{IJ,j}e^{-2i m_I t}
                    +\bar{\Psi}_{IJ,i}\bar{\Psi}_{IJ,j}e^{2i m_I t}\}.
 \label{eq:12}
 \end{eqnarray}

\section{Gauge or Coordinate Conditions}

In finding solutions to the Einstein-Klein-Gordon equations, one must
make a choice of coordinates or gauge for the gravitational field.  I
shall restrict consideration to 3+1 dimensional spacetime.  There
there are four coordinates to be chosen, giving the freedom of four
free functions over spacetime for the gauge group of coordinate
transformations.

I generally find it convenient to use these four degrees of freedom to
set the time-space components of the metric to be zero, $g_{0i}=0$ for
the three $i$, and to set $g_{00}$ to be independent, or nearly
independent, of the time coordinate $t$.  This then implies that the
hypersurfaces of constant $t$ are orthogonal to the worldlines of
constant spatial coordinates $x^i$, and that along each such
worldline, the proper time is nearly proportional to the coordinate
time $t$ (with a space-dependent constant of proportionality).  For
example, if the metric is periodic in time, we can choose $g_{00}$ to
be precisely independent of the time coordinate $t$.  However, this
still leaves the freedom to make arbitrary spatial coordinate
transformations that are independent of $t$.

%We shall be considering cases in which the fractional amount of energy
%radiated away from an oscillaton or other self-gravitating
%configuration of one or more scalar fields is small on the timescale
%set by the scalar field frequencies.  Therefore, if we initially
%ignore the resulting small secular change in the metric, we can
%approximate the metric by one that is periodic in time (if the masses
%$m_I$ are commensurate) or quasi-periodic (if they are incommensurate).

If we wish to pin down the spatial coordinates, we could, for example,
choose them so that the time average of the spatial metric, $\langle
g_{ij} \rangle$, over a time that is long with respect to the
reciprocal of the smallest natural frequency difference $|m_I-m_{I'}|$,
is as nearly proportional to the $3\times 3$ identity matrix as
possible.  More explicitly, if we take $\langle g_{ij} \rangle$ to be
a spatially dependent $3\times 3$ matrix, we could choose spatial
coordinates so that the integral over all space of the square of the
traceless part of this matrix is minimized.

This would still not pin down the spatial origin or angular
orientation of the resulting quasi-Cartesian spatial coordinate
system, but we could choose the spatial origin to be that which gives
the center of mass of the asymptotic form of the time-averaged metric.
If it is necessary to fix the orientation, we could, for example, fix
it so that the asymptotic quadrupole moment has its principle axes
lying along the three coordinate axes in some order determined by,
say, the ordering of the eigenvalues of the quadrupole moment.  Of
course, this specification is degenerate in the spherically symmetric
case, but then the angular orientation about the center of mass makes
no difference.

For our purposes below it is not necessary to be so precise, but I am
just illustrating how for a generic nearly periodic metric, it appears
to be possible to fix all of the coordinates completely.  Of course,
there is some arbitrariness in the procedure chosen for this (e.g.,
whether to take the time average of the spatial metric or of its
inverse or of some other matrix function of the spatial metric, and
how to define the time average over a finite time if the
non-oscillatory part of the metric is slowly varying).  But once a
sufficiently precise procedure is chosen, the coordinates are in
principle rigidly given, and hence so are the metric components for a
given spacetime in that coordinate system.  That is, the procedure
makes the resulting coordinates and metric components become
procedure-dependent but gauge-invariant functions over the spacetime.

As a result of all but the last parts of the procedure outlined above,
one can write a generic periodic or approximately periodic metric in
the form
 \begin{equation}
 ds^2 = - e^{2U(x^k)}dt^2 
  + e^{-2U(x^k)+2V(x^k)}\{[1+2W(t,x^k)]\delta_{ij}
                         +\sigma_{ij}(x^k)+h_{ij}(t,x^k)\}dx^idx^j,
 \label{eq:13}
 \end{equation}
where the time averages of the time-dependent quantities, that is the
scalar $W(t,x^k)$ and the traceless symmetric tensor $h_{ij}(t,x^k)$,
are all zero, and where the spatial coordinates are chosen to minimize
the integral over all space of the trace of the square of the
time-independent traceless symmetric tensor $\sigma_{ij}(x^k)$.

(If the metric is only approximately periodic, it may be able to be
written exactly in the form above for only a limited amount of time.
Alternatively, to be applicable for longer times, either the form
above may be only approximate, or else one might need to give $U$,
$V$, and $\sigma_{ij}$ some slow time dependence to deal with slow
nonperiodic changes in the geometry.)

Of course, there are many other similar forms for the metric from
similar procedures, such as having $W(t,x^k)$, $h_{ij}(t,x^k)$, and/or
$\sigma_{ij}(x^k)$, or suitable multiples of these quantities, as
arguments of an exponential, so I am not claiming that there is a
unique preferred form for a periodic or approximately periodic metric,
but only that the form above, or its slight generalization to the case
when $U$, $V$, and $\sigma_{ij}$ have some slow temporal variation, is
sufficient for our purposes.

Note that in the spherically symmetric case, to which I shall turn
later, the time-averaged spatial metric is necessarily conformally
flat by its spherical symmetry, so $\sigma_{ij}(x^k)=0$.  Both
time-dependent quantities, $W(t,x^k)$ and $h_{ij}(t,x^k)$, are
generically nonzero.  The spatial dependence of any scalar quantity is
a function of the one spatial function $r^2 = \delta_{ij}x^ix^j$, so
$U$ and $V$ are functions purely of the `radius' $r$, and $W$ is a
function of $t$ and of $r$.  With spherical symmetry, the traceless
tensorial quantity has the form $h_{ij} dx^i dx^j = h(t,r)(\delta_{ij}
dx^i dx^j - 3 dr^2)$ for some function $h(t,r)$ of both time and
radius whose time average is zero.

\section{Nearly-Newtonian Scalar Field Configurations}

In this paper I shall focus on self-gravitating configurations of one
or more massive scalar fields in which the gravitational field is very
weak (given to an adequate approximation by the linearized Einstein
equations), and the scalar fields have a very slow spatial dependence
(so the dominant piece of $\Psi_{IJ}$ has a very slow spacetime
dependence, though $\phi_{IJ}$ does have a temporal oscillation of
frequency roughly $m_I$ that is not considered slow, since slowness is
taken to be relative to these frequencies).  See \cite{HBG, GUL} for
previous analyses in this limit, which have been a motivation for some
of my choices of variables.

In this limiting case, the metric functions $U(x^k)$ and $W(t,x^k)$
are much smaller in magnitude than unity (but not negligible), and
$V(x^k)$, $\sigma_{ij}(x^k)$, and $h_{ij}(t,x^k)$ are negligibly
small.  Therefore, the metric (\ref{eq:13}) takes the form
 \begin{equation}
 ds^2 \approx - [1+2U(x^k)]dt^2 
  + [1-2U(x^k)+2W(t,x^k)]\delta_{ij}dx^idx^j.
 \label{eq:14}
 \end{equation}

If $W(t,x^k)=0$, then the approximate metric (\ref{eq:14}) would be
truly Newtonian, but the temporal oscillations of the scalar fields
give oscillating components of their stress-energy tensor and hence of
the Ricci tensor components (\ref{eq:10})-(\ref{eq:12}) and of the
metric (mainly at twice the frequencies of the fields themselves), so
$W(t,x^k)$ is nonzero even at the linearized gravity level.

As we shall see below, since the scalar fields are assumed to have
slow spatial dependences, a typical magnitude of $W(t,x^k)$ (say its
rms value at some spatial location where that is maximized, but of
course not its time average, which is zero at all spatial locations by
definition) is of the same order of magnitude as a typical magnitude
of the square of $U(x^k)$, which we have dropped in expanding the
exponential.  It is also of the same order of magnitude as a typical
value of $V(x^k)$, which we have also dropped.  Therefore, it may be
thought a bit of a cheat to include the $W(t,x^k)$ term in the metric
above but not the $U(x^k)^2$ and $V(x^k)$ terms, which are similar in
magnitude.

However, the point is that $W(t,x^k)$ is the largest time-dependent
term and is responsible for the dominant contribution to the classical
decay of oscillatons and of other self-gravitating real scalar field
configurations in the nearly-Newtonian limit.  The $U(x^k)^2$ and
$V(x^k)$ terms that have been dropped are smaller than the $U(x^k)$
time-independent terms that have been kept, and none of those terms
directly contributes to the classical decay.  Thus the philosophy is
that the metric (\ref{eq:14}) includes the dominant time-independent
corrections to the flat Minkowski metric (the $U(x^k)$ terms) and the
dominant time-dependent corrections to the flat metric (the $W(t,x^k)$
term, in the gauge in which $g_{00}$ is independent of $t$ by
construction).

Before going to periodic configurations (in the approximation of
neglecting the scalar field emission), let us consider the slight
generalization to the metric
 \begin{equation}
 ds^2 \approx - (1+2U)dt^2 
  + (1-2U+2W)\delta_{ij}dx^idx^j
 \label{eq:15}
 \end{equation}
in which $W$ has a time dependence at frequencies that are roughly
twice that of the $m_I$, but $U$ is now allowed to have some time
dependence that is even much slower than its slow spatial dependence.

Now, instead of Eqs. (\ref{eq:7})-(\ref{eq:9}), I shall take
 \begin{equation}
 \phi_{IJ} \approx \psi_{IJ}e^{-i m_I t} + \bar{\psi}_{IJ} e^{i m_I t}
 \label{eq:16}
 \end{equation}
without the restriction
$\dot{\psi}_{IJ}e^{-im_It}+\dot{\bar{\psi}}_{IJ}e^{im_It}=0$ that is
true for $\Psi_{IJ}$ from Eqs. (\ref{eq:7})-(\ref{eq:9}).
Instead, I shall assume that each $\phi_{IJ}$ is such that $\psi_{IJ}$
can be chosen to give $\phi_{IJ}$ approximately and also give
 \begin{equation}
 |\ddot{\psi}_{IJ}| \ll m_I|\dot{\psi}_{IJ}| \ll m_I^2|\psi_{IJ}|.
 \label{eq:17}
 \end{equation}

Then the Klein-Gordon equation in the metric (\ref{eq:15}) with $|W|
\ll |U| \ll 1$, which is
 \begin{equation}
 \ddot{\phi}_{IJ} \approx -m_I^2\phi_{IJ}-2m_I^2 U\phi_{IJ}
      + c^2\nabla^2 \phi_{IJ}
 \label{eq:KG}
 \end{equation}
when for the moment we ignore the $W$ term,
implies that each $\psi_{IJ}$ approximately obeys the
Schr\"{o}dinger equation
 \begin{equation}
 \dot{\psi}_{IJ}
 \approx {ic^2\over 2m_I} \nabla^2 \psi_{IJ} -i m_I U \psi_{IJ},
 \label{eq:18}
 \end{equation}
where $\nabla^2$ is the flat-space Laplacian, $\nabla^2 \psi_{IJ}\equiv
\delta^{ij}\psi_{IJ,ij}$.

Note that it is $m_{*I}c^2 U = \hbar m_I U$ that is the Newtonian
potential energy of the particle of `mass' $m_I$, and not $U$ itself
(which is dimensionless).  Also, by having $m_I$ have units of
frequency rather than conventional mass units, the explicit appearance
of $\hbar$ is avoided in Eq. (\ref{eq:18}).  This is what one would
expect, since this Schr\"{o}dinger equation came from the purely
classical Klein-Gordon equation for the real scalar field, rather than
from any quantum equation.  Note that I have chosen $\psi_{IJ}$ to be
dimensionless rather than, say, having the spatial integral of its
absolute square be unity (or perhaps some other positive integer), as
one would normally normalize the wavefunction of a truly quantum
Schr\"{o}dinger equation.

%One does need to set $c=1$ or else measure spatial distances in time
%units in order that the Laplacian acting on the dimensionless
%$\psi_{IJ}$ have the units of temporal frequency squared to match the
%other terms in the equation (with $U$ being dimensionless).  Or, one
%could insert a factor of $c^2$ in front of the Laplacian if it is taken
%with its conventional meaning of having the units of reciprocal
%distance squared, but I shall generally not bother doing that here.

It is most straightforward to regard the approximate equivalence
between the real second-order Klein-Gordon equation and the complex
first-order Schr\"{o}dinger equation (\ref{eq:18}) as a procedure that
works when one starts with a solution of the Schr\"{o}dinger equation
(\ref{eq:18}) (with a weak gravitational potential, $|U| \ll 1$) for
which $|\nabla^2 \psi_{IJ}| \ll m^2_I |\psi_{IJ}|$ (except near
possible zeros of $\psi_{IJ}$) and then uses Eq. (\ref{eq:18}) to
construct from it an approximate solution of the Klein-Gordon
equation.

In the reverse direction it is a bit more subtle.  If one uses
Eq. (\ref{eq:7}) to define $\Psi_{IJ}$ in terms of $\phi_{IJ}$ and its
time derivative, $\dot{\phi}_{IJ}$, this $\Psi_{IJ}$ itself will be
close to the solution $\psi_{IJ}$ of the Schr\"{o}dinger equation
(\ref{eq:18}).  However, since $\Psi_{IJ}$ will generically have a
small term roughly proportional to $e^{2im_It}$ as well as its dominant
term with a much slower time variation, the time derivative
$\dot{\Psi}_{IJ}$ will pick up a relatively significant contribution
from the term that is roughly proportional to $e^{2im_It}$ and so will
be significantly different from $\dot{\psi}_{IJ}$.  Thus $\Psi_{IJ}$
defined by Eq. (\ref{eq:7}) will not satisfy the Schr\"{o}dinger
equation (\ref{eq:18}).

However, one can instead define
 \begin{eqnarray}
 \psi_{IJ} &=& \Psi_{IJ} +{i\over 2m_I}\dot{\Psi}_{IJ} \equiv 
 {1\over 4m_I^2}e^{i m_I t}
        (m_I^2\phi_{IJ}+2im_I\dot{\phi}_{IJ}-\ddot{\phi}_{IJ})
 \nonumber \\
 &\approx& {1\over 2}e^{i m_I t}(\phi_{IJ}+{i\over m_I}\dot{\phi}_{IJ}
                     +U\phi_{IJ}-{c^2\over 2m_I^2}\nabla^2 \phi_{IJ}),
 \label{eq:19}
 \end{eqnarray}
where for the last expression I have used the approximate form
(\ref{eq:KG}) of the Klein-Gordon equation in the Newtonian part of
the metric to evaluate the second time derivative of $\phi_{IJ}$ in
terms of its value and its spatial Laplacian.  This $\psi_{IJ}$ then
obeys the Schr\"{o}dinger equation (\ref{eq:18}) when $U$ is small and
slowly varying and when $\phi_{IJ}$ is oscillating at nearly its
natural frequency $m_I$ and has a slow spatial variation in units of
$m_I$.

With this definition of the complex $\psi_{IJ}$ in terms of the real
$\phi_{IJ}$ and its derivatives, Eq. (\ref{eq:16}) is still a fairly
good approximation for $\phi_{IJ}$ in terms of $\psi_{IJ}$, but an
inversion of Eq. (\ref{eq:19}) that is accurate to one higher order is
 \begin{eqnarray}
 \phi_{IJ} &\approx& (\psi_{IJ}-{i\over m_I}\dot{\psi})e^{-i m_I t}
                + (\bar{\psi}_{IJ}+{i\over
                m_I}\dot{\bar{\psi}})e^{im_It}
 \nonumber \\
 &\approx& (1-U)[(\psi_{IJ}+{c^2\over 2m_I^2}\nabla^2\psi_{IJ})e^{-im_It}
   +(\bar{\psi}_{IJ}+{c^2\over 2m_I^2}\nabla^2\bar{\psi}_{IJ})e^{im_It}].
 \label{eq:20}
 \end{eqnarray}
Nevertheless, although this formula for $\phi_{IJ}$ is a more accurate
inversion of Eq. (\ref{eq:19}) than is Eq. (\ref{eq:16}), I don't know
that it really gives a more accurate solution of the Klein-Gordon
equation than Eq. (\ref{eq:16}) does from a solution of the
Schr\"{o}dinger equation (\ref{eq:18}).

After getting an approximate solution of the Klein-Gordon equation in
the nearly-Newtonian metric (\ref{eq:15}) with $|W| \ll |U| \ll 1$
(and temporarily ignoring the small effect of the tiny but
rapidly-time-varying $W$ term, which shall be discussed later), we
need to solve the Einstein equation for the effect of the
stress-energy tensor of the scalar fields on the metric.  To leading
order, the resulting Ricci-tensor components from the stress-energy
tensor and the Einstein equation are
 \begin{equation}
 R_{00} \approx \sum_{IJ}\{\dot{\phi}_{IJ}^2 -{1\over 2}m_I^2\phi_{IJ}^2\}
 \approx \sum_{IJ} \{m_I^2 [\psi_{IJ}\bar{\psi}_{IJ}
     -{3\over 2}(\psi_{IJ}^2 e^{-2i m_I t}
                           +\bar{\psi}_{IJ}^2 e^{-2i m_I t})]\},
 \label{eq:21}
 \end{equation}
 \begin{equation}
 R_{0i} \approx 0,
 \label{eq:22}
 \end{equation}
 \begin{equation}
 R_{ij} \approx \sum_{IJ}{1\over 2}m_I^2\delta_{ij}\phi_{IJ}^2
 \approx \sum_{IJ} m_I^2 \delta_{ij}[\psi_{IJ}\bar{\psi}_{IJ}
                           +{1\over 2}(\psi_{IJ}^2 e^{-2i m_I t}
                           +\bar{\psi}_{IJ}^2 e^{2i m_I t})].
 \label{eq:23}
 \end{equation}

The corresponding Einstein tensor components are simpler,
 \begin{equation}
 G_{00} \approx \sum_{IJ} 2m_I^2 \psi_{IJ}\bar{\psi}_{IJ},
 \label{eq:21e}
 \end{equation}
 \begin{equation}
 G_{0i} \approx 0,
 \label{eq:22e}
 \end{equation}
 \begin{equation}
 G_{ij} \approx -\sum_{IJ} m_I^2 \delta_{ij}(\psi_{IJ}^2 e^{-2i m_I t}
                           +\bar{\psi}_{IJ}^2 e^{2i m_I t}).
 \label{eq:23e}
 \end{equation}

This corresponds to an energy density that has only whatever slow time
variation the mass-squared-weighted sum of the squares of the absolute
values of the $\psi_{IJ}$'s may have, and an isotropic pressure that
oscillates at the frequencies $2m_I$ and has a time average that is
zero to this order of approximation (though at the next order there
are small time-independent pressure gradients that hold up the
self-gravitating energy density in the approximately periodic or
quasi-periodic cases).

Directly from the nearly-Newtonian metric (\ref{eq:15}) itself, with
the spatial derivatives of $W$ being negligibly small, the
leading-order linearized Einstein tensor components are
 \begin{equation}
 G_{00} \approx 2 c^2 \nabla^2 U,
 \label{eq:24}
 \end{equation}
 \begin{equation}
 G_{0i} \approx 0,
 \label{eq:25}
 \end{equation}
 \begin{equation}
 G_{ij} \approx -2\ddot{W}\delta_{ij}.
 \label{eq:26}
 \end{equation}

Thus the Einstein equation in the nearly-Newtonian case becomes
 \begin{equation}
 c^2 \nabla^2 U \approx \sum_{IJ} m_I^2 \psi_{IJ}\bar{\psi}_{IJ},
 \label{eq:27}
 \end{equation}
 \begin{equation}
 W \approx -{1\over 8}\sum_{IJ}(\psi_{IJ}^2 e^{-2i m_I t}
                           +\bar{\psi}_{IJ}^2 e^{2i m_I t}).
 \label{eq:28}
 \end{equation}

In summary, for nearly-Newtonian self-gravitating configurations of
real self-gravitating minimally coupled massive scalar fields, the
Einstein-Klein-Gordon equations become the coupled approximate partial
differential equations (\ref{eq:18}) and (\ref{eq:27}), which are the
time-dependent Newton-Schr\"{o}dinger or Schr\"{o}dinger-Newton
equations \cite{FLP,P,MPT,TM}, plus the additional algebraic equation
(\ref{eq:28}) for the small rapidly varying term $W$ in the
nearly-Newtonian metric (\ref{eq:15}).  The conditions for these
nearly-Newtonian equations to be valid are that $\sum_{IJ}|\psi_{IJ}|^2
\ll 1$, $|U| \ll 1$, and that the spatial derivatives of the
$\psi_{IJ}$'s are small in comparison with their typical values
multiplied by their natural frequencies $m_I$.  [Then the
Schr\"{o}dinger equation (\ref{eq:18}) implies that the time
derivatives of the $\psi_{IJ}$'s are also small in comparison with
their typical values multiplied by their natural frequencies $m_I$.]

Unless explicitly stated otherwise, in this paper we shall assume that
the metric is asymptotically flat with asymptotically Lorentzian
coordinates (except for a possible rescaling).  That is, we assume
that $U$ goes to a time- and direction-independent constant at spatial
infinity.  Eq. (\ref{eq:27}) then implies that the $\psi_{IJ}$'s must
all aymptotically tend to zero at spatial infinity.
Eqs. (\ref{eq:18}) and (\ref{eq:27}) are invariant under shifting $U$
by a constant or a function purely of $t$, provided the $\psi_{IJ}$'s
are shifted by the appropriate phase factor that is also a function
purely of $t$.  (This is simply the gauge transformation of replacing
the time coordinate $t$ with a new time coordinate $t'$ that is purely
a function of the old time coordinate $t$, and of rescaling the
spatial coordinates appropriately.)  One could thus set the asymptotic
value of $U$ to be zero, making the coordinates asymptotically
Lorentzian without any scaling factors.  However, in some cases it is
more convenient not to make this restriction, such as when the time
dependence of the $\psi_{IJ}$'s is purely by a time-dependent phase
factor, in which case one can cancel the phase factor and make the
$\psi_{IJ}$'s independent of time by an appropriate nonzero but
time-independent asymptotic value of $U$.

For the rest of this section, we shall take the approximate
Newton-Schr\"{o}dinger equations (\ref{eq:18}) and (\ref{eq:27}) as
exact and so write $=$ signs rather than $\approx$ equal signs.
However, we must bear in mind that these are actually only
approximations, valid in the nearly-Newtonian limit, for the actual
Einstein-Klein-Gordon equations taken as fundamental in this paper.

The time-dependent Newton-Schr\"{o}dinger equations (\ref{eq:18}) and
(\ref{eq:27}), now temporarily re-interpreted as exact equations, may
be derived from the classical action (cf. \cite{C})
 \begin{eqnarray}
 I_{NS}&=&{1\over 8\pi c^3}\int dt d^3x \sum_{IJ}
 [im_I(\bar{\psi}_{IJ}\dot{\psi}_{IJ}-\psi_{IJ}\dot{\bar{\psi}}_{IJ})
  -c^2(\nabla U)^2-c^2|\nabla\psi_{IJ}|^2-2m_I^2 U|\psi_{IJ}|^2]
 \nonumber \\
 &=&\int dt L_{NS}
 \label{eq:27b}
 \end{eqnarray}
(in units of time squared, to be multiplied by $c^5/\hbar G$ if one
wants a dimensionless action), where the classical Lagrangian (with
units of time) is
 \begin{eqnarray}
 L_{NS}&=&{1\over 8\pi c^3}\int d^3x \sum_{IJ}
 im_I(\bar{\psi}_{IJ}\dot{\psi}_{IJ}-\psi_{IJ}\dot{\bar{\psi}}_{IJ})
  -E_U-E_K-E_V
 \nonumber \\
 &=&{1\over 8\pi c^3}\int d^3x \sum_{IJ}
 im_I(\bar{\psi}_{IJ}\dot{\psi}_{IJ}-\psi_{IJ}\dot{\bar{\psi}}_{IJ})
  -MU_{\infty}-E_U-E_K-2E_P
 \nonumber \\
 &=&{1\over 8\pi c^3}\int d^3x \sum_{IJ}
 im_I(\bar{\psi}_{IJ}\dot{\psi}_{IJ}-\psi_{IJ}\dot{\bar{\psi}}_{IJ})
  -MU_{\infty}-E.
 \label{eq:27c}
 \end{eqnarray}

Here the asymptotic mass of the configuration (essentially the total
rest mass, uncorrected for gravitational binding energy), in time
units, is
 \begin{equation}
 M={1\over 4\pi c^3}\int d^3x \sum_{IJ} m_I^2 |\psi_{IJ}|^2
  ={G\over c^3}\int d^3x \rho
 \label{eq:27d}
 \end{equation}
with the rest mass density, in conventional units, being
 \begin{equation}
 \rho={1\over 4\pi G} \sum_{IJ} m_I^2 |\psi_{IJ}|^2.
 \label{eq:27dd}
 \end{equation}

When the Newton-Schr\"{o}dinger action $I_{NS}$ given by
Eq. (\ref{eq:27b}) is extremized, so that the Newton-Schr\"{o}dinger
equations (\ref{eq:18}) and (\ref{eq:27}) are satisfied, one can
readily see that the asymptotic form of the Newtonian potential is
 \begin{equation}
 U \sim U_{\infty} - {Mc\over r},
 \label{eq:27ddd}
 \end{equation}
where $M$ is the asymptotic value of the mass, given by
Eq. (\ref{eq:27d}) above (in time units, $G/c^3$ times the mass $M_*$
in conventional mass units, so $Mc$ has units of length and is half
the Schwarzschild radius corresponding to the mass), and $r$ is the
radial distance from the center of mass, also in units of length.

By Eq. (\ref{eq:18}) and the asymptotic boundary conditions given, the
mass $M$ is an approximately conserved quantity [approximate only to
the extent that Eq. (\ref{eq:18}) is approximate], corresponding under
the approximations used to the exactly conserved ADM mass of the
spacetime.  In some sense it is more nearly the rest mass of the
matter (in time units), but since only the zeroth-order approximation
is being used for the rest mass density and for the spatial volume
element, $M$ is only a zeroth-order approximation for the rest mass as
well.

The various other energies appearing in the Newton-Schr\"{o}dinger
Lagrangian (\ref{eq:27c}) (smaller than $M$ by factors that in
equilibrium are of the order of a typical value of $|U|$, which must
be much less than unity for the nearly-Newtonian approximation to be
valid) are the following quantities (in time units, inserting the
factor of $1/c$ into the right hand side of each of the equations with
squares of spatial gradients, and the factor of $1/c^3$ into those
without them, under the assumption that the spatial distances and
gradients are being measured in length units rather than in the time
units that would avoid the need for all factors of $c$): the
(positive) Newtonian potential gradient energy
 \begin{equation}
 E_U={1\over 8\pi c}\int d^3x (\nabla U)^2,
 \label{eq:27e}
 \end{equation}
the (positive) scalar field gradient energy or matter kinetic energy
 \begin{equation}
 E_K={1\over 8\pi c}\int d^3x \sum_{IJ}|\nabla\psi_{IJ}|^2,
 \label{eq:27f}
 \end{equation}
the (indefinite in sign, though negative in static equilibrium) matter
potential energy
 \begin{equation}
 E_V={1\over 8\pi c^3}\int d^3x \sum_{IJ}2m_I^2 U|\psi_{IJ}|^2
    ={G \over c^3}\int d^3 x \, \rho \, U,
 \label{eq:27g}
 \end{equation}
the (indefinite in sign, though also negative in static equilibrium)
rescaled gravitational potential energy
 \begin{eqnarray}
 E_P&=&{1\over 2}E_V-{1\over 2}MU_{\infty}
    ={1\over 8\pi c^3}\int d^3x \sum_{IJ} m_I^2
    (U-U_{\infty})|\psi_{IJ}|^2
 \nonumber \\
    &=&{G \over 2c^3}\int d^3x \,\rho\,(U-U_{\infty}),
 \label{eq:27h}
 \end{eqnarray}
and the total Newtonian energy
 \begin{eqnarray}
 E&=&E_U+E_K+E_V-MU_{\infty}=E_U+E_K+2E_P
 \nonumber \\
 &=&{1\over 8\pi c^3}\int d^3x \sum_{IJ}
  [c^2(\nabla U)^2+c^2|\nabla\psi_{IJ}|^2+2m_I^2 (U-U_{\infty})|\psi_{IJ}|^2].
 \label{eq:27i}
 \end{eqnarray}

The extrema of the Newton-Schr\"{o}dinger action $I_{NS}$ given by
Eq. (\ref{eq:27b}), with $U$ fixed to a constant value $U_{\infty}$ at
spatial infinity, and with the $\psi_{IJ}$ functions falling off
sufficiently rapidly at spatial infinity, are solutions of the
Newton-Schr\"{o}dinger equations (\ref{eq:18}) and (\ref{eq:27}).

When these equations of motion are satisfied, one can readily show (by
integration by parts, etc., using the boundary conditions given in the
previous sentence) that $M$ is a constant of motion (as mentioned
above), that
 \begin{equation}
 E_V=MU_{\infty}-2E_U,
 \label{eq:27j}
 \end{equation}
that
 \begin{equation}
 E_P=-E_U,
 \label{eq:27k}
 \end{equation}
that
 \begin{equation}
 E=E_K+E_P=E_K-E_U,
 \label{eq:27l}
 \end{equation}
and that the various energies (all of which have absorbed a factor of
$G/c^3$ to have units of time) have the following time derivatives:
 \begin{equation}
 \dot{E}_P={1\over 2}\dot{E}_V=-\dot{E}_U=-\dot{E}_K
 ={G\over c^3}\int d^3x \,\mathbf{J}\!\cdot\!\mathbf{\nabla}U,
 \label{eq:27m}
 \end{equation}
where
 \begin{equation}
 \mathbf{J}=\sum_{IJ}
 [-{im_Ic^2\over 8\pi G}
 (\bar{\psi}_{IJ}\mathbf{\nabla}\psi_{IJ}
  -\psi_{IJ}\mathbf{\nabla}\bar{\psi}_{IJ})]
 \label{eq:27n}
 \end{equation}
is the mass-current flux vector in the conventional units of mass per
area per time (when we remember that $m_I$ has units of inverse time
and that the $\psi_{IJ}$'s are dimensionless).

One can then see from Eqs. (\ref{eq:27l}) and (\ref{eq:27m}) that the
Newton-Schr\"{o}dinger equations (\ref{eq:18}) and (\ref{eq:27}),
along with the asymptotic boundary conditions, imply that the total
nonrelativistic Newtonian energy $E$ is conserved, or constant in
time.  Using the expressions (\ref{eq:27e}) and (\ref{eq:27f}) above,
and using Eq. (\ref{eq:27l}), we can see that Eq. (\ref{eq:27i}) may
be written, for $E$ in time units (or $E_* = c^5 E/G$ in conventional
energy units), as
 \begin{equation}
 E \equiv {GE_*\over c^5} = {1\over 8\pi c}
 \int d^3x \left(\sum_{IJ}|\nabla \psi_{IJ}|^2 -(\nabla U)^2\right)
 = \mathrm{const}.
 \label{eq:30}
 \end{equation}
This is a first-order correction to the rest mass energy of the
configuration in the total ADM mass-energy.  [But since
Eq. (\ref{eq:27d}) for $M$ is only correct for the rest mass to zeroth
order, the correct first-order expression for the ADM mass is not
simply $M+E$.]

If one has a static solution of the Newton-Schr\"{o}dinger equations
(\ref{eq:18}) and (\ref{eq:27}), in which $U$ and each $\psi_{IJ}^2$
is constant in time, then these are extrema of $E_U+E_K+E_V =
E+MU_{\infty}$.  To put it another way, they are extrema of the total
Newtonian energy $E$ with the constraint of fixed $M$.  If one finds
these extrema by the method of Lagrange multipliers, extremizing
$E-\lambda M$, then the Lagrange multiplier is $\lambda = -
U_{\infty}$, which thus is $dE/dM$ for a continuous sequence of such
extrema.

If one takes such an extremum static spatial configuration and
replaces $U(\mathbf{x})$ by
$\tilde{U}(\mathbf{x})=U(e^{\alpha}\mathbf{x})$ and each
$\psi_{IJ}(\mathbf{x})$ with
$\tilde{\psi}_{IJ}(\mathbf{x})=e^{\beta}\psi_{IJ}(e^{\alpha}\mathbf{x})$
for small constants $\alpha$ and $\beta$, then for the variation of
$E_U+E_K+E_V = E+MU_{\infty}$ to vanish to first order in both
$\alpha$ and $\beta$, and with the use also of
Eqs. (\ref{eq:27j})-(\ref{eq:27l}), one can readily see that the
static equilibrium configurations have
 \begin{equation}
 E_U=2E_K=-2E_V=-E_P={2\over 3}MU_{\infty}=-2E.
 \label{eq:27o}
 \end{equation}

The relation that the gravitational potential energy is twice the
negative of the kinetic energy, $E_P=-2E_K$, and hence that $E=-E_K$,
is just the usual virial relation for an attractive inverse-square
force.  The relation that $E=-(1/3)MU_{\infty}$ involves the scaling
behavior of the Schr\"{o}dinger equation and implies that for each
$\psi_{IJ}^2$ to be static (e.g., not to have any time-dependent phase
factor), one must have $U_{\infty} > 0$ (essentially for the system to
be bound).

A solution of the approximate time-dependent Newton-Schr\"{o}dinger
equations (\ref{eq:18}) and (\ref{eq:27}) is determined by an initial
spatial configuration of the $\psi_{IJ}$'s (and, for the quantities in
those equations, but not for the gauge-invariant physical quantities,
by the gauge choice of the asymptotic value of $U$).  If the scalar
fields all disperse indefinitely, then $U$ will tend to $U_{\infty}$
everywhere in space asymptotically with time so that the integral of
the negative $-(\nabla U)^2$ term in the integral (\ref{eq:30}) goes to
zero, and thus the Newtonian energy $E$ must be non-negative.

But if Eq. (\ref{eq:30}) gives a negative Newtonian energy $E$, then
the scalar fields cannot all disperse.  To the extent that
Eqs. (\ref{eq:18}) and (\ref{eq:27}) are accurate, at least some of
the scalar field energy must remain gravitationally bound
indefinitely.  Unless the scalar fields collapse gravitationally into
configurations violating the nearly-Newtonian approximation being used
(and perhaps leading to continued gravitational collapse into one or
more black holes), the scalar fields will continue to oscillate,
giving an oscillaton.

Negative Newtonian energy $E$ is thus a sufficient condition for an
oscillaton (or for gravitational collapse as a possible alternative)
in the nearly-Newtonian limit, though it is not a necessary condition,
as one can have an initial configuration which asymptotically tends to
a bound part with negative Newtonian energy and a dispersing part with
a larger positive Newtonian energy.  (See \cite{SS2} for an example of
this.)

In any case, we see that, at least in the nearly-Newtonian case,
gravitationally bound oscillatons are quite generic (unless all but a
set of measure zero collapse gravitationally into black holes, which
na\"{\i}vely seems unlikely), occurring if the single inequality $E <
0$ is true (and also in other cases when just part of the system is
bound).  That is, any perturbation of a nearly-Newtonian oscillaton
within a sufficiently small neighborhood in the space of perturbations
gives another oscillaton (unless it collapses).  In this sense
nearly-Newtonian oscillatons are apparently stable under small
perturbations, to the degree that Eqs. (\ref{eq:18}) and (\ref{eq:27})
are accurate and continue to remain valid (e.g., when one ignores the
scalar field radiation considered below, and when no continued
gravitational collapse occurs that takes one outside the validity of
these equations).

Of course, if one requires an oscillaton of fixed total mass (which
itself is a continuous variable classically, though quantized when one
includes the fact that the scalar field particles are quantized) to be
periodic with some definite period (say in proper time at spatial
infinity, to make the period gauge invariant), then one gets a
nonlinear eigenvalue problem with presumably only a discrete set of
eigensolutions for each mass (modulo gauge transformations, including
spatial translations and rotations and the gauge transformation of
shifting $U$).  Thus I would expect there to be discrete periodic
oscillatons embedded in an open set of nonperiodic oscillatons for a
given total mass.

However, if we include the effect of the oscillating term of the
nearly-Newtonian metric (\ref{eq:14}), the $W$ term given by
Eq. (\ref{eq:28}), we shall see in the next Section that it leads to
classical emission of each scalar field of mass $m_I$ at frequencies
roughly $2m_K\pm m_I$, so that generic oscillatons are {\it not}
classically stable but radiate away their energy and scalar fields.
The only exceptions appear to be field configurations that have a
$U(1)$ invariance, so that their total stress-energy tensor is
independent of time and the $W$ term vanishes.

\section{Classical Emission of Scalar Fields for Generic Situations}

The oscillating $W$ term of the nearly-Newtonian metric (\ref{eq:14})
gives, in the Klein-Gordon equation of the scalar field $\Phi_{IJ}$,
transitions from its dominant frequency near $m_I$ to frequencies near
$2m_K\pm m_I$ that radiate away, carrying off energy and causing a
generic oscillaton to decay.  Here we assume that the terms given in
Eq. (\ref{eq:16}) describe the non-radiating scalar field oscillations
of an oscillaton.

To describe the radiation, extend Eq. (\ref{eq:16}) to include scalar field
oscillations at these emitted frequencies, so the dimensionless
rescaled scalar field $\phi_{IJ} \equiv \sqrt{8\pi G/c^2}\,\Phi_{IJ}$
has the form
 \begin{eqnarray}
 \phi_{IJ}\!\! &\approx&\! \psi_{IJ}e^{-i m_I t}
 + \bar{\psi}_{IJ} e^{i m_I t}
 \nonumber \\
 &+&\!\!\sum_{K}(\chi_{IJK}e^{-i(2m_K+m_I)t}\!
                 +\! \bar{\chi}_{IJK}e^{i(2m_K+m_I)t}
       \!+\!\chi'_{IJK}e^{-i(2m_K-m_I)t}
       \!+\! \bar{\chi'}_{IJK} e^{i(2m_K-m_I)t}),
 \nonumber \\
 & &
 \label{eq:31}
 \end{eqnarray}
where not only the $\psi_{IJ}$'s and their complex conjugates, but
also the $\chi_{IJK}$'s and $\chi'_{IJK}$'s and their complex
conjugates, are functions varying much more slowly than the
frequencies $m_I$ and $m_K$.  Since fields that can radiate away must
have frequencies larger than $m_I$, in the sum above we can omit the
$\chi'_{IJK}$'s which have $m_K \leq m_I$.

Of course, the sum over $K$ includes $I$, so the radiation field
occurs even if there is only a single $m_I$ or even just a single real
scalar field (one choice for the indices ${IJ}$ denoting the field).
(Then we can omit the $\chi'_{IJK}$ and $\bar{\chi}'_{IJK}$ terms,
since their frequencies would be just $m_I$, and they would just give
small corrections to the $\psi_{IJ}$ and $\bar{\psi}_{IJ}$ terms that
are not radiating away by assumption.)

In principle one should include a whole infinite series of frequencies
by adding to $m_I$ all possible positive and negative integer multiples
of all $m_K$'s, but the additional terms will generally be smaller
yet, and so to a good approximation it is sufficient to consider only
the $\psi_{IJ}$, $\chi_{IJK}$, and $\chi'_{IJK}$ terms and their
complex conjugates.

Now the term with the $e^{-i(2m_K+m_I)t}$ time dependence in the
Klein-Gordon equation (\ref{eq:0}) for the field $\phi_{IJ}$ in the
metric (\ref{eq:15}) with the oscillating $W$ term included, given by
Eq. (\ref{eq:28}), yields
 \begin{equation}
 [\nabla^2+4m_K(m_K+m_I)]\chi_{IJK}
  \approx -{3\over 4}m_Im_K\sum_L \psi_{KL}^2\psi_{IJ},
 \label{eq:32}
 \end{equation}
and the term with the $e^{-i(2m_K-m_I)t}$ time dependence gives
 \begin{equation}
 [\nabla^2+4m_K(m_K-m_I)]\chi'_{IJK}
  \approx {3\over 4}m_Im_K\sum_L \psi_{KL}^2 \bar{\psi}_{IJ}.
 \label{eq:33}
 \end{equation}

We shall assume, except where explicitly discussed below, that there
are no scalar waves incoming from infinity, so we shall impose
outgoing wave boundary conditions on the $\chi_{IJK}$'s and
$\chi'_{IJK}$'s.  Then they have the form
 \begin{equation}
 \chi_{IJK}(\mathbf{x}) \approx {3m_Im_K\over 16\pi}
 \int d^3\mathbf{x}'
 {e^{2i\sqrt{m_K(m_K+m_I)}|\mathbf{x-x'}|}\over |\mathbf{x-x'}|}
 \sum_L \psi_{KL}^2(\mathbf{x}')\psi_{IJ}(\mathbf{x}'),
 \label{eq:34}
 \end{equation}
 \begin{equation}
 \chi'_{IJK}(\mathbf{x}) \approx -{3m_Im_K\over 16\pi}
 \int d^3\mathbf{x}'
 {e^{2i\sqrt{m_K(m_K-m_I)}|\mathbf{x-x'}|}\over |\mathbf{x-x'}|}
 \sum_L \psi_{KL}^2(\mathbf{x}')\bar{\psi}_{IJ}(\mathbf{x}').
 \label{eq:35}
 \end{equation}

These represent scalar waves that are propagating outward at
asymptotic speeds
 \begin{equation}
 v_{IJK}={2\sqrt{m_K(m_K+m_I)} \over 2m_K+m_I}
 \label{eq:36}
 \end{equation}
and
 \begin{equation}
 v'_{IJK}={2\sqrt{m_K(m_K-m_I)} \over 2m_K-m_I}
 \label{eq:37}
 \end{equation}
respectively, which generically are within a factor of the order of
unity of the speed of light (taken to be unity in these equations).

In contrast, the $\psi_{IJ}$'s are assumed to be localized almost
entirely within some region that we shall call the system.  Since the
$\psi_{IJ}$'s are assumed to be slowly varying with respect to the
natural frequencies $m_I$, we assume that the $\psi_{IJ}$'s within the
system do not change much during the light travel time across the
system, or during the time it takes for the waves represented by the
$\chi_{IJK}$'s and $\chi'_{IJK}$'s to traverse the system.  That is
why we can use an instantaneous approximation for the propagators in
the formulas (\ref{eq:34}) and (\ref{eq:35}) for the $\chi_{IJK}$'s
and $\chi'_{IJK}$'s.

If we surround the system by a sphere much larger than the dominant
region over which the $\psi_{IJ}$'s are significant, then we can
calculate the flux of mass out through that sphere in the scalar waves
represented by the $\chi_{IJK}$'s and $\chi'_{IJK}$'s given above.
When we average this over the oscillation periods and do a bit of
algebra that is not repeated here, we get a classical mass loss rate
of the oscillaton that is
 \begin{eqnarray}
 -{dM\over dt}\!\!&\approx &\!\! \sum_{IJKLM} {9m_I^2m_K^2 \over 256\pi^2}
 \int {d^3\mathbf{x} d^3\mathbf{x'}\over |\mathbf{x-x'}|}
 \nonumber \\
 &\times&\!\!\left[(2m_K\!+\!m_I)
    \sin(2\sqrt{m_K(m_K\!+\!m_I)}|\mathbf{x\!-\!x'}|)
             \psi_{KL}^2(\mathbf{x})\psi_{IJ}(\mathbf{x})
             \bar{\psi}_{KM}^2(\mathbf{x}')\bar{\psi}_{IJ}(\mathbf{x}')\right.
 \nonumber \\
      &+&\!\!\left.(2m_K\!-\!m_I)
          \sin(2\sqrt{m_K(m_K\!-\!m_I)}|\mathbf{x\!-\!x'}|)
             \psi_{KL}^2(\mathbf{x})\bar{\psi}_{IJ}(\mathbf{x})
             \bar{\psi}_{KM}^2(\mathbf{x}')\psi_{IJ}(\mathbf{x}')\right].
 \nonumber \\
      & &
 \label{eq:38}
 \end{eqnarray}
In the second of the two terms inside the integrand, whenever
$m_K-m_I$ is negative, that term is to be omitted, since the
corresponding $\chi'_{IJK}$ is spatially exponentially damped rather
than having the oscillatory behavior representing an outgoing wave.
(Of course, if $m_K-m_I=0$, this second term is just zero, so we need
not consider it in that case either.)

Remembering that I am using units and conventions in which $M$ has the
units of time, in which the $m_I$'s have the units of temporal
frequency (inverse time), in which the $\psi_{IJ}$'s and their complex
conjugates are dimensionless, and in which either the spatial
coordinates $x^i$ (represented above by the 3-vector $\mathbf{x}$)
have the units of time or else the speed of light is set equal to
unity, it is easy to see that both sides of Eq. (\ref{eq:38}) are
dimensionless.

Since the classical mass loss rate formula (\ref{eq:38}) is rather
complicated when there are several real massive scalar fields of
different masses, it may help to give it explicitly when there is only
one real scalar field of mass (or actually natural frequency) $m$:
 \begin{equation}
 -{dM\over dt} \approx {27 m^5 \over 256\pi^2}
 \int {d^3\mathbf{x} d^3\mathbf{x'}\over |\mathbf{x-x'}|}
 \sin(\sqrt{8}m|\mathbf{x-x'}|)\psi^3(\mathbf{x})\bar{\psi}^3(\mathbf{x}').
 \label{eq:39}
 \end{equation}

When there are two scalar fields of the same mass $m$, then one gets
 \begin{eqnarray}
 -{dM\over dt} &\approx& {27 m^5 \over 256\pi^2}
 \int {d^3\mathbf{x} d^3\mathbf{x'}\over |\mathbf{x-x'}|}
 \sin(\sqrt{8}m|\mathbf{x-x'}|)
 \nonumber \\
 &\times&[\psi_1^2(\mathbf{x})+\psi_2^2(\mathbf{x})]
 [\bar{\psi}_1^2(\mathbf{x}')+\bar{\psi}_2^2(\mathbf{x}')]
 [\psi_1(\mathbf{x})\bar{\psi}_1(\mathbf{x}')
 +\psi_2(\mathbf{x})\bar{\psi}_2(\mathbf{x}')].
 \label{eq:40}
 \end{eqnarray}
In particular, when $\psi_1$ and $\psi_2$ each have the same magnitude
everywhere and are everywhere $90^{\circ}$ out of phase, or $\psi_1=\pm
i\psi_2$ (so the two real fields can be interpreted as forming a
single complex field with a global $U(1)$ symmetry), then the mass
loss rate is zero.  This is a case in which the stress-energy tensor
of the scalar fields does not have an oscillating component, and so
there is no oscillating $W$ term in the metric (\ref{eq:15}).

There is also the analogous case in which one has an arbitrary number
(greater than one) of fields of each mass $m_I$, when one has zero
everywhere for the sum of the squares of the $\psi_{IJ}$'s (not the
squares of the absolute values of these quantities) for each fixed $I$
(for each different mass $m_I$, the sum being over $J$ that labels the
different fields with fixed $I$ and hence with fixed mass $m_I$).
Then again the stress-energy tensor and the metric (\ref{eq:15}) has
no oscillating term, and so there is no generation of outgoing
$\chi_{IJK}$ or $\chi'_{IJK}$ waves.

In these cases, if the $\psi_{IJ}$'s are not stationary (or stationary
up to a time-dependent phase factor) but are slowly changing their
form, then although there may be no oscillations at the frequencies
$m_I$ or their sums or differences, the metric would still have a slow
time dependence, and this would presumably lead to some scalar field
radiation, though at an amount presumably considerably reduced from
what it would be the case if the sum of the squares of the
$\psi_{IJ}$'s for at least one $I$ were different from zero so that
there would be the more rapid oscillations in the metric.

The only case in which I would expect absolutely no scalar radiation
would be the case in which the metric is absolutely stationary.
Otherwise it would seem extremely unlikely that the outgoing radiation
at all possible multiples of the metric oscillation frequency, plus or
minus the natural frequency of the fields that can potentially be
emitted, would be zero.  However, I have not tried to find a rigorous
proof that there are not exceptional counter-examples to this
conjecture.

\section{Classical Emission of Scalar Fields with \\ Spherical Symmetry}

A simple subset of the set of all oscillatons is the set in which the
metric and all of the scalar fields have spherical symmetry.  In the
nearly-Newtonian limit (which I am taking to exclude gravitational
waves), the spherical symmetry of the metric follows from the
spherical symmetry of the scalar fields, and the spherical symmetry of
the scalar fields (including the outgoing waves) follows from the
spherical symmetry of the $\psi_{IJ}$'s.  Therefore, we basically just
need to assume that each $\psi_{IJ}=\psi_{IJ}(t,r)$.

The Klein-Gordon equation (\ref{eq:0}) for the scalar field implies
that $\psi_{IJ}$ obeys the Schr\"{o}dinger equation (\ref{eq:18}),
which for the spherical symmetric case becomes
 \begin{equation}
 \dot{\psi}_{IJ} \approx {ic^2\over 2m_Ir} (r\psi_{IJ})'' -i m_I U \psi_{IJ},
 \label{eq:41}
 \end{equation}
where, except for the prime on $\chi'_{IJK}$ and on dummy variables
inside integrals, a prime henceforth denotes a partial derivative with
respect to $r$ (or later, with respect to a rescaled radial variable
$x=kr/c$).  One can avoid the explicit $r$'s in this equation by
defining
 \begin{equation}
 f_{IJ}(t,r) \equiv r\psi_{IJ}(t,r),
 \label{eq:42}
 \end{equation}
which makes the Schr\"{o}dinger equation take the form
 \begin{equation}
 \dot{f}_{IJ} \approx {ic^2\over 2m_I} f_{IJ}'' -i m_I U f_{IJ}.
 \label{eq:43}
 \end{equation}

The Newtonian part of the Einstein equations, Eq. (\ref{eq:27}), can
be integrated in the spherically symmetric case to give at each time
 \begin{equation}
 U \approx U_{\infty}
 -{1\over c^2 r}\int_0^r dr' \int_{r'}^{\infty}dr''r''
   \sum_{IJ}m_I^2|\psi_{IJ}(r'')|^2.
 \label{eq:44}
 \end{equation}
Here of course the primes and double primes on the $r$'s inside the
integrals just denote dummy variables to be integrated over, and not
derivatives with respect to $r$.

The spherically symmetric analogues of (\ref{eq:34}) and (\ref{eq:35})
for the $\chi_{IJK}$'s and $\chi'_{IJK}$'s obeying Eqs. (\ref{eq:32})
and (\ref{eq:33}) (to solve the Klein-Gordon equation) are (with $c=1$
for simplicity here and in many formulas following)
 \begin{eqnarray}
 \chi_{IJK}(r)&\approx& {3m_Im_K\over 8\sqrt{m_K(m_K+m_I)}\,r}
 \nonumber \\
 &\times&\left[e^{2i\sqrt{m_K(m_K+m_I)}\,r}
 \int_0^{\infty}dr'r'\sin{(\sqrt{m_K(m_K+m_I)}\,r)}
 \sum_L \psi_{KL}^2(r')\psi_{IJ}(r')\right.
 \nonumber \\
 &-&\left.\int_r^{\infty}dr'r'\sin{(\sqrt{m_K(m_K+m_I)}\,(r'-r))}
 \sum_L \psi_{KL}^2(r')\psi_{IJ}(r')\right],
 \label{eq:45}
 \end{eqnarray}
 \begin{eqnarray}
 \chi'_{IJK}(r)&\approx& {3m_Im_K\over 8\sqrt{m_K(m_K-m_I)}\,r}
 \nonumber \\
 &\times&\left[e^{2i\sqrt{m_K(m_K-m_I)}\,r}
 \int_0^{\infty}dr'r'\sin{(\sqrt{m_K(m_K-m_I)}\,r)}
 \sum_L \psi_{KL}^2(r')\bar{\psi}_{IJ}(r')\right.
 \nonumber \\
 &-&\left.\int_r^{\infty}dr'r'\sin{(\sqrt{m_K(m_K-m_I)}\,(r'-r))}
 \sum_L \psi_{KL}^2(r')\bar{\psi}_{IJ}(r') \right].
 \label{eq:46}
 \end{eqnarray}

The classical mass loss rate becomes
 \begin{eqnarray}
 -{dM \over dt} &\approx& \sum_{IJK} {9m_I^2m_K^2 \over 16}
 \nonumber \\
 &\times&\left[{2m_K+m_I \over 2\sqrt{m_K(m_K+m_I)}}
 \left|\int_0^{\infty}dr r \sin{(2\sqrt{m_K(m_K+m_I)}r)}
 \sum_L \psi_{KL}^2(r)\psi_{IJ}(r) \right|^2\right.
 \nonumber \\
 &+&\left.{2m_K-m_I \over 2\sqrt{m_K(m_K-m_I)}}
 \left|\int_0^{\infty}dr r \sin{(2\sqrt{m_K(m_K-m_I)}r)}
 \sum_L \psi_{KL}^2(r)\bar{\psi}_{IJ}(r) \right|^2 \right].
 \nonumber \\
 & &
 \label{eq:47}
 \end{eqnarray}
As in Eq. (\ref{eq:38}), in the second term that has the factors of
$\sqrt{m_K(m_K-m_I)}$, if $m_K-m_I$ is negative, the corresponding
term in the sum over modes is to be omitted, since it corresponds to a
$\chi'_{IJK}$-mode frequency $2m_K-m_I$ that is below the natural
frequency $m_I$ and hence to a mode that is not freely propagating at
large radial distance to carry off mass.

When only one real scalar field of mass $m$ is present,
the classical mass loss rate becomes
 \begin{equation}
 -{dM \over dt} \approx {27m^4 \over 2^{11/2}}
 \left|\int_0^{\infty}dr \, r \, \sin{(\sqrt{8}mr)} \, \psi^3(r) \right|^2.
 \label{eq:48}
 \end{equation}

With two scalar fields of identical masses present, one gets
 \begin{equation}
 -{dM \over dt} \!\approx\! {27m^4 \over 2^{11/2}}
 \left[
 \left|\int_0^{\infty}\!dr r
 \sin{(\!\sqrt{8}mr)}(\psi_1^2\!+\!\psi_2^2)\psi_1\right|^2
 \!+\!\left|\int_0^{\infty}\!dr r
 \sin{(\!\sqrt{8}mr)}(\psi_1^2\!+\!\psi_2^2)\psi_2\right|^2
 \right].
 \label{eq:49}
 \end{equation}
Again one can readily see that if the sum of the squares of the
complex $\psi$ fields with identical masses are zero, then there is no
classical mass loss, at least at this level of approximation.  With
just two fields of identical masses, this condition is that $\psi_1 =
\pm i\psi_2$.

\section{The Simplest Spherical Oscillatons}

Now let us focus on the simplest case, in which there is a single
massive scalar field in a finite-mass spherically symmetric
nearly-Newtonian configuration that is very nearly periodic in time
and has no nodes.  In particular, require that $\psi$ be spherically
symmetric, independent of time (except for a possible slowly varying
phase factor), and nowhere zero (no nodes, though asymptotically zero
at spatial infinity).

By a suitable choice of the hypersurfaces of constant time, one can
cancel the phase factor to make $\psi$ real and positive everywhere,
which is what I shall assume, at the cost of having $U$ approach a
nonzero constant at spatial infinity.  Then the time-dependent
Newton-Schr\"{o}dinger equations (\ref{eq:18}) and (\ref{eq:27}) for a
single scalar field with a single real $\psi$ become the
time-independent Newton-Schr\"{o}dinger equations \cite{FLP,P,MPT,TM}
(where for simplicity I am using units in which $c=1$ so that I can
drop many occurrences of factors of $c$ that one can easily put back
in by dimensional analysis if needed with other choices of units),
 \begin{equation}
 \nabla^2 \psi \approx 2m^2 U \psi,
 \label{eq:50}
 \end{equation}
and
 \begin{equation}
 \nabla^2 U \approx m^2 \psi^2.
 \label{eq:51}
 \end{equation}

In the case of spherical symmetry which I am now also assuming, the
equations take the form (with $\psi$ real) of two coupled second-order
ordinary differential equations,
 \begin{equation}
 \psi''+{2\over r}\psi' \approx 2m^2 U \psi,
 \label{eq:52}
 \end{equation}
and
 \begin{equation}
 U''+{2\over r}U' \approx m^2 \psi^2.
 \label{eq:53}
 \end{equation}

As noted previously \cite{FLP,MPT}, this system of equations has the
scale invariance
 \begin{equation}
 (\psi,U,r) \mapsto (\lambda^2\psi,\lambda^2 U, \lambda^{-1}r).
 \label{eq:54}
 \end{equation}
Inserting explicitly the speed of light $c$ so that the radius $r$ can
have units of length instead of time as it implicitly does above, I
shall set
 \begin{equation}
 r = {cx \over k},
 \label{eq:55}
 \end{equation}
 \begin{equation}
 \psi = {k^2 S \over \sqrt{2} m^2},
 \label{eq:56}
 \end{equation}
 \begin{equation}
 U = {-k^2 V \over 2 m^2}
 \label{eq:57}
 \end{equation}
so with $k$ having the units of frequency or inverse time, $x$, $S$,
and $V$ are the dimensionless variables used by \cite{MPT} (except
that what I now call $x$, they call $r$).  Note that this $V$, which
is just a rescaling of the Newtonian potential $U$ with its sign
reversed, is not to be confused with the $V(t,x^k)$ in the metric
(\ref{eq:13}), which is negligible in the nearly-Newtonian metric.

Although so far I have used a prime generally to denote a derivative
with respect to the radius $r$, when I am using the dimensionless
rescaled radial variable $x$ instead as the independent variable, I
shall use a prime to denote a derivative with respect to $x$.  Then
Eqs.(\ref{eq:52}) and (\ref{eq:53}) (where the prime denoted $d/dr$)
become the dimensionless time-independent Newton-Schr\"{o}dinger
equations
 \begin{equation}
 (xS)''=-xSV
 \label{eq:58}
 \end{equation}
and
 \begin{equation}
 (xV)''=-xS^2.
 \label{eq:59}
 \end{equation}
Note that I have now replaced the $\approx$ signs with $=$ signs, even
though these equations are only an approximation to the actual
Einstein-Klein-Gordon equations, though an approximation that becomes
arbitrarily good in the nearly-Newtonian limit.

These time-independent Newton-Schr\"{o}dinger equations are the same
as Eqs. (6a) and (6b) of \cite{MPT}, except for the replacement of
their radius $r$ by my dimensionless rescaled $x$.  The fact that the
arbitrary constant $k$ does not appear in these equations illustrates
their scale invariance.

Another form of the equations that is helpful for some of the analysis
below is to use
 \begin{equation}
 p \equiv {1\over 2}(S-V)
 \label{eq:60}
 \end{equation}
instead of $V$ in Eqs. (\ref{eq:58}) and (\ref{eq:59}), which then
become
 \begin{equation}
 (xS)''=-xS(S-2p)
 \label{eq:61}
 \end{equation}
and
 \begin{equation}
 (xp)''=xSp.
 \label{eq:62}
 \end{equation}

Yet another set of variables to use that were particularly convenient
for numerical integration of the equations and for the interpretation
of the results are
 \begin{equation}
 X \equiv xS,
 \label{eq:63}
 \end{equation}
 \begin{equation}
 \mathcal{M} \equiv -x^2 V',
 \label{eq:64}
 \end{equation}
 \begin{equation}
 w \equiv -(xV)' = {\mathcal{M}\over x} - V.
 \label{eq:65}
 \end{equation}
The variable $X$ is the same as that used in \cite{TM}.  The
asymptotic value of $\mathcal{M}(x)$, say $\mathcal{M}_{\infty}$, is
what is called $B$ in \cite{MPT} and what is called $I$ in Eq. (2.19)
of \cite{TM}.  The asymptotic value of $w(x)$, say $w_{\infty}$, which
is positive, is by Eq. (\ref{eq:65}) the same as the asymptotic value
of $-V(x)$, say $-V_{\infty}$, which is the same as $-A$ in
\cite{MPT}.

$\mathcal{M}(x)$ has the interpretation of the rescaled mass interior
to the sphere at $x$.  From the fact that asymptotically one has $U
\sim U_{\infty}-M/r$ and $V \sim V_{\infty}+\mathcal{M}/x$ (assuming a
finite-mass oscillaton in which the mass-energy density, proportional
to $S^2$, asymptotically rapidly approaches zero), one can see that
the mass (in units of time) interior to a sphere of radius $r=cx/k$ is
 \begin{equation}
 M(r)={k\over 2 m^2}\mathcal{M}(x).
 \label{eq:66}
 \end{equation}

The variable $w$ can be written as
 \begin{equation}
 w={2 m^2\over k^2}(U+rg),
 \label{eq:67}
 \end{equation}
where $g$ is the acceleration of gravity at the same radius $r$ where
the gravitational potential $U$ is evaluated,
 \begin{equation}
 g = {dU\over dr} = {M(r)\over r^2}.
 \label{eq:68}
 \end{equation}
Thus $w$ can be interpreted as a rescaled value that the gravitational
potential $U$ would have at twice the radius $r$ if $U$ had a uniform
gradient from $r$ to $2r$, with this uniform gradient having the same
value of the actual gradient at radius $r$.  If the mass-energy
density dropped precisely to zero outside some radius, $w$ would be
constant outside this radius, at the value
$w=-V_{\infty}=(2m^2/k^2)U_{\infty}$.  In the actual case in which the
mass-energy density, proportional to $S^2$, drops exponentially toward
zero, $w$ is exponentially close to $-V_{\infty}$ and is therefore a
good integration variable to use to evaluate the asymptotic value of
$V$, that is $V_{\infty}$.

The differential equations in terms of these variables are
 \begin{equation}
 X'' = (w-{\mathcal{M}\over x})X \equiv -VX,
 \label{eq:69}
 \end{equation}
 \begin{equation}
 \mathcal{M}'=X^2,
 \label{eq:70}
 \end{equation}
 \begin{equation}
 w'={X^2 \over x}.
 \label{eq:71}
 \end{equation}

The initial conditions for these variables (initial in $x$, of course,
not in time, since all of the quantities being considered presently
are independent of time in the approximation that we initially ignore
the mass decay rate) are that at $x=0$, we have $X=0$, $X'=S_0$,
$\mathcal{M}=0$, and $w=-V_0$.

As discussed in \cite{MPT,TM}, for an everywhere regular static
spherically symmetric solution to the Newton-Schr\"{o}dinger
equations, $S$ and $V$ must be smooth everywhere, and $S$ must be
decreasing exponentially at spatial infinity ($x \rightarrow \infty$).
At the origin ($x=0$), $S$ and $V$ must have finite values, $S_0$ and
$V_0$ respectively, and must have zero dimensionless radial
derivatives, $S'=0$ and $V'=0$.

Because of the scale invariance, the only independent nontrivial
parameter for a solution regular at the origin is the ratio $S_0/V_0$.
Integrating out from the origin gives a solution that diverges at
finite radius (with $S$ going to $+\infty$ and with $V$ going to
$-\infty$ there) if $V_0 \leq 0$, so we shall choose $V_0 > 0$.  Using
the scale invariance, without loss of generality we can and shall set
$V_0=1$, leaving the nontrivial parameter to be $S_0$.  If $S_0=0$,
then we just get the trivial solution $S=0$, $V=V_0$, which is flat
spacetime with no matter, a solution we shall discard as previously
studied by other people.  By the symmetry of the equations under $S
\mapsto -S$, we can thus choose $S_0 > 0$.

If $S_0$ is too large, Eq. (\ref{eq:59}) implies that $xV$ (initially
growing as $x$) bends down rapidly, so that $V$ goes negative while
$X=xS$ is still growing. (Initially $X=xS$ also grows linearly with
$x$, as $S_0x$, but, like $xV$, it also bends down.  However,
Eq. (\ref{eq:58}) implies that it does not bend down so fast as $xV$
bends down, for $S>V>0$.)  Then when $V$ becomes negative, $X'$ grows
with $x$, and so $X$ grows faster and faster, and eventually so does
$S$, by Eq. (\ref{eq:58}) or the equivalent Eq. (\ref{eq:69}).
Eq. (\ref{eq:59}) imples that then $xV$ and eventually also $V$ gets
more and more negative.  In fact, then $S$ goes to $+\infty$ and $V$
goes to $-\infty$ at a singularity of infinite mass at finite $x$.  Of
course, we shall discard these solutions.

On the other hand, if $S_0$ is positive but too small (e.g., less than
$V_0=1$ \cite{TM}), Eq. (\ref{eq:59}) implies that $V$ will stay
positive long enough for Eq. (\ref{eq:58}) to imply that $xS$ will
oscillate (with characteristic period in the $x$-variable of
$2\pi/\sqrt{V}$ if $V$ were constant).  However, we want the regular
solution with no nodes, the solution with the largest value of $S_0$
that does not lead to a singularity.  This value of $S_0$ is an
eigenvalue for the system.

At this eigenvalue, $X=xS$ will bend over from increasing at $x=0$
($X'=S_0$ there) to decreasing again toward zero value asymptotically,
but never crossing zero.  $xV$ will also bend over from increasing (at
unit rate) at $x=0$ (since we have chosen $V_0=1$) and will cross zero
to become negative and will eventually keep decreasing at the
asymptotically rate given by the asymptotic negative value of $V$, say
$V_{\infty}$.  That is, $S$ will start out at $S_0$ with zero
derivative with respect to $x$ ($S'=0$ initially) but will then bend
down to approach zero asymptotically at large $x$ (while then bending
back upward just enough to keep from crossing zero, but never quite
leveling out, except asymptotically).  Similarly, as a function of $x$
from $x=0$ to $x=\infty$, $V$ will start out at $V_0=1$ also with zero
derivative with respect to $x$ ($V'=0$ initially) but will then bend
down and cross zero before bending back up to level off toward some
negative constant asymptotic value $V_{\infty}$.

For this eigensolution, since $X$ starts at zero and initially
increases linearly with $x$, and since $\mathcal{M}'=X^2$ with
$\mathcal{M}(0)=0$, and since $w$ starts at $-V_0=-1$ with $w'=X^2/x$,
we have initially (near the origin, $x \ll 1$),
 \begin{equation}
 \mathcal{M} \sim {1\over 3}S_0^2 x^3 - {1\over 15}S_0^2 x^5,
 \label{eq:72}
 \end{equation}
 \begin{equation}
 w \sim -1 + {1\over 2}S_0^2 x^2,
 \label{eq:73}
 \end{equation}
 \begin{equation}
 X \sim S_0 x - {1\over 6}S_0 x^3,
 \label{eq:74}
 \end{equation}
 \begin{equation}
 S \sim S_0 - {1\over 6}S_0 x^2,
 \label{eq:75}
 \end{equation}
 \begin{equation}
 V \sim 1 - {1\over 6}S_0^2 x^2,
 \label{eq:76}
 \end{equation}
 \begin{equation}
 p \sim {1\over 2}(S_0-1)(1 + {1\over 6}S_0 x^2).
 \label{eq:77}
 \end{equation}

Then as $x$ is increased to $\infty$, $X$ will at first increase,
while bending downward and eventually passing a maximum and then
decreasing.  While decreasing, $X$ will pass an inflection point (at
the point at which $V$ crosses below zero) and will then bend upward
to level out asymptotically as it also approaches zero asymptotically.
At the same time (here meaning during the same evolution in $x$ from 0
to $\infty$), $\mathcal{M}$ will start from 0 with zero slope and
curvature and initially grow as the cube of $x$ (i.e., as the volume
interior to the sphere of radius $r=cx/k$) but eventually will reach
an inflection point (at the point at which $X$ reaches its maximum)
and then gradually level off to approach its asymptotic value
$\mathcal{M}_{\infty}$.  The variable $w$ will start at $-1$ at $x=0$
with zero slope and will bend up to cross zero, before bending
downward to level off asymptotically and approach its asymptotic value
$w_{\infty}=-V_{\infty} > 0$.

For use below in calculating the quantum decay rate of this type of
oscillaton, it is also of interest to integrate the variable $D(x)$
given by the differential equation
 \begin{equation}
 D' = x^2 S^4 = {X^4 \over x^2}
 \label{eq:78a}
 \end{equation}
with the boundary condition $D(0)=0$.  Since the mass density in
conventional units is (with $m$ and $k$ in frequency units)
 \begin{equation}
 \rho = {m^2|\psi|^2 \over 4\pi G} = {k^4 S^2 \over 8\pi G m^2},
 \label{eq:79a}
 \end{equation}
the asymptotic value of $D$,
namely $D_{\infty}$, is proportional to the integrated square of the
mass-energy density (and hence to the total annihilation rate of two
scalarons into two gravitons):
 \begin{equation}
 D_{\infty} = {16\pi m^4 \over k^5} \int (G\rho)^2 d^3 x.
 \label{eq:80a}
 \end{equation}

I have used the differential equation routine of Maple 8 to evaluate
to high accuracy the eigenvalue $S_0$ and the asymptotic values
$\mathcal{M}_{\infty}$ and $w_{\infty}$, making many overnight and
several-day runs on a Dell 2350 PC with the Maple ``Digits'' parameter
set to values as high as 62 (but more often 40 or 50) and the
``relerr'' and ``abserr'' parameters set as small as $10^{-35}$.  The
calculations performed roughly $10^8$ function evaluations in a
one-day run, so several billion function evaluations were performed.

I did face the annoyance that after completing a calculation, Maple 8
would often crash when I attempted to save the file and would erase
the entire file (even if I had saved the pre-run version).  I had to
get used to renaming and saving files before making runs and to
copying down key results by hand, and then copying and pasting small
pieces of the files before attempting to save the entire file, so that
I would have something left in case the entire file crashed and
disappeared.  This meant that by the end of the calculations, I had
thousands of digits copied down by hand, and hundreds of different
files saved.

The differential equation routine handled the regular singular point
of the differential equations above at $x=0$, so I could simply set
the initial conditions there.  However, it could not handle the
essential singularity at $x=\infty$, so I could only do the
integration out to a finite value of $x$.  Any errors in $S_0$ and in
the finite-differencing technique would lead to an $S(x)$ that did not
continue to approach zero indefinitely but rather would diverge away
from zero and eventually go to $+\infty$ (say if $S_0$ were set too
large) or to $-\infty$ (say if $S_0$ were set too small).  Figure 1 of
\cite{MPT} illustrates this phenomenon.

Therefore, for large values of the radial variable $x$, I developed an
asymptotic series solution in terms of decaying and growing
exponentials and in terms of an expansion in terms of $1/x_f$, where
$x_f$ is the final value integrated to (generally in the range 30-50),
for estimating the exponentially-growing error terms (from an
inaccurate $S_0$ or from numerical errors at smaller $x$), for getting
rid of their dominant effects, and for estimating what the corrected
solution would give (through five orders in $1/x_f$) for
$\mathcal{M}_{\infty}$, $w_{\infty}$, and $D_{\infty}$.

Once I had found $S_0$ to a fairly high degree of accuracy, I assumed
that the error in $X$ that I could thus estimate at some $x_f$ was
linear in the error of $S_0$, an assumption that fit the data very
well when I tried four slightly different values of $S_0$ (e.g.,
differing at the 35th digit).  Then I could interpolate to a value of
$S_0$ that would be predicted to give no error.  The interpolations to
four predicted values of $S_0$, from four of the six pairs of four
trial integrations at the highest accuracy settings, agreed to 45
digits, and I shall give my best estimate for $S_0$ to 42 digits,
using a well-known answer from elsewhere in choosing this number of
digits.  (The high agreement of the interpolated values for $S_0$
suggested that the random numerical errors from the numerical
differencing were much smaller than the values of $10^{-35}$ I set in
these multi-day computer calculations for the numerical routine
parameters ``relerr'' and ``abserr.'')

However, this agreement does not take into account a possible
systematic error from the numerical differencing.  From other trials
at larger values of ``relerr'' and ``abserr,'' I crudely estimated
that the resulting interpolation for $S_0$ might have an error around
$10^{-35}$, though it might be smaller.  Therefore, although I am
giving my estimate of $S_0$ to 42 digits below and believe that value
of $S_0$ is the best value (to that number of digits) to use in the
routine I used in Maple 8 with the accuracy parameters as I set them,
I am not very confident about the accuracy of the last few digits
given if one were able to integrate the differential equations without
numerical errors.  Therefore, the last few digits should be taken as
merely conjectural.

Then from the estimates from each of the four most accurate trials (at
very slightly different trial values of $S_0$ near the true value) for
$\mathcal{M}_{\infty}$, $w_{\infty}$, and $D_{\infty}$ (each of which
attempted to correct for the exponentially growing error term, which
was a very small fraction of the total for $X$ at the values $x_f=35$,
$x_f=40$, $x_f=45$, and $x_f=50$ that I used, at least for the trial
values of $S_0$ that were sufficiently close to my final estimate), I
could also do linear interpolations to what their values would be
expected to be at the correct value of $S_0$ (again neglecting a
possible systematic error from the numerical differencing).  The four
interpolations agreed for $\mathcal{M}_{\infty}$ through 36 digits,
for $w_{\infty}$ through 38 digits, and for $D_{\infty}$ through 44
digits (more digits in this case because the integrand for $D$ falls
exponentially twice as fast as the integrands for
$\mathcal{M}_{\infty}$ and $w_{\infty}$).  However, because of a
possible systematic error from the numerical differencing, these
interpolated asymptotic values are not likely to be accurate to these
numbers of digits.  Based on a comparison with trials at larger values
of ``relerr'' and ``abserr,'' I suspect that they are accurate through
about 32 digits.  Nevertheless, I shall give them to 36 digits, again
with the last few digits being just my best conjectures from the long
calculations.

The number of runs at small values of ``relerr'' and ``abserr'' was
limited by the large amount of running time for the Maple 8 program to
achieve its acuracy criteria.  For example, one of the final four runs
I did, all four with ``Digits'' set to 50 and ``relerr'' and
``abserr'' each set to $10^{-35}$, took 3.8 days to get just to
$x_f=35$ (where I terminated the calculation), though I had asked it
to give the results at a sequence of $x_f$'s out to 50.  (A final run,
using a value of $S_0$ that agreed to 40 digits with my best estimate
after using the results of that run, did get to $x_f=50$ within 1.6
days.  I'm not sure why it went further and faster.)  So on my PC it
would be too time-consuming for me to try to give the results to much
more accuracy than what I am giving below.

The values I obtained were
 \begin{equation}
 S_0
 \approx 1.088\,637\,079\,429\,044\,995\,770\,660\,314\,259\,096\,461\,243\,20,
 \label{eq:79}
 \end{equation}
 \begin{equation}
 \mathcal{M}_{\infty}
 \approx 3.618\,701\,237\,823\,656\,810\,212\,940\,073\,900\,601\,34,
 \label{eq:80}
 \end{equation}
 \begin{equation}
 w_{\infty}
 \approx 1.065\,731\,278\,365\,451\,058\,714\,678\,859\,310\,396\,16,
 \label{eq:81}
 \end{equation}
 \begin{equation}
 D_{\infty} \approx 1.320\,680\,334\,028\,957\,063\,595\,721\,629\,052\,089\,05.
 \label{eq:82}
 \end{equation}

As noted above, I think these numbers are likely to be correct to at
least 30-35 digits, but I am not really sure about the last several
digits, which should be interpreted as just my best current guess for
what they are.

One can see that my value for $S_0$ confirms all but the last of the
15 digits given for this quantity by \cite{MPT}.  (And if $S_0$ were
larger by just $4.23 \times 10^{-18}$ than it seems to be, the last
digit of the best 15-digit approximation would have indeed rounded up
to the 5 that \cite{MPT} gave.  However, modulo some highly unexpected
conspiracy in numerical errors, I am fairly confident that my value is
correct to at least 30 digits, since I could get agreement to more
digits even when I changed from numerically integrating one form of
the set of differential equations to numerically integrating another
form.)

From these numbers, one can of course construct various combinations
of them, such as the scale-invariant quantity $A/B^2$ that \cite{MPT}
discuss:
 \begin{equation}
 {A\over B^2} = -{w_{\infty}\over \mathcal{M}_{\infty}^2}
 \approx -0.081\,384\,603\,921\,072\,995\,209\,339\,465\,504\,462\,035\,7.
 \label{eq:84}
 \end{equation}
The first two nonzero digits of this quantity seems to agree with the
value \cite{MPT} plotted in their Figure 4, but they do not list its
numerical value.

From the scaling relation given by Eq. (\ref{eq:66}), we can get the
small dimensionless mass parameter of the nearly-Newtonian oscillaton,
 \begin{equation}
 \mu \equiv Mm ={k\over 2m} \mathcal{M}_{\infty}.
 \label{eq:85}
 \end{equation}
Since in the end we want to express other properties of the oscillaton
in terms of $\mu$, we shall actually invert this to get the scaling
parameter $k$ (which has units of frequency, as does $m$) as
 \begin{equation}
 k = {2m\mu\over \mathcal{M}_{\infty}}.
 \label{eq:86}
 \end{equation}

For example, the value of the Newtonian potential at infinity is
 \begin{equation}
 U_{\infty} = -{k^2\over 2m^2}V_{\infty}
 = {2 w_{\infty}\over \mathcal{M}^2_{\infty}}\mu^2
 \approx 0.162\,769\,207\,842\,145\,990\,418\,678\,931\,008\,924\,071\,\mu^2
 \label{eq:87}
 \end{equation}

We can then express the fractional binding energy of a
nearly-Newtonian oscillaton as
 \begin{equation}
 {E\over M} = -{1\over 3}U_{\infty} = -{k^2\over 6m^2}V_{\infty}
 = -{2 w_{\infty}\over 3\mathcal{M}^2_{\infty}}\mu^2 = -\epsilon\mu^2
 \label{eq:88}
 \end{equation}
with
 \begin{equation}
 \epsilon = {2 w_{\infty}\over 3\mathcal{M}^2_{\infty}}
 \approx 0.054\,256\,402\,614\,048\,663\,472\,892\,977\,002\,974\,690\,5.
 \label{eq:89}
 \end{equation}

If we take the number of scalar particles to be
 \begin{equation}
 N = {M_* \over m_*} = {c^5 M \over \hbar G m},
 \label{eq:90}
 \end{equation}
so the conventional rest mass of the oscillaton is $N$ times the
conventional rest mass of the scalar field quantum, $M_* = N m_*$,
then the conventional total energy of the oscillaton is, to first
order in $\mu^2$,
 \begin{equation}
 E_{\mathrm{tot}} = N m_* c^2(1+E/M)
 = N m_* c^2 - N^3 m_*^5 \epsilon G^2/\hbar^2.
 \label{eq:91}
 \end{equation}

This would be the same value for a boson star with a complex scalar
field having a $U(1)$ symmetry.  For that problem the value was
calculated by Ruffini and Bonazzola \cite{RB} over 34 years ago,
getting $\epsilon = 0.1626$.  Except for the last digit, this
corresponds to three times the value above.

One can also compare my numerical results with the 5-place results for
a boson star by Friedberg, Lee, and Pang \cite{FLP} (for $n=0$ nodes).
In terms of my calculated parameters and my numerical results, their
calculated parameters would be
 \begin{equation}
 \hat{\gamma}_0 = -V_0/S_0 = 1/S_0
 \approx 0.918\,579\,771\,804\,638\,252\,351\,006\,743\,940\,091\,299\,098\,45,
 \label{eq:92}
 \end{equation}
 \begin{equation}
 \hat{\gamma}_{\infty} = w_{\infty}/S_0
 \approx 0.978\,959\,194\,486\,001\,441\,125\,618\,083\,933\,781\,532,
 \label{eq:93}
 \end{equation}
 \begin{equation}
 \hat{\gamma}_1 = \mathcal{M}_{\infty}/S_0^{1/2}
 \approx 3.468\,256\,171\,397\,572\,159\,904\,340\,931\,700\,591\,17.
 \label{eq:94}
 \end{equation}
They got $\hat{\gamma}_0=-0.91858$, $\hat{\gamma}_{\infty}=0.97896$,
and $\hat{\gamma}_1=3.46826$, in perfect agreement with my results
rounded to five digits after the decimal point.

\section{Classical Emission from the Simplest Spherical Oscillatons}

Now that the nodeless spherically symmetric nearly-Newtonian
configurations have been found (determined by the single scaling
parameter $\mu = Mm$), we need to use Eq. (\ref{eq:48}) to evaluate the
classical mass loss rate.  By using Eqs. (\ref{eq:55}), (\ref{eq:56}),
and (\ref{eq:86}), we can write the classical mass loss rate, or power
emitted in classical scalar radiation (dimensionless when the mass $M$
is in time units, i.e., $M=GM_*/c^2$ in terms of the mass $M_*$ in
conventional mass units), as
 \begin{equation}
 P_c \equiv -\dot{M}_{\mathrm{classical}}
 \approx {27\mu^8 \over \sqrt{2}\mathcal{M}_{\infty}^8} F^2,
 \label{eq:95}
 \end{equation}
 \begin{equation}
 F = \int_0^{\infty}dx \, x \, S^3(x) \sin{(ax)},
 \label{eq:96}
 \end{equation}
 \begin{equation}
 a = {\sqrt{2}\mathcal{M}_{\infty} \over \mu}.
 \label{eq:97}
 \end{equation}

Because the nearly-Newtonian configurations have $\mu \ll 1$, the
$\mu$-dependent constant $a$ is very large, $a \gg 1$.  Therefore, the
$\sin{(ax)}$ factor in the integral (\ref{eq:96}) for $F$ oscillates
very rapidly and nearly washes out the integral for large $a$, causing
$F$ to be very small.

We can estimate the value of $F$ by the following method of contour
integration: Since $S(x)$ is an even function of $x$, the integrand is
an even function of $x$, and so the integral along the real axis from
0 to $\infty$ may be replaced by half of the integral along the real
axis from $-\infty$ to $+\infty$.  Then the real variable $x$ may be
extended to the complex variable $z$, as a function of which the
integrand is analytic except at poles of $S(z)$.  One may split
$\sin{(az)} = 0.5i e^{-iaz} - 0.5i e^{iaz}$ into the first
exponential, which drops exponentially along the negative imaginary
axis for the complex $z$, and the second exponential, which drops
exponentially along the positive imaginary axis for the complex $z$.
By splitting up the integral into the corresponding two pieces, the
first piece may be replaced by a contour integral making a clockwise
loop around the lower half plane (Im $z < 0$), and the second piece
may be replaced by a contour integral making a counterclockwise loop
around the upper half plane (Im $z > 0$), which gives an equal
contribution.

Now one of these contour integrals, say the second one (since they
each give equal contributions), may be evaluated by finding the
residues at each of the poles of the integrand.  There are a series of
poles of $S(z)$ running up the imaginary $z$ axis.  Because of the
$e^{iaz}$ factor from $\sin{(az)}$, the dominant residue will come
from the pole closest to the real azis, say at $z=iy_0$.

From an analysis \cite{MPT,TM} of the time-independent
Newton-Schr\"{o}dinger equations in their dimensionless form
(\ref{eq:58}) and (\ref{eq:59}), one can see that the solutions are
analytic over the complex $z$-plane, except for movable (moving if
$S_0$ were changed) double poles with coefficients $-6$.  In
particular, near the pole at $z=iy_0$, $S$ has the asymptotic form
 \begin{equation}
 S(z) \sim {-6 \over (z-iy_0)^2}.
 \label{eq:98}
 \end{equation}

Because it is $S^3(z)$ that appears in the integral (\ref{eq:96}) for
$F$, which thus has a 6th-order pole at $z=iy_0$, one must integrate
this factor by parts five times, giving the 5th derivative of the
$e^{iaz}$ factor, and hence 5 powers of $a$, in the dominant term of
the result.  (I shall drop terms with lower powers of $a$ that arise
from differentiating the $z$ factor rather than the $e^{iaz}$ factor,
since they involve higher powers of the small quantity $\mu$.)  As a
result, one finds that
 \begin{equation}
 F \approx -1.8\pi a^5 y_0 e^{-ay_0}.
 \label{eq:99}
 \end{equation}

Thus there is one more parameter that must be determined numerically
before we can evaluate explicitly the dominant term (for $\mu \ll 1$)
in the classical mass loss rate, namely the value $y_0$ that locates
the pole in $S(z)$ at $z=iy_0$ that is above the real axis but nearest
to it.

To find $y_0$, one can integrate the time-independent
Newton-Schr\"{o}dinger equations (\ref{eq:58}) and (\ref{eq:59}) up
the imaginary $z$-axis (after replacing the real radial variable $x$
with the complex radial variable $z$).  That is, one can set $z=iy$
and integrate these equations from $y=0$ to the singularity at
$y=y_0$.  If one now wants a prime to denote a derivative with respect
to $y$ rather than with respect to $x$ as it does in
Eqs. (\ref{eq:58}) and (\ref{eq:59}), one merely needs to reverse the
sign on the right hand sides of these equations.  To have only real
quantities in the resulting equations, one could also replace each $x$
that occurs once on each side of each of these equations with $y$, so
that one has purely real equations to integrate along the $y$-axis.
Because both $S(z)$ and $V(z)$ are even functions of their argument,
they both remain real when one evaluates them along the imaginary
$z$-axis.

An alternative way to do this, getting a single real equation for both
real $z$ and purely imaginary $z$, would be to define $v=-z^2$ to use
instead of $x$ or $y$ as the independent variable.  Then to get the
nearly-Newtonian oscillaton configuration for real radius, one would
integrate from $v=0$ along the negative real $v$-axis, whereas to find
the first pole at the imaginary value of the radius, one would
integrate from $v=0$ along the positive real $v$-axis.  The dependent
variables $S(v)$ and $V(v)$ would both be real along both the negative
and positive branches of the real $v$-axis, though they are not even
functions of $v$ as they are of $z$.

Of course, one cannot integrate numerically precisely to the pole, so
one must instead integrate to a point near it and then use the
behavior there to estimate where the pole is.  To minimize the
numerical error in this process, I thought it would be better to
redefine the dependent variables so that most of them (all but $r$
below, where this dependent variable $r$ is not to be confused with
the radial distance $r$ that is imaginary for this part of the
calculation) do not blow up at the movable singularity.  I used
$v=-z^2$ as the independent variable and the following four dependent
variables:
 \begin{equation}
 p = {1 \over 2}(S-V),
 \label{eq:100}
 \end{equation}
 \begin{equation}
 q = \sqrt{6 \over S},
 \label{eq:101}
 \end{equation}
 \begin{equation}
 r = {S'-V' \over 4z},
 \label{eq:102}
 \end{equation}
 \begin{equation}
 s = -{\sqrt{6}S' \over 4 z S^{3/2}}.
 \label{eq:103}
 \end{equation}

Then the time-independent Newton-Schr\"{o}dinger equations
(\ref{eq:58}) and (\ref{eq:59}) (with $x$ replaced by $z$ and a prime
then denoting $d/dz$) give the following four coupled
first-order differential equations:
 \begin{equation}
 {dp\over dv} = -r,
 \label{eq:104}
 \end{equation}
 \begin{equation}
 {dq\over dv} = -s
 \label{eq:105}
 \end{equation}
 \begin{equation}
 v{dr\over dv} = {3\over 2} \left({p\over q^2}-r \right),
 \label{eq:106}
 \end{equation}
 \begin{equation}
 v{ds\over dv} = {3 - 12 v s^2 \over 4 q} - {3\over 2}s - {1 \over 4}pq.
 \label{eq:107}
 \end{equation}

The initial conditions for these equations are that at $v=0$, one has
 \begin{equation}
 p(0) = {S_0-1 \over 2},
 \label{eq:108}
 \end{equation}
 \begin{equation}
 q(0) = \sqrt{6 \over S_0},
 \label{eq:109}
 \end{equation}
 \begin{equation}
 r(0) = {S_0(S_0-1) \over 12},
 \label{eq:110}
 \end{equation}
 \begin{equation}
 s(0) = {1 \over \sqrt{24S_0}}.
 \label{eq:111}
 \end{equation}

These dependent variables were chosen so that $q$ goes to zero at the
pole, linearly in $v$, and so that $p$ and $r$ have a negligible
effect on the evolutions of $q$ and its negative derivative $s$ near
the pole.

By defining the variable $u = q/\sqrt{v}$, which like $q$ goes to zero
at the pole, and developing series solutions in terms of $u$ at points
sufficiently near the pole that $u$ is very small, one can find that
the location of the pole at $v=y_0^2$ is given to high accuracy (when
$u \ll 1$) by
 \begin{equation}
 y_0 \approx \sqrt{v}[1+u+0.2u^2-0.02u^3+0u^4-0.0018u^5-0.00256u^6+O(u^7)].
 \label{eq:112a}
 \end{equation}

Therefore, I continued to use the Maple 8 differential equation
routine to integrate Eqs. (\ref{eq:104})-(\ref{eq:107}) from $v=0$,
where I used the initial conditions (\ref{eq:108})-(\ref{eq:111}), out
to various values very near $v=y_0^2$ where $u$ was quite small, and
then I used the values of $u$ and $v$ there to estimate $y_0$, the
location of the pole, by Eq. (\ref{eq:112a}).

Actually, unlike the case of Eqs. (\ref{eq:69})-(\ref{eq:71}), which
Maple 8 handled with the initial values given at the regular singular
point $x=0$, Maple 8 gave me trouble trying to start integrating
Eqs. (\ref{eq:104})-(\ref{eq:107}) from their regular singular point
at $v=0$, so I had to develop series solution near that point and
start the numerical integration from a slightly nonzero value of $v$,
say $v_1$.  I needed to choose $v_1$ small enough that the series
solution was sufficiently accurate at this starting point, and yet not
too small, or else the effects of the regular singular point increased
the numerical error in integrating from that tiny value of $v$ to a
more reasonable value.  I found that the precise value of $v_1$ had
very little effect on the first 36 digits of the result for $y_0$ if
it were around $v_1 = 10^{-16}$, so I generally used that starting
value for the independent variable $v$.

I also had to choose a final value of $v$, say $v_2$, very near the
singularity, so that Eq. (\ref{eq:112a}) would be accurate in
estimating the precise location of the singularity, and yet not so
near that the numerical errors from integrating too near the
singularity would be significant.  I tried about 20 values for $v_2$
and chose the value for $v_2$ so that the resulting $y_0$ did not
not vary much with shifting $v_2$ a small amount either way from the
value I thus chose for it.  Again the result for $y_0$ showed
stability to about 36 digits when $v_2$ was chosen in a suitable
range, such as between 14.843\,684\,26 and 14.843\,684\,27.

Of course, there is likely to be some systematic error beyond the
random error that caused the results to fluctuate in roughly the 37th
digit when I changed $v_1$ and $v_2$ slightly, so I am not confident
that I really have determined $y_0$ to 36-digit accuracy, but since
I got a preferred estimate to this many digits from my numerical
calculations, let me give the pole location $y_0$, like
$\mathcal{M}_{\infty}$, $w_{\infty}$, and $D_{\infty}$, to 36 digits:
 \begin{equation}
 y_0 \approx 3.852\,750\,221\,596\,529\,691\,939\,909\,038\,965\,636\,80.
 \label{eq:112}
 \end{equation}
Let me repeat my warning that the last few digits of this result are
uncertain.

Now that we have the last parameter, $y_0$, that we need to evaluate
$F$ by Eq. (\ref{eq:99}), we can go back and plug the result into
Eqs. (\ref{eq:95}) and (\ref{eq:97}) to give the classical mass loss
rate as
 \begin{equation}
 P_c \equiv -\dot{M}_{\mathrm{classical}}
 \, \approx \, {C \over \mu^2} e^{-{\alpha/\mu}},
 \label{eq:113}
 \end{equation}
where
 \begin{equation}
 \alpha = \sqrt{8}\mathcal{M}_{\infty} y_0
 \approx 39.433\,795\,197\,160\,163\,093\,719\,521\,465\,426\,369\,6
 \label{eq:114a}
 \end{equation}
and where
 \begin{equation}
 C = {2^{3/2} 3^7 \over 5^2} \pi^2 \alpha^2
 \approx 3\,797\,437.776\,333\,014\,909\,099\,808\,643\,511\,322\,13
 \label{eq:114b}
 \end{equation}
As with the other numerical quantities given above, the last few of
the 36 digits given for $\alpha$ and $C$ are actually uncertain.

I should emphasize that this is what I believe to be just the dominant
term in the classical emission of scalar waves from a nearly-Newtonian
oscillaton when $\mu \ll 1$.  I would expect this expression to have a
relative error of the order of $\mu$.

To put it another way, one can presumably write
 \begin{eqnarray}
 \ln{1 \over P_c} &=& {\alpha \over \mu} -2\ln{1 \over \mu} -\ln{C}
 +O(\mu)
 \nonumber \\
 &\approx& {39.433\,795\,197\,160\,163\,093\,719\,521\,465\,426\,369\,6 \over \mu}
 -2\ln{1 \over \mu}
 \nonumber \\
 && \ - 15.149\,837\,127\,888\,728\,198\,841\,232\,066\,408\,315\,377 +O(\mu),
 \label{eq:115}
 \end{eqnarray}
so that the calculations performed here have given the three leading
terms in an expansion for $\ln{(1/P_c)}$.

It is beyond the scope of this paper to do the nonlinear gravitational
calculations to find the classical mass loss rate when $\mu$ is not
small, but if one neglects the $O(\mu)$ corrections, one can make a
very crude estimate for the mass loss rate even up to the maximum
value of $\mu$, say $\mu_{\rm max}$.

For what are generally called boson stars (stationary spherical
configurations of a complex massive scalar field whose phase rotates
in a circle in the complex plane, equivalent to two real scalar fields
oscillating $90^{\circ}$ out of phase), the maximum value of $\mu$ is
0.633 \cite{FLP}.  For oscillatons of a real massive scalar field, the
initial calculations \cite{SS} gave $\mu_{\rm max} \approx 0.6$.
Ure\~{n}a-L\'{o}pez \cite{U-L} has given $\mu_{\rm max} \simeq 0.5522$,
but Alcubierre {\it et al} \cite{ABGMNUL} have given $\mu_{\rm max} =
0.607$, so I do not know what the correct answer is, and I have not
attempted to calculate it myself.  However, this uncertainty is no
doubt small in comparison with the error of applying
Eq. (\ref{eq:113}) or (\ref{eq:115}) to $\mu = \mu_{\rm max}$.

For example, if the coefficient of the $O(\mu)$ term of
Eq. (\ref{eq:115}) were the same magnitude (but of uncertain sign) as
the coefficient of the $1/\mu$ term, namely $\alpha \approx 39.4338$, then
just taking this single term with, say $\mu = 0.633$, would change
$\ln{(1/P_c)}$ by roughly $\pm 25.0$, or a total range of roughly 50.0
for this quantity, giving an uncertainty in the mass decay rate by a
factor of about $e^{50.0} \sim 5\times 10^{21}$.  One might hope that
the uncertainty is a lot less, but without calculating the $O(\mu)$
and higher terms, I don't see how one can be sure.

Despite this proviso, if we did naively insert the boson star maximum
mass parameter $\mu_{\rm max} \approx 0.633$ into Eq. (\ref{eq:113}),
we would get a mass loss rate (dimensionless, since our masses denoted
by $M$ have the factor of $G/c^3$ inserted to give them the dimension
of time) of about $8\times 10^{-21}$.  However, if it could be larger
or smaller by a factor of roughly $e^{25.0} \sim 7\times 10^{10}$, the
dimensionless mass loss rate could be as large as roughly $6\times
10^{-10}$ or as small as roughly $10^{-31}$.

Similarly, if $\mu_{\rm max} \approx 0.607$, one would get a crude
estimates of the mass loss rate as $6\times 10^{-22}$, and a crude
guess that it might be as large as $1.6\times 10^{-11}$ or as small as
$2.5\times 10^{-31}$; if $\mu_{\rm max} \approx 0.5522$, one would
get a crude estimates of the mass loss rate as $1.2\times 10^{-24}$,
and a crude guess that it might be as large as $3\times 10^{-15}$ or
as small as $4\times 10^{-34}$.

In any case, unless the coefficient of $\mu$ in the $O(\mu)$ term of
Eq. (\ref{eq:115}) (or actually this entire correction term divided by
$\mu$) were negative and had a larger magnitude than the coefficient
of the $1/\mu$ term, it seems that the dimensionless mass loss rate is
always less than about $10^{-9} \mu$.  This means that during a
one-radian change in the phase of the scalar field oscillation (a time
$t = 1/m$), the oscillaton would have lost less than one-billionth of
its mass.  If instead the correction to applying Eq. (\ref{eq:113}) to
$\mu = \mu_{\rm max}$ were negligible, then during the $2\pi$-longer
period of a full scalar field oscillation, even a maximum-mass
oscillaton would have lost less than a billionth of a billionth of its
mass.

This result shows why the numerical analyses to date have not shown
any instability of the oscillaton, since its mass loss rate is so low.

Figure 6 of \cite{ABGMNUL} for $\mu = 0.5726$ shows an apparent numerical
mass loss rate of about $3\times 10^{-9} \sim 5\times 10^{-9}\mu$
in dimensionless units, but since this is several times higher than
the crude guess above for the upper limit for the mass loss rate, and
is $\sim 2\times 10^{14} \sim e^{33} \sim e^{58\mu}$ times larger
than what my rashly-interpolated formula would predict for that $\mu$,
namely $P_c \sim 1.4\times 10^{-23}$, I suspect that the authors are
indeed correct in attributing this to ``a small amount of numerical
dissipation still present in our numerical method.''  However, I can't
completely rule out the possibility that the $O(\mu)$ term in
Eq. (\ref{eq:115}) is roughly $-58\mu$ at $\mu=0.5726$, so that
there conceivably might be mass loss comparable to that given by
Figure 6 of \cite{ABGMNUL}.

Later we shall find that for sufficiently small $\mu$, the classical
mass loss rate (\ref{eq:113}) is dominated by a quantum mass loss
rate.  However, when the classical mass loss rate dominates, and when
$\mu \ll 1$, the time $t_2-t_1$ to evolve from $\mu_1$ to $\mu_2$ is
approximately
 \begin{equation}
 t_2-t_1 \approx {\mu_2^4 \over \alpha C m} e^{\alpha/\mu_2}
                -{\mu_1^4 \over \alpha C m} e^{\alpha/\mu_1}.
 \label{eq:116}
 \end{equation}

For example, we could define $t(\mu)$ to be the time to decay from
$\mu=\mu_{\rm max}$ to some smaller value of $\mu$.  Then if $\mu_{\rm
max}-\mu \gg \mu^2_{\rm max}/\alpha \sim 0.00934$, then the magnitude
of the first term on the right hand side of Eq. (\ref{eq:116}) (with
$\mu_2 = \mu$) is much greater than the magnitude of the second term
(say with $\mu_1=\mu_{\rm max} \approx 0.607$).  Since this is
necessarily the case for $\mu \ll 1$ where Eq. (\ref{eq:116}) is
applicable, we may then drop the second term and say that the time to
decay down to $\mu \ll 1$ from $\mu_{\rm max}$ by the classical
emission of scalar radiation is
 \begin{equation}
 t(\mu) \approx {\mu^4 \over \alpha C m} e^{\alpha/\mu}.
 \label{eq:117}
 \end{equation}

\section{Precisely-Periodic but Infinite-Mass Oscillatons}

Although it is beyond the scope of the present paper, which just gives
numerical results for $\mu \ll 1$, it would be of interest to be able
to calculate the function $t(\mu)$ for all $\mu < \mu_{\rm max}$.  To
calculate `exact' results (i.e., exact up to numerical errors in
solving the differential equations), one needs a precise definition of
$\mu_{\rm max}$ and of $\mu(t)$ for the time thereafter.

It is rather hard to define $\mu_{\rm max}$ precisely (and the initial
oscillaton configuration that gives this maximum mass), since any
initial configuration is losing mass (assuming boundary conditions of
no incoming scalar waves), so one could start with a wide variety of
initial configurations.  But essentially one would like to start with
one out of a set of initial configurations that lose mass as slowly as
possible for each initial mass, and then choose the maximum-mass
element from the set members that do not have rapid mass loss (e.g.,
at a rate roughly given by the dynamical timescale $1/m$).  If the
initial decay timescale is of the order of $[\mu_{\rm max}^4/(\alpha C
m)] e^{\alpha/\mu_{\rm max}}$, as the weak-field formulas would
suggest, then to the degree that this is much larger than $1/m$, one
can define the maximum-mass initial configuration to that accuracy,
i.e., with a relative error that would be expected to be of the order
of $(\alpha C/\mu_{\rm max}^4) e^{-\alpha/\mu_{\rm max}}$.  However,
to define it to more accuracy (e.g., `precisely') would require more
care.

Although I do not see any highly preferred way to make the definition
precise, I can propose the following {\it ad hoc} method, which leads
to a consideration of precisely-periodic but infinite-mass oscillatons:

Temporarily relax the condition of no incoming waves that has been
fundamental to the discussion so far.  Then it appears that one can
have precisely periodic oscillatons, though now with infinite mass, so
that the spacetime is not quite asymptotically flat (although having
curvatures falling off fairly rapidly, so, for example, the spatial
integral of the Kretschman invariant, the total 4-fold contraction of
the square of the Riemann tensor, is finite at any time).

The idea of these precisely periodic (though infinite mass)
oscillatons is that not only does the oscillation of the scalar field
at frequency $m$ (with respect to a suitably scaled coordinate time
$t$ in a gauge or choice of coordinates in which $g_{0i}=0$ and
$g_{00}$ is independent of $t$) have precisely the right phase to drop
exponentially to zero at large spatial distance by the gravitational
binding of that mode, but also the oscillations at all odd multiples
of $m$ also have precisely the right phases to drop exponentially to
zero at large spatial distance by the gravitational binding of those
modes as well, when one takes into account the gravitational field not
only of the mode at frequency $m$ but also the gravitational field of
all the higher-frequency modes.

That is, one has a sequence of relations for the modes at each
frequency that start as follows for the lowest mode:  The mode with
frequency $m$ is chosen to have the right phase (e.g., the right
relation of its initial-in-$r$ value at $r=0$, determined by the
quantity $S_0$ above, to the the value of $g_{00}$ there, determined
by the quantity $V_0$ above in the weak-field or nearly-Newtonian
limit), so that once the stress-energy tensor of the scalar field
(predominantly from this mode at small radii) causes $-g_{00}$ to rise
above unity at sufficiently large radius $r$ (so that this mode there
has a proper-time frequency that is less than its natural frequency
$m$ and hence has a concave or exponential radial dependence for
larger values of $r$), one has only the asymptotically exponentially
decaying behavior of the mode at larger values of $r$.

However, the oscillation of the $g_{ij}$ components of the metric (at
frequencies that are even multiples of $m$) couples, via the
Klein-Gordon equation for the minimally coupled massive scalar field,
the modes of the scalar field that have frequencies that are different
odd multiples of $m$.  Thus, for example, the mode with frequency $3m$
is excited and propagates out from the region of the oscillaton where
the mode with frequency $m$ dominates the stress-energy tensor.

So far, this is just like the decaying oscillaton described above.  If
there are no incoming waves at frequency $3m$, the outgoing waves
carry off energy from the oscillaton, which decays (and hence is not
precisely periodic in time).

On the other hand, for the precisely periodic but infinite-mass
oscillaton, there are both incoming and outgoing waves at frequency
$3m$, and their stress-energy tensor (which at sufficiently large $r$
dominates over that of the mode of frequency $m$, since that
$m$-frequency mode is asymptotically exponentially decaying at
sufficiently large $r$) eventually causes $-g_{00}$ to rise above 9,
so that then the mode with frequency $3m$ also has an approximately
exponential radial dependence at larger $r$.  With the right choice of
phase of this mode (i.e., of how big is the part that goes roughly as
$\sin{(\sqrt{8}mr)}/r$ at small $r$), only the decaying exponential is
present, and the mode at frequency $3m$ also goes to zero
exponentially rapidly at sufficiently large $r$.  (The mode at
frequency $3m$ starts to fall off exponentially in radius at a much
larger radius than the mode at frequency $m$ does.)  That is, the
phase is chosen so that at each radius where this mode is oscillating
radially, there are equal parts of outgoing and incoming waves, and
then the rising gravitational potential from the stress-energy tensor
of mostly that mode makes the mode gravitationally bound so that the
outgoing waves reflect off the potential barrier and become the
incoming modes at smaller radii.

Of course, there is also coupling to modes with frequency $5m$, and
these modes must also be given the right phase to have equal parts of
outgoing and incoming waves for radii at which $-g_{00} < 25$, so that
when one gets so far out that the energy density of that mode, mostly,
causes $-g_{00}$ to rise above 25 and that mode also to develop an
exponential behavior, one has only the approximately exponentially
decaying piece.

And so it goes for all higher modes with frequencies that are odd
multiples of $m$, say $(2n+1)m$.  They must each be chosen so that
have equal amounts of outgoing and incoming waves for $-g_{00} <
(2n+1)^2$, where that mode oscillates in the radial direction, and
then that mode is totally reflected by the gravitational potential for
$-g_{00} > (2n+1)^2$.

Since this process must continue indefinitely in order that the
oscillaton be precisely periodic in time and not be losing energy to
outgoing waves of any mode, one must have $-g_{00}$ rising
indefinitely, so the metric is not asymptotically flat and indeed has
$M(r)$, the mass interior to radius $r$, rising indefinitely with $r$:
these are infinite-mass oscillatons.  (They are also almost certainly
unstable to small perturbations, but that is another story.)

These precisely-periodic but infinite-mass oscillatons are of course
not physically realistic, but they do make interesting theoretical
solutions of the coupled Einstein-Klein-Gordon equations that one may
use for such purposes as defining a precise canonical (though rather
{\it ad hoc}) maximum mass for a finite-mass oscillaton without nodes.

For example, one could take a precisely-periodic oscillaton and then
evaluate $M(t,r)$ at a time $t$ when the scalar field at that $r$ is
passing through zero, and at a value of $r$ where the time average of
the energy density of the mode with frequency $m$ is equal to the time
average of the energy density of the mode with frequency $3m$.  (If
the scalar field passes though zero more than twice in each coordinate
time period $2\pi/m$, then choose one of the times at which the field
at that $r$ is zero but the magnitude of the time derivative of the
field is a minimum among all of these times when the field is zero
there.  If that still does not uniquely specify the time within one
period of the metric, which is a half-period $\pi/m$ of the scalar
field, then go to the minimum of the magnitude of the second time
derivative of the field among those times, etc, until the degeneracy
is broken.  If the degeneracy never is broken, then presumably the
mass does not depend on which of those times it is evaluated.)

The value of $M(t,r)$ at that value of $t$ and of $r$ could then be
said to be a canonical (though {\it ad hoc}) finite precise value for
the mass of a decaying finite-mass oscillaton without nodes that is in
some sense represented by the precisely-periodic infinite-mass (at
$r=\infty$) oscillaton.

The idea of the representation is that in the precisely-periodic
oscillaton, the value of $r$ determined above would be where the mode
with frequency $m$ has decayed exponentially to a very small value,
where its time-averaged energy density has dropped to that of the tiny
amount of outgoing and incoming waves at frequency $3m$.  For the
finite-mass decaying oscillaton, if one does not include the energy
density of the modes with frequency $3m$ or higher outside this radius
(e.g., if one assumes that they are just starting to be emitted and so
at that time are negligible outside this radius), the energy density
of the mode with frequency $m$ is so small at that radius (and
dropping roughly exponentially with radius) that there is a negligible
addition to the mass from that energy density in the finite-mass
oscillaton.

That is, it is negligible unless we are wanting to get some
absolutely-precise number to assign as the mass of the oscillaton, in
which case we must go from the slightly-poorly-defined initial
configuration of a finite-mass decaying oscillaton to the precisely
defined (but unphysical) infinite-mass oscillaton.

To get a representation of the maximum mass of a decaying finite-mass
oscillaton by this sort of precise mathematical definition, one must
choose the right precisely-periodic oscillaton.  There can be an
arbitrary number of nodes in the mode with frequency $m$ before it
decays approximately exponentially in the radial direction, and to
model the decaying finite-mass oscillaton which has no nodes in its
mode with frequency $m$, we want the precisely-periodic oscillaton
also to have no nodes in that mode.  On the other hand, we want the
modes at frequencies that are higher odd multiples of $m$, say
$(2n+1)m$ for each positive integer $n$, to have as low a value of
energy density possible, which would mean they should each have the
largest number of nodes possible before $-g_{00}$ rises above
$(2n+1)^2$ [at a rate with respect to radius that is given mainly by
the energy density in the outgoing and incoming waves at that
frequency, once $-g_{00}$ rises above $(2n-1)^2$ and the mode at the
next lower frequency, $(2n-1)m$, is reflected back inward].

Smaller numbers of nodes are possible for each of the higher-frequency
modes if there is an extra magnitude of outgoing and incoming waves at
that frequency, so there appears to be a whole infinite sequence of
integers to be specified for the generic precisely-periodic
oscillaton, even after specifying, say, the time-averaged energy
density at the center to give the one continuous parameter of the
classical oscillaton that is the nonlinear-gravity analogue of the
scale of the nearly-Newtonian oscillaton.  This means that there
should be uncountably many oscillatons at the same value of the
central time-averaged density.

It is plausible that they form a fractal set of perpetually
oscillating spatially inhomogeneous but spherically symmetric
periodic-in-time solutions of the Einstein-Klein-Gordon equation,
somewhat analogous to the apparently fractal set of homogeneous but
nonperiodic cosmological solutions of the same equations
\cite{Pagfrac}.

However, here we want just the simplest example, with the fastest
possible falloff of the energy density, to use to define a precise
value for the maximum mass of a finite-mass oscillaton with no nodes
for the mode with frequency $m$ and only outgoing components for the
modes with higher frequencies---these frequencies being only
approximate in the slightly-non-periodic case in which the finite-mass
oscillaton is actually slowly decaying with time.

Once we have chosen the simplest precisely-periodic oscillaton for
each possible value of the central time-averaged energy density, we
still need to choose the one that gives the maximum value of the mass
defined by the procedure above.  That will be the unique
precisely-periodic oscillaton that we will use to represent in some
sense the initial state of a finite-mass decaying oscillaton, and it
will give us a precise (though rather {\it ad hoc}) mathematical
definition of $\mu_{\rm max}$.

Now to get a definite initial state [say for giving a precise
definition of $\mu(t)$ and of its inverse, $t(\mu)$] for a finite-mass
decaying oscillaton that is represented by the unique
precisely-periodic oscillaton above, one could take the initial metric
and scalar field values and time derivatives of the precisely-periodic
infinite-mass oscillaton at the time $t$ defined above for determining
the value of $M(t,r)$ that was used to represent the maximum mass of a
decaying oscillaton, out to the value of $r$ that was also determined
by the procedure above (where the time average of the energy density
of the mode with frequency $m$ had dropped to the same time average of
the mode with frequency $3m$).  One could then truncate the
precisely-periodic oscillaton at that radius, replacing its initial
data on that hypersurface with initial data that agreed for smaller
values of $r$ but which for larger values of $r$ has zero for both the
scalar field and its time derivative, and the Schwarzschild values for
the initial data of the metric.  (That is, the solution can be taken
to be vacuum Schwarzschild on the initial time surface for larger
values of $r$.)

Next, simply let this truncated oscillaton initial data evolve by
solving the Einstein-Klein-Gordon equations, to represent the decay of
an initially maximal-mass oscillaton.

As energy flows out from the decaying oscillaton in the form of scalar
waves moving slower than the velocity of light, the ADM mass and also
the mass at future null infinity stay constant, so we need a
different definition of the mass of the decaying oscillaton to define
a nontrivial time dependence for it.  For that, we can simply choose
$M(t,r)$, with the coordinate value of $r$ kept fixed at the value
where, in the periodic oscillaton that provided the initial data out
to that radius, the time-averaged energy density of the mode with
frequency $m$ equaled that in the mode with frequency $3m$.

Here the coordinate value of $r$, once determined for the
precisely-periodic oscillaton, is assumed to be kept rigid, with no
gauge freedom, say by continued use in the decaying oscillaton of the
gauge choice that $g_{0i}=0$ and that $g_{00}$ has no periodic
component.  However, since $g_{00}$ has a slow secular evolution for
the decaying oscillaton, I'm not quite sure how to fix the gauge
absolutely precisely in this case, since I don't see how uniquely to
disentangle this slow secular evolution from a periodic component.
Perhaps if one wanted an absolutely precise mathematical definition,
which has been the aim of this long discussion, one should use instead
as the sphere where $M(t,r)$ is to be evaluated as a function of $r$, a
sphere that has a constant circumference (the Schwarzschild
$r$-coordinate, after taking $r$ to be the circumference divided by
$2\pi$, though not the $r$ coordinate that makes $g_{00}$ independent
of time for the precisely-periodic oscillaton).

There is also a slight ambiguity in defining $t$ absolutely precisely
if $g_{00}$ has a slow secular evolution rather than being precisely
independent of $t$ as it can be in the precisely-periodic oscillaton
with a suitable choice for the time and spatial coordinates.  One
commonly-used precise way to define it would be to define the
constant-$t$ hypersurfaces to be orthogonal to the worldlines at
constant angular coordinates on the spheres of constant circumference,
which leads to so-called Schwarzschild coordinates.

One disadvantage of the Schwarzshild $(t,r)$ coordinates is that inside
the oscillaton, the worldtubes of constant Schwarzschild-$r$
(circumference divided by $2\pi$) oscillate in and out, relative to
coordinates in which not only is $g_{0i}=0$ (as is true also for the
Schwarzschild coordinates by construction) but also $g_{00}$ is
constant (for the precisely-periodic oscillaton).  Then since the
hypersurfaces of constant $t$ in Schwarzschild coordinates are by
constuction orthogonal to the worldlines of constant Schwarzschild $r$
(and of contant angular position on the 2-spheres), these
hypersurfaces also bend forward and back by a rather considerable
amount in the interior of the oscillaton, giving periodic effects that
depend on this global choice of Schwarzschild coordinates and are
actually considerably larger than any effect in the local geometry.

However, at the value of $r$ defined above, where the time-average of
the energy density of the mode with frequency $m$ has dropped down to
equal the corresponding value for the tiny outgoing and incoming wave
modes with frequency $3m$, one is where the energy density of the
oscillaton is so low that the radial oscillations of the worldtubes of
constant Schwarzschild-$r$ are very small and should not have a
significant effect.

One still has to define the $t$-labelling of this foliation into
constant-$t$ hypersurfaces.  When the metric is asymptotically flat,
as it would be for the decaying oscillaton, then one can define it so
that $t$ is the proper time along the worldtube of infinite
circumference.  However, with this normalization, the coordinate-time
period of the approximately-periodic decaying oscillaton would be
shifted from the value $2\pi/m$ that it is given by construction
(normalization of $t$) in the precisely-periodic oscillaton.

Another simple but inequivalent choice would be to choose $t$ to be
proportional to proper time of the central worldline at $r=0$.  Then
one can have the approximate period of the approximately-periodic
decaying oscillaton be very near $2\pi/m$ by choosing the constant of
proportionality between proper time and coordinate time to be the same
as it is in the precisely-periodic oscillaton at $r=0$ in the gauge
for that solution in which $g_{00}$ is independent of time.  However,
one should note that as the oscillaton decays and the gravitational
potential at the center changes relative to that at infinity, the
nearly-periodic oscillations of the scalar field will have its
approximate period of oscillation secularly shifted from the value
$2\pi/m$ that it has initially by this choice of $t$.

A third choice for the labeling of the constant-Schwarzschild-$t$
hypersurfaces would be to choose at $r=0$ the time coordinate $t$ so
that the quantity $\psi$ defined by Eq. (\ref{eq:19}) would be real
and positive for all $t$.  This essentially forces the coordinate $t$
to be chosen so that, at the center at least, the coordinate period of
oscillation of the scalar field $\Phi$ (or of the rescaled
dimensionless field $\phi$) is fixed to be $2\pi/m$.  This is perhaps
the best choice if one wants to count the number of oscillations of
the scalar field, and to make a precise count even when that number is
not an integer.  Henceforth we shall assume that we have made this
choice for the Schwarzschild time coordinate $t$ (which will {\it not}
be the same choice that would make $t$ equal to proper time at radial
infinity).

Once a suitably time-evolving sphere is defined to represent the outer
surface of the decaying oscillaton (with the waves that go outside
that radius being considered outgoing waves rather than part of the
oscillaton), and once one has a precise time coordinate $t$, one can
in principle solve numerically for the mass as a function of this $t$.
Multiplying the mass $M(t)$ (in time units) by the constant $m$ (the
natural frequency of the scalar field, in inverse time units) gives
$\mu(t)$, the dimensionless measure of the evolving mass of the
oscillaton.

Assuming that $\mu(t)$ is monotonically decreasing with $t$ (which it
conceivably need not be within each period, since the waves need not
be purely outgoing over the entirety of a period, though they should
be when averaged over an integral number of periods), one can invert
this relation to get $t(\mu)$ to see the time needed for an oscillaton
to decay from the maximum-mass configuration to a smaller value of
$\mu$.  In the nearly-Newtonian limit in which $\mu \ll 1$, one would
expect that this time should asymptotically approach that given by
Eq. (\ref{eq:117}), not in the sense that the difference between the
actual $t(\mu)$ and that given by this formula would go to zero, but
in the sense that the ratio of the actual $t(\mu)$ to that given by
Eq. (\ref{eq:117}) would tend to unity as $\mu$ tends to zero in this
purely classical calculation.

Of course, it would be interesting to calculate numerically the value
of $\mu_{\rm max}$ and of the ratio of the actual $t(\mu)$ to that given by
Eq. (\ref{eq:117}), the latter being the function
 \begin{equation}
 R(\mu) \equiv {\alpha C m \over \mu^4} e^{-\alpha/\mu} t(\mu).
 \label{eq:118}
 \end{equation}

In particular, it would be interesting to calculate the large
dimensionless number
 \begin{equation}
 q \equiv mt(e^{-1}\mu_{\rm max}),
 \label{eq:119}
 \end{equation}
which is the angle by which the phase of the oscillating scalar field
advances ($2\pi$ times the number of scalar field oscillations) as the
oscillaton decays from its maximum mass to a mass that is smaller by a
factor of $e$.  The numerical calculations done above for the
nearly-Newtonian limit imply that
 \begin{equation}
 q \approx R(e^{-1}\mu_{\rm max}){\mu_{\rm max}^4\over e^4\alpha C}
   e^{e\alpha/\mu_{\rm max}}.
 \label{eq:120}
 \end{equation}

This dimensionless number $q$, which is somewhat of an analogue of an
average $Q$-value of the oscillaton system for masses within a factor
of $e$ of the maximum mass, is a pure mathematical number determined
purely by the Einstein-Klein-Gordon equations (with no dependence upon
the scale set by the scalar field natural frequency $m$) and by the
mathematically-precise (but admittedly rather {\it ad hoc}) procedure
given above.

It is interesting that $q$ is apparently quite large (at least if the
correction factor $R(e^{-1}\mu_{\rm max})$ is not too many orders of
magnitude smaller than unity), a counterexample to the folklore that
numbers defined purely mathematically tend to be within a few orders
of magnitude of unity.  For example, if $\mu_{\rm max}=0.607$ and if
$R(e^{-1}\mu_{\rm max})$ were unity, then one would get $q \approx
8.20\times 10^{65}$.  (However, there are other counterexamples that
are even more extreme, such as the unknown prime $n$ that is the first
positive integer greater than one such that the number of primes less
than or equal to $n$, namely $\pi(n)$, is greater than ${\rm Li}(n)$,
the principal value of the integral of $1/\ln{x}$ from $x=0$ to $x=n$,
which forms a good asymptotic estimate for $\pi(n)$.)

It would certainly be worthwhile to calculate numerically the values
of $\mu_{\rm max}$ and of $q$, as well as the function $R(\mu)$ for
values of $\mu$ up to $\mu_{\rm max}$, but since that cannot be done
within the nearly-Newtonian calculation reported here, it is beyond
the scope of this paper and will have to wait for future research.

It would also be amusing mathematically to get a quantitative
description of the particular precisely-periodic oscillaton described
qualitatively above that I have used to represent (by its part within
the radius $r$ defined above) the maximum-mass decaying oscillaton.
In particular, what is the quantitative asymptotic behavior of the
metric and of the scalar field [each mode of which, of frequency
$(2n+1)m$, eventually decays approximately exponentially in the radial
direction as the integrated gravitational effect of this and other
modes causes $-g_{00}$ to rise above $(2n+1)^2$ so that the mode
becomes gravitationally bound]?  However, this will also be left to
future work.

\section{Quantum Decay of Single-Field Oscillatons}

Besides the classical decay of finite-mass oscillatons to outgoing
scalar radiation, there are also quantum decay processes that appear
to be dominated by the annihilation of two scalar particles into two
gravitons.  This rate also goes to zero as $\mu$ is taken to zero, but
only as a power-law in $\mu$, so for sufficiently small $\mu$
(depending on the ratio of $m$ to the Planck value), it actually
dominates over the classical decay into scalar radiation.

The annihilation cross section for two nonrelativistic scalar
particles to annihilate into two gravitons has been given by DeWitt
\cite{DeW} to be (where I have inserted the factors of the speed of
light, $c$, that he and I usually set equal to unity)
 \begin{equation}
 \sigma_{\rm NR} = {2\pi G^2 m_*^2 \over c^3 v}
                 = {2\pi \hbar^2 G^2 m^2\over c^7 v},
 \label{eq:121}
 \end{equation}
where the first form is in terms of the scalar particle mass in
conventional mass units, which I have been calling $m_*$, and the
second form is in terms of the scalar field natural frequency in
inverse time units, which I have been calling $m$, which is given by
Eq. (\ref{eq:1}) as $m = m_{*}c^2/\hbar$.

Therefore, if we have a number density $n$ of scalar field particles
of one species (more than one species will be considered in the next
Section), and hence a conventional mass density
 \begin{equation}
 \rho = m_* n = (\hbar m/c^2) n,
 \label{eq:122}
 \end{equation}
the annihilation rate (per time) for one particle passing through is
 \begin{equation}
 R = n\sigma_{\rm NR}v = (2\pi \hbar^2 G^2 m^2/c^7) n.
 \label{eq:123}
 \end{equation}

Two scalar particles annihilate in each such process, but when one
takes the square of the number density, there is also a factor of 2
overcounting the number of pairs of identical particles, so these two
factors of 2 cancel each other and give the number rate per time per
volume by which scalar particles annihilate as
 \begin{equation}
 -{dN\over dt d({\rm vol})} = R n = (2\pi G^2 m_*^2/c^3) n^2
 = (2\pi \hbar^2 G^2 m^2/c^7) n^2 = 2\pi G^2 \rho^2/c^3.
 \label{eq:124}
 \end{equation}

In the nearly-Newtonian limit (which is where the nonrelativistic
annihilation cross section formula would apply), the mass density with
one scalar field present, and represented by the dimensionless complex
field $\psi$ given by Eq. (\ref{eq:19}), is
 \begin{equation}
 \rho = {m^2|\psi|^2 \over 4\pi G}.
 \label{eq:125}
 \end{equation}

If one integrates over an oscillaton with one real scalar field (one
complex $\psi$), the total annihilation rate per
time is
 \begin{equation}
 -{dN\over dt} = {1\over 8\pi c^3} \int m^4 |\psi|^4 d^3x.
 \label{eq:126}
 \end{equation}
In particular, for a spherical oscillaton with real $\psi$, one gets
 \begin{equation}
 -{dN\over dt} = {1\over 2 c^3} \int m^4 \psi^4 r^2 dr.
 \label{eq:127}
 \end{equation}

Now if we use the fact that by Eq. (\ref{eq:2}), the total mass in
time units of the oscillaton is $M \equiv GM_{*}/c^3$, and if we use
the fact that the mass in conventional units is the conventional mass
$m_{*}$ of each scalar particle times the number $N$ of such particles
(neglecting the small correction due to the kinetic energy of the scalar
particles and the gravitational binding energy), $M_{*} = m_{*}N =
(\hbar m/c^2)N$, then we get the rate per time at which the total mass
$M$ in time units decays away, the dimensionless rate
 \begin{equation}
 P_q \equiv -\dot{M}_{\mathrm{quantum}}
 = -{d\mu\over d(mt)}=-{\hbar G\over c^5}m{dN\over dt}
 \approx {\hbar G\over 8\pi c^8} \int m^5 \psi^4 d^3x 
 = {\hbar G\over 2 c^8} \int m^5 \psi^4 r^2 dr.
 \label{eq:128}
 \end{equation}

For the nearly-Newtonian spherical oscillaton analyzed above, by using
Eqs. (\ref{eq:55}), (\ref{eq:56}), (\ref{eq:63}), (\ref{eq:78a}),
(\ref{eq:79a}), (\ref{eq:80a}), and (\ref{eq:85}), one can show that
the quantum decay rate (per time) in the mass (in time units) is
 \begin{equation} 
 P_q \equiv -\dot{M}_{\mathrm{quantum}}
 \approx 4{\hbar G\over c^5}m^2{D_{\infty}\over \mathcal{M}_{\infty}^5} \mu^5
 = Q{m^2 \over m_{\rm Pl}^2} \mu^5
 = Q t_{\rm Pl}^2 m^2 \mu^5,
 \label{eq:129}
 \end{equation}
where
 \begin{equation} 
 m_{\rm Pl} = \sqrt{c^5\over\hbar G}
 \approx 1.855\times 10^{43} {\rm s}^{-1}
 \label{eq:130}
 \end{equation}
is the Planck frequency,
 \begin{equation} 
 t_{\rm Pl} = \sqrt{\hbar G \over c^5}
 \approx 5.391\times 10^{-44} {\rm s},
 \label{eq:130b}
 \end{equation}
is the Planck time, the reciprocal of the Planck frequency, and where
 \begin{equation} 
 Q = {4D_{\infty}\over \mathcal{M}_{\infty}^5}
 \approx  0.008\,513\,223\,934\,732\,691\,876\,362\,000\,723\,963\,272\,14,
 \label{eq:131}
 \end{equation}
using the numerical results given in Eqs. (\ref{eq:80}) and (\ref{eq:82}).

When both the classical decay rate given by Eq. (\ref{eq:113}) and the
quantum decay rate given by (\ref{eq:129}) are both significant, one
has the total decay rate being given by
 \begin{equation}
 -{dM\over dt}=-{d\mu\over d(mt)} = P_c + P_q 
 \,\, \approx \,\, {C \over \mu^2} e^{-{\alpha \over \mu}}
 + Q t_{\rm Pl}^2 m^2 \mu^5.
 \label{eq:132}
 \end{equation}

It may also be of interest to calculate the expected number of
scalarons and gravitons emitted during one period of oscillation of
the oscillaton scalar field, which is a time $2\pi/m$.  Since by far
the most dominant scalaron emission is at frequency $3m$, the energy of almost
every scalaron emitted is $3\hbar m$, so the expected number of
scalarons emitted in one period is
 \begin{equation}
 N_s = -{2\pi/m\over 3\hbar m}\left({dM_{*}c^2\over dt}\right)_{\rm classical}
 \approx {2\pi\over 3}\left({m_{\rm Pl}\over m}\right)^2
     {C \over \mu^2} e^{-{\alpha \over \mu}}.
 \label{eq:132b}
 \end{equation}

Similarly, the most dominant graviton emission is at frequency $m$ (as
two scalarons of this frequency annihilate into two gravitons of the
same frequency), so the expected number of gravitons emitted in one
period is
 \begin{eqnarray}
 N_g &=& -{2\pi/m\over \hbar m}\left({dM_{*}c^2\over dt}\right)_{\rm quantum}
 \approx 2\pi Q \mu^5
 \nonumber \\
 &\approx&
 0.053\,490\,163\,543\,442\,036\,463\,830\,655\,145\,560\,622\,6\,\,
   \mu^5.
 \label{eq:132bb}
 \end{eqnarray}

It is interesting that although it is the classical emission power
(into scalar waves or scalarons) $P_c$ that depends only on $\mu$,
with the quantum emission power (into gravitons) depending also on
$m/m_{\rm Pl}$, for the expected number of particles emitted in one
oscillaton period, it is the quantum emission into gravitons that
depends only on $\mu$ (with an expected number of gravitons per period
never larger than unity, and in fact never larger than roughly 0.0044
if the maximum value for $\mu$ is 0.607 and if the formula above
indeed applies to this large a value of $\mu$), whereas the emission
into scalarons (classical) gives a number depending also on $m/m_{\rm
Pl}$ (and which can be larger than unity for sufficiently small mass
$m$).

The expected number of scalarons emitted per oscillaton period is
unity, $N_s = 1$, for
 \begin{eqnarray}
 &&{m \over m_{\rm Pl}} \approx 
 {\sqrt{2\pi C/3} \over \mu} e^{-{\alpha \over 2\mu}}
 \nonumber \\ &&
 \approx {2\,820.165\,789\,522\,802\,746\,688\,836\,089\,134\,773\,52 \over \mu}
  e^{-{19.716\,897\,598\,580\,081\,546\,859\,760\,732\,713\,184\,8
 \over \mu}}.
 \nonumber \\ &&
 \label{eq:132c}
 \end{eqnarray}
If this formula were to apply for $\mu=0.607$, which is one numerical
result for the maximum mass parameter of an oscillaton \cite{ABGMNUL}, then
this would occur for
 \begin{equation}
 m_{*} \approx 3.63\times 10^{-11} m_{*\rm Pl}
 \approx 4.43\times 10^8 {\rm GeV}/c^2.
 \label{eq:132d}
 \end{equation}
Then there would be some allowed oscillaton mass value at which the
oscillaton would emit an expected number of one scalaron per
oscillation period, for any scalaron mass less than some scalaron mass
value that is given crudely by Eq. (\ref{eq:132d}) that uses
Eq. (\ref{eq:113}) outside its $\mu \ll 1$ domain of validity.

If we define for use here (not the same $x$ as used previously for the
rescaled radial variable)
 \begin{equation}
 x \equiv {\alpha\over\mu}
 \approx {39.433\,795\,197\,160\,163\,093\,719\,521\,465\,426\,369\,6\over\mu}
 \label{eq:133}
 \end{equation}
and
 \begin{equation}
 \gamma = {C\over \alpha^7 Q}
 \approx 0.003\,008\,268\,339\,955\,585\,528\,843\,272\,770\,555\,262\,66,
 \label{eq:134}
 \end{equation}
then the ratio of the classical mass-loss rate to the quantum
mass-loss rate is, for $\mu \ll 1$,
 \begin{equation}
 {P_c\over P_q} = \gamma {m_{\rm Pl}^2 \over m^2} x^7 e^{-x},
 \label{eq:135}
 \end{equation}
which is unity for
 \begin{equation}
 x - 7\ln{x} = \ln{\left(\gamma{m_{\rm Pl}^2 \over m^2}\right)},
 \label{eq:136}
 \end{equation}
or
 \begin{eqnarray}
 x &=& \ln{\left(\gamma{m_{\rm Pl}^2 \over m^2}\right)} + 7\ln{x}
   = \ln{\left(\gamma{m_{\rm Pl}^2 \over m^2}\right)}
    +7\ln{\left(\ln{\left(\gamma{m_{\rm Pl}^2 \over m^2} \right)}
    +7\ln{x}\right)}
 \nonumber \\
   &=& \ln{\left(\gamma{m_{\rm Pl}^2 \over m^2}\right)}
    +7\ln{\left(\ln{\left(\gamma{m_{\rm Pl}^2 \over m^2} \right)}
      +7\ln{\left(\ln{\left(\gamma{m_{\rm Pl}^2 \over m^2} \right)}
    +7\ln{x}\right)}\right)},
% \nonumber \\
%  &\ &
 \label{eq:137}
 \end{eqnarray}
to give the first three steps in an iterative procedure for solving
for the value of $x=\alpha/\mu$ that gives $P_c=P_q$.

The decay time for an oscillaton to have its dimensionless decay
parameter decay from $\mu_1$ to a smaller value of $\mu$ is given by
 \begin{equation}
 m(t-t_1)=\int_{\mu}^{\mu_1} {d\mu \over P_c + P_q}.
 \label{eq:137b}
 \end{equation}

Define the $m$-dependent constant $r$ [for the use in this
immediate section; not the same $r$ as used elsewhere for the radial
variable or with yet another meaning as the dependent variable defined
in Eq. (\ref{eq:102})] as the largest real solution to the equation
 \begin{equation}
 {r \over (\ln{r})^3} = s \equiv {1\over \alpha^4 Q t_{\rm Pl}^2 m^2},
 \label{eq:137c}
 \end{equation}
where the $m$-dependent constant $s$ here is also not the dependent
variable defined in Eq. (\ref{eq:103}).  For brevity also define here
$q$ [now neither the dependent variable defined in Eq. (\ref{eq:101})
nor the large dimensionless constant angle defined in
Eq. (\ref{eq:119})] to be the function of the classically
dimensionless mass $\mu$ that is
 \begin{equation}
 q \equiv {\mu^4 \over \alpha C} e^{\alpha/\mu}.
 \label{eq:137d}
 \end{equation}

Then when $\ln s = -\ln{(\alpha^4 Q t_{\rm Pl}^2 m^2)} \gg 1$ (which
implies that $\ln{r} \gg 1$), and when Eq. (\ref{eq:132}) also holds,
one can show that the integral on the right hand side of
Eq. (\ref{eq:137b}) can be approximated by the following explicit
function of $r$, $q$, and $q_1$ (the value of $q$ when $\mu$ has its
initial value $\mu_1$):
 \begin{equation}
 m(t-t_1) \approx {r \over 4(\ln{r})^3}
       \{[\ln{(r+q)}]^4 - [\ln{(r+q_1)}]^4\}.
 \label{eq:137e}
 \end{equation}

We can use Eq. (\ref{eq:137c}) to rewrite this equation in the form
 \begin{equation}
 4\alpha^4 Q t_{\rm Pl}^2 m^3 (t-t_1)
 \approx [\ln{(r+q)}]^4 - [\ln{(r+q_1)}]^4.
 \label{eq:137f}
 \end{equation}

This gives an approximate formula for the decay time, $t-t_1$, for an
oscillaton in terms of its mass $M$ and the scalar field mass $m$
(which then determine the dimensionless mass parameter $\mu$),
assuming that one knows the initial value $\mu_1$ of the dimensionless
mass parameter.

For other purposes, one might know the decay time and from it want to
get a relationship between the oscillaton mass $M$ and the
dimensionless mass parameter $\mu$.  For this purpose, it is helpful
to define from the decay time the large dimensionless parameter [not
the same $a$ as the $\mu$-dependent constant defined in Eq. (\ref{eq:97})]
 \begin{equation}
 a \equiv \left({16\over \alpha^4 Q}\right)^{1/3}
          \left({t-t_1\over t_{\rm Pl}}\right)^{2/3} \gg 1.
 \label{eq:137g}
 \end{equation}
It is also helpful to define a new dimensionless mass parameter
 \begin{equation}
 \nu \equiv {M \over [4\alpha^4 Q t_{\rm Pl}^2 (t-t_1)]^{1/3}},
 \label{eq:137h}
 \end{equation}
which depends on the oscillaton mass $M$ and on the decay time $t-t_1$
but not on the mass $m$ of the scalar field as the other dimensionless
mass parameter $\mu = Mm$ does.

Then, for example, Eq. (\ref{eq:137f}) becomes
 \begin{equation}
 \mu \approx \nu
       \{[\ln{(r+q)}]^4 - [\ln{(r+q_1)}]^4\}^{1/3},
 \label{eq:137i}
 \end{equation}
and Eq. (\ref{eq:137c}) becomes
 \begin{equation}
 {r \over (\ln{r})^3} = s = {a \nu^2 \over \mu^2}.
 \label{eq:137j}
 \end{equation}
For $s \gg 1$, this solution to this equation is very roughly
 \begin{equation}
 r \sim s(\ln s)^3
 = {a \nu^2 \over \mu^2}\left(\ln{a \nu^2 \over \mu^2}\right)^3.
 \label{eq:137k}
 \end{equation}

Now we can insert $r$ from Eq. (\ref{eq:137k}) and the definition of
$q$ from Eq. (\ref{eq:137d}) into Eq. (\ref{eq:137i}) to get the
following rough explicit algebraic relation between $\mu$,
$\mu_1$, and $\nu$ for fixed $t-t_1$, and hence for fixed $a$ defined
by Eq. (\ref{eq:137g}):
 \begin{equation}
 \left({\mu\over\nu}\right)^3 \sim 
 \left\{\ln{\left[{a \nu^2 \over \mu^2}\left(\ln{a \nu^2 \over \mu^2}\right)^3
   +{\mu^4\over\alpha C}e^{\alpha/\mu}\right]}\right\}^4
  -\left\{\ln{\left[{a \nu^2 \over \mu^2}\left(\ln{a \nu^2 \over \mu^2}\right)^3
   +{\mu_1^4\over\alpha C}e^{\alpha/\mu_1}\right]}\right\}^4.
 \label{eq:137l}
 \end{equation}
However, I should note that although this relation is a good
approximation to Eq. (\ref{eq:137i}) when $\mu$ is very small, it can
be off by about $20-30\%$ when $\mu$ is large.

For $\alpha/\mu_1 < \alpha/\mu \ll \ln a$, so that the classical decay
dominates, this relationship can be solved explicitly for $\nu$ to
give
 \begin{equation}
 \nu \approx 
  {\alpha C a \mu\over 4(\mu^4 e^{\alpha/\mu}-\mu_1^4 e^{\alpha/\mu_1})}.
 \label{eq:137m}
 \end{equation}
In this limit, this Eq. (\ref{eq:137m}) is actually a better
approximation to Eq. (\ref{eq:137i}) than is Eq. (\ref{eq:137l}).  At
the other extreme, for $\alpha/\mu \gg \ln a$, so that the quantum
annihilation dominates during most of the decay, this relationship can
also be solved explicitly for $\nu$ to give a fairly accurate
approximation to Eq. (\ref{eq:137i}), namely
 \begin{equation}
 \nu \approx \alpha^{-4/3}\mu^{7/3}
\left(1-{\mu^4\over\mu_1^4}-4{\mu\over\alpha}
  \ln{\mu^4\over\alpha C}\right)^{-1/3}.
 \label{eq:137n}
 \end{equation}
In between these two extremes, i.e., for $\alpha/\mu \sim \ln a$, I
don't see how to give any simple expression that would solve
explicitly for either $\mu$ or $\nu$ in terms of the other (say for
fixed $a$ and $\mu_1$), though of course one could solve
Eqs. (\ref{eq:137i}) and (\ref{eq:137j}) numerically for either $\mu$
or $\nu$ in terms of a specific value of the other, for fixed decay
time $t-t_1$ and hence for fixed $a$ given by Eq. (\ref{eq:137g}).

Let us put in some numbers for these quantities.  If we take $\mu_1$
to be the maximum mass of an oscillaton given by \cite{ABGMNUL}, namely
$\mu_1 = 0.607$, then we get
 \begin{equation}
 q_1 \equiv {\mu_1^4 \over \alpha C}e^{\alpha/\mu_1}
     \approx 1.48\times 10^{19}.
 \label{eq:137o}
 \end{equation}
If we take the oscillaton decay time $t-t_1$ to be the present age of
the universe, about 13.7 billion years or $8.02\times 10^{60} t_{\rm
Pl}$ in Planck units, then
 \begin{equation}
 a \equiv \left({16\over \alpha^4 Q}\right)^{1/3}
          \left({t-t_1\over t_{\rm Pl}}\right)^{2/3}
          \approx 3.68\times 10^{39},
 \label{eq:137p}
 \end{equation}
or $\ln{a} \approx 91.10$.

In the previous paragraph it was noted that Eq. (\ref{eq:137m})
applies for $\alpha/\mu \ll \ln a$ and that Eq. (\ref{eq:137n})
applies for $\alpha/\mu \gg \ln a$, so these inequalities are
saturated and one has $\alpha/\mu = \ln a$ at $\mu \approx 0.433$.
Therefore, since $\mu$ cannot be much larger than this value, one
never really has the validity of the inequality $\alpha/\mu \ll \ln
a$, but in actuality Eq. (\ref{eq:137m}) is a good approximation to
Eq. (\ref{eq:137i}) for $e^{\alpha/\mu} \ll a$, and indeed at $\mu =
\mu_1$, one has $e^{\alpha/\mu}/a \approx 4.44\times 10^{-12} \ll 1$.

However, one must still remember that Eq. (\ref{eq:137i}), or
equivalently Eq. (\ref{eq:137e}) or Eq. (\ref{eq:137f}), are valid
approximations only to the extent that Eqs. (\ref{eq:113}) and
(\ref{eq:129}) are valid for the classical and quantum decays rates
$P_c$ and $P_q$ respectively.  In this paper these formulas were
derived under the assumption that $\mu \ll 1$, so they are not likely
to be accurate for the small $\mu$ values where $e^{\alpha/\mu} \ll
a$, particularly for $P_c$ with its very strong dependence on $\mu$.
Hence the formulas used here for when the classical decay appears to
dominate should be taken with a big grain of salt, as merely
provisional formulas that might give some rough qualitative indication
of the true quantitative behavior.

Nevertheless, to get some idea of this rough qualitative behavior, let
us for now assume that Eqs. (\ref{eq:113}) and (\ref{eq:129}) are
valid for all $\mu$ less than its maximum value, at, say, $\mu = \mu_1
= 0.607$, and put in some various possible values for $m$, $\mu$,
and/or $M$.

For example, taking the example used by Seidel and Suen \cite{SS} in
which the scalar field mass is typical of that of an axion, $m_{*} =
10^{-5} {\rm eV}/c^2$ or $m \approx 1.519\times 10^{10} s^{-1} \approx
8.19\times 10^{-34} m_{\rm Pl}$, then $\ln{(\gamma_{\rm Pl}^2/m^2)}
\approx 146.564$, so one finds that $P_c=P_q$ at $x \approx 183.031$,
or $\mu \approx 0.2154$ (if the formulas above really apply to this
large a value of $\mu$), which corresponds to an oscillaton mass in
time units of $M \approx 1.418\times 10^{-11} {\rm s}$ or an
oscillaton mass in conventional units of $M_{*} \approx 5.726\times
10^{27} g = 0.9585 M_{\oplus}$ (about $96\%$ of the mass of the
earth).

For this oscillaton, assuming that the equations above did apply for
$\mu \approx 0.2154$ even though this is not much smaller than unity,
one gets that the total power emitted would be $P=P_c+P_q \approx
5.3\times 10^{-72}$, and the logarithmic rate of decrease of the mass
would be $-d\ln{M}/dt \approx 3.7\times 10^{-61} {\rm s}^{-1}$.  The
oscillaton would contain about $N=M_{*}/m_{*}=(m_{\rm Pl}/m)^2 \mu
\approx 3.2\times 10^{65}$ scalarons, and so in each period of
oscillation of the oscillaton, there would be about $(2\pi/3)(m_{\rm
Pl}/m)^2 P_c \approx 8.3\times 10^{-6}$ scalarons emitted (each of
energy roughly $3m_{*}c^2$) and about $(2\pi)(m_{\rm Pl}/m)^2 P_q
\approx 2.5\times 10^{-5}$ gravitons emitted (each of energy roughly
$m_{*}c^2$).  That is, one would need to wait on average about
$120\,000$ periods of oscillation between the emission of successive
scalarons, and about $80\,000$ periods of oscillation between the
emission of successive pairs of gravitons (since they come out
predominantly in pairs, with the pair having two-thirds the energy of
a typical scalaron that is emitted).  Therefore, although this
oscillaton is not absolutely stable, for astronomical purposes it is very
nearly stable.

If this oscillaton with $\mu \approx 0.2154$ had actually decayed from
$\mu = \mu_1 = 0.607$, that would have taken a time $t-t_1 \sim
10^{51} {\rm yr}$, again using Eqs. (\ref{eq:113}) and (\ref{eq:129})
outside their true range of validity just to give a qualitative
answer.

On the other hand, if an oscillaton with this value of the scalar
field mass, $m_{*} = 10^{-5} {\rm eV}/c^2$, were to have started at
$\mu_1 = 0.607$ at the beginning of the universe and had decayed up
until its present age, it would now have $\mu \approx 0.459$ (again
taking the classical decay rate formula outside its range of validity,
$\mu \ll 1$).  But even though this calculated value for $\mu$ today
is not likely to be actually correct, it is interesting that it is
significantly below the initial value.  Thus if this result is at
least qualitatively correct, an initially maximum-mass oscillaton with
this value of the scalar field mass would have decayed by a
significant amount during a time comparable to the age of the universe.

To take a more extreme example, if we imagine that there is a scalar
field (quintessence?) of natural frequency $m$ that has the value of
the current Hubble expansion rate, $H_0 \approx 2.3\times 10^{-18}
s^{-1} \approx 1.24\times 10^{-61} m_{\rm Pl}$, which corresponds to
$m_{*} c^2 \approx 1.51\times 10^{-33} {\rm eV}$, then
$\ln{(\gamma_{\rm Pl}^2/m^2)} \approx 274.678$, so then $P_c=P_q
\approx 4\times 10^{-129}$ at $x \approx 314.945$, or $\mu \approx
0.1252$, giving an oscillaton mass in time units of $M \approx
5.4\times 10^{16} {\rm s} \approx 1.7$ billion years, or an oscillaton
mass in conventional units of $M_{*} \approx 2.2\times 10^{55} g =
1.1\times 10^{22} M_{\odot}$ (of the same order of magnitude as that
of all the observable galaxies in the universe, by the same
coincidence that this mass is roughly within a factor of 10 or so of
what is needed to close the universe).  For this example, one would
need to wait on average about $1\,800\,000$ periods of oscillation
between the emission of successive scalarons, and about $1\,200\,000$
periods of oscillation between the emission of successive pairs of
gravitons.

If we took an oscillaton with a scalar field of this $m = H_0$,
started it with $\mu_1 = 0.607$, and applied the formulas above, we
would find that during the age of the universe, it would have had
$\mu$ decay only by about $6.3\times 10^{-22}$, an insignificant
reduction in its value.  Besides the usual caveat about the
inapplicability of Eq. (\ref{eq:116}) to this large $\mu$ value---note
that Eq. (\ref{eq:117}) does not apply here, since in this case the
two terms in Eq. (\ref{eq:116}) are very nearly equal---there is also
the error from the fact that the age of the universe corresponds to
only about 0.99 of a radian of the phase of the oscillation of an
oscillaton with a scalar field mass equalling the current value of the
Hubble constant, whereas the formulas above apply only for time
periods containing many oscillations.

For values of $x$ smaller than the solution of Eq. (\ref{eq:136}) or
(\ref{eq:137}), so that $\mu$ is larger than the corresponding critical
$\mu$ value for that $m/m_{\rm Pl}$, then the classical decay rate
dominates (i.e., for large oscillaton masses).  On the other hand, for
values of $x$ larger than the solution of Eq. (\ref{eq:136}) or
(\ref{eq:137}), so that $\mu$ is smaller than the corresponding critical
$\mu$ value for that $m/m_{\rm Pl}$, then the quantum decay rate
dominates (i.e., for small oscillaton masses).

It is interesting that even with perhaps about the smallest
conceivable value of the scalar field mass in the present universe,
that of the Hubble constant, the quantum emission dominates over the
classical emission when $\mu$ is only as small as about 1/8 (and of
course for all smaller values of $\mu$).  That is, for almost any
conceivable oscillaton in the present universe, if the dimensionless
mass parameter $\mu$ is smaller than roughly 1/8, the classical
emission of scalar waves would be even less than the tiny quantum
emission of gravitons from the annihilation of pairs of scalar
particles in the oscillaton.  This illustrates how rapidly the
classical emission drops as $\mu$ is made small.

When the mass of the scalar field is much smaller than the Planck
mass, as it must be for one to have nearly-Newtonian oscillatons
containing a large number of scalar particles, as is implicitly
assumed in the analysis above, then by Eq. (\ref{eq:136}) or
(\ref{eq:137}) one finds that $x$ is large in comparison with unity for
$P_c = P_q$.  However, because $\alpha$ is itself rather large, it is
not necessarily the case that this value of $x$ corresponds to a small
value of $\mu$ or even a value of $\mu$ less than $\mu_{\rm max}$.

For example, if Eq. (\ref{eq:135}) were valid not just for very small
$\mu$ but also for values of $\mu$ up to $\mu_{\rm max}$, and if we
take $\mu_{\rm max}=0.607$, then we would find that $P_c = P_q$ at $\mu
= \mu_{\rm max}=0.607$ for $m \approx 9.47\times 10^{-10} m_{\rm Pl}$ or
$m_{*} \approx 9.47\times 10^{-10} m_{*\rm Pl} \approx 1.16\times 10^{10}
{\rm GeV}/c^2$.  Assuming that this is correct, then for larger values
of the scalar field mass, the quantum decay would dominate ($P_q >
P_c$) for all values of $\mu$ up to and including $\mu_{\rm max}$.  On
the other hand, for smaller values of the scalar field mass, which is
more realistic if the scalar field is, say, an axion, then there is
always a mass range (say $\mu_c < \mu < \mu_{\rm max}$ with $\mu_c$
being given roughly by $\alpha/x$ with $x$ the solution of
Eq. (\ref{eq:136}) or (\ref{eq:137}) when it gives a small $\mu_c$),
where the classical emission dominates for the mass loss rate, though
for $\mu < \mu_c < \mu_{\rm max}$ the quantum mass-loss decay rate
would dominate.

It may be of interest to estimate the present upper bound on $\mu$ for
oscillatons with scalar field masses other than the two examples
above, assuming that the oscillatons formed in the early universe and
have been decaying for a time comparable to the age of the universe.
The maximum present value of $\mu$ that they would have would be what
they would have if they started with the maximum allowed initial value
of $\mu$, which here I shall take to be $\mu_1 = 0.607$, as above.  I
shall also assume that the scalar field mass-energy is $m_{*} c^2 \ll
10^{10}\ {\rm GeV}$, so that the decay from $\mu = \mu_1$ within the
present lifetime of the universe would be in the regime where the
classical decay dominates (see below for more details on this) and
where I shall assume that Eq. (\ref{eq:116}) holds.

Then if I use $x \equiv \alpha/\mu$ defined by Eq. (\ref{eq:133}), one gets
(with a new use for $y$)
 \begin{equation}
 {e^x\over x^4} \approx y
 \label{eq:137q}
 \end{equation}
where for this Section I shall define
 \begin{eqnarray}
 y &\equiv& {C\over\alpha^3}[m(t-t_1)+q_1]
  \nonumber \\
  &\approx& 4.067\,595\times 10^{34}
  \left({m_{*} c^2\over {\rm eV}}\right)
  \left({t-t_1\over 1.37 \times 10^{10} {\rm yr}}\right)
  +9.188\,619\times 10^{20}
 \label{eq:137qq}
 \end{eqnarray}
(not the same $y$ that denoted the rescaled imaginary radial
 coordinate in Section 8).

An explicit approximation that solves Eq. (\ref{eq:137q}) for $x$ in
terms of $y$ to at least 8-digit accuracy is
 \begin{equation}
 x \approx
 \ln{y}+4\ln{(\ln{y}+4\ln{(\ln{y}+4\ln{(\ln{y}+4\ln{(\ln{y}+4\ln{x_1})})})})},
 \label{eq:137r}
 \end{equation}
where
 \begin{equation}
 x_1 \equiv {\alpha\over\mu_1} \approx 64.965,
 \label{eq:137rr}
 \end{equation}
 \begin{equation}
 4\ln{x_1} =4\ln{(\alpha/\mu_1)} \approx 16.695.
 \label{eq:137rrr}
 \end{equation}
Then we get
 \begin{equation}
 \mu \approx {39.433\,795 \over 
 \ln{y}+4\ln{(\ln{y}+4\ln{(\ln{y}+4\ln{(\ln{y}+4\ln{(\ln{y}+16.695)})})})}}.
 \label{eq:137s}
 \end{equation}

However, Eq. (\ref{eq:137q}) itself is not that accurate, since it was
derived on the assumption of the accuracy of Eq. (\ref{eq:116}), which
is in doubt, since that equation was derived for $\mu \ll 1$, whereas
here for $m_{*} c^2 \ll 10^{10}\ {\rm GeV}$, one gets $\mu$ in the
range roughly between 0.3 and 0.6, which is not much less than unity.

Once we have an estimate for $\mu$ (no doubt rather crude, since it
does not give $\mu \ll 1$ where it would be valid), or for the maximum
value of $\mu$, as a function of the scalar field mass, we can easily
get the oscillaton mass $M=\mu/m$ in time units or $M_{*}=\hbar c
\mu/(G m_{*})$ in conventional mass units.  In terms of the solar mass
$M_{\odot} \approx 1.989\times 10^{33}\, g \approx 0.9137\times
10^{38}\, m_{*{\rm Pl}}$, one can use Eq. (\ref{eq:3}) to write
 \begin{equation}
 M_{*}
 \approx 1.336\,337\,63\times 10^{-10} M_{\odot}
 \left({1\ {\rm eV}\over m_{*} c^2}\right)\mu.
 \label{eq:137t}
 \end{equation}
Combining this with Eq. (\ref{eq:137s}) then gives
 \begin{equation}
 M_{*}
 \approx {5.269\,686\,43\times 10^{-9} M_{\odot}
   [(1\ {\rm eV})/(m_{*} c^2)] \over
 \ln{y}+4\ln{(\ln{y}+4\ln{(\ln{y}+4\ln{(\ln{y}+4\ln{(\ln{y}+16.695)})})})}}.
 \label{eq:137u}
 \end{equation}
where $y$ is given by Eq. (\ref{eq:137qq}).  This would be the
estimated value for the oscillaton mass if it started at $\mu_1 =
0.607$ and would be an upper limit for the mass if 0.607 were the
maximum value of $\mu$ at which it could have started.  Again, this
formula is applicable for $m_{*} c^2 \ll 10^{19} {\rm eV}$ (where the
classical decay dominates for an oscillaton starting with $\mu_1 =
0.607$ and decaying for up to 13.7 billion years), and it also assumes
the dubious correctness of Eq. (\ref{eq:113}) for the resulting fairly
large values of $\mu$ as given by Eq. (\ref{eq:137s}).

For example, if we let $\mu(m_{*} c^2)$ be the value of $\mu$ that an
oscillaton of of scalaron mass-energy $m_{*} c^2$ would decay to, from
$\mu_1 = 0.607$, in a time of 13.7 billion years, then we had shown
above that $\mu(10^{-5} {\rm eV}) \approx 0.459$.  We can also readily
calculate the following values of $\mu(m_{*} c^2)$ for other values of
$m_{*} c^2$:
 \begin{eqnarray}
 \mu(10^{-35} {\rm eV}) &\approx& \mu_1-4.41\times 10^{-24},
 \nonumber \\
 \mu(10^{-30} {\rm eV}) &\approx& \mu_1-4.41\times 10^{-19},
 \nonumber \\
 \mu(10^{-25} {\rm eV}) &\approx& \mu_1-4.41\times 10^{-14},
 \nonumber \\
 \mu(10^{-20} {\rm eV}) &\approx& \mu_1-4.41\times 10^{-9},
 \nonumber \\
 \mu(10^{-15} {\rm eV}) &\approx& \mu_1-4.41\times 10^{-4},
 \nonumber \\
 \mu(10^{-10} {\rm eV}) &\approx& 0.534,
 \nonumber \\
 \mu(10^{-5} {\rm eV}) &\approx& 0.459,
 \nonumber \\
 \mu(1\ {\rm eV}) &\approx& 0.402,
 \nonumber \\
 \mu(10^5 {\rm eV}) &\approx& 0.358,
 \nonumber \\
 \mu(10^{10} {\rm eV}) &\approx& 0.323,
 \nonumber \\
 \mu(10^{15} {\rm eV}) &\approx& 0.295.
 \label{eq:137v}
 \end{eqnarray}

Similarly, we can calculate $M_{*}(m_{*} c^2)$, the value of $M_{*}$ that an
oscillaton of scalaron mass-energy $m_{*} c^2$ would decay to, from
$\mu_1 = 0.607$ and
 \begin{equation}
 M_{*1}={\hbar c\mu_1\over G m_{*}}
 \approx 8.111\times 10^{-11} M_{\odot} {1\ {\rm eV} \over m_{*} c^2},
 \label{eq:137w}
 \end{equation}
in a time of 13.7 billion years, as having the following values:
 \begin{eqnarray}
 \mu(10^{-35} {\rm eV}) &\approx& 8.111\times 10^{24} M_{\odot},
 \nonumber \\
 \mu(10^{-30} {\rm eV}) &\approx& 8.111\times 10^{19} M_{\odot},
 \nonumber \\
 \mu(10^{-25} {\rm eV}) &\approx& 8.111\times 10^{14} M_{\odot},
 \nonumber \\
 \mu(10^{-20} {\rm eV}) &\approx& 8.111\times 10^{9} M_{\odot},
 \nonumber \\
 \mu(10^{-15} {\rm eV}) &\approx& 81\,055 M_{\odot},
 \nonumber \\
 \mu(10^{-10} {\rm eV}) &\approx& 0.7133 M_{\odot},
 \nonumber \\
 \mu(10^{-5} {\rm eV}) &\approx& 6.128\times 10^{-6} M_{\odot},
 \nonumber \\
 \mu(1\ {\rm eV}) &\approx& 5.375\times 10^{-11} M_{\odot},
 \nonumber \\
 \mu(10^5 {\rm eV}) &\approx& 4.790\times 10^{-16} M_{\odot},
 \nonumber \\
 \mu(10^{10} {\rm eV}) &\approx& 4.322\times 10^{-21} M_{\odot},
 \nonumber \\
 \mu(10^{15} {\rm eV}) &\approx& 3.938\times 10^{-26} M_{\odot}.
 \label{eq:137x}
 \end{eqnarray}

These are either the estimates for the masses, if the oscillatons
started 13.7 billion years ago with $\mu = 0.607$, or are estimates
for the upper bounds of the oscillaton masses, if 0.607 is
merely an upper bound on the initial value of $\mu$.

\section{Quantum Decay When the Particle Number Gets Small}

When $\mu \ll 1$ and when Eq. (\ref{eq:135}) gives $P_c/P_q \ll 1$,
then the quantum mass-loss rate dominates and is given to good
accuracy by Eq. (\ref{eq:129}), but only so long as the number of
scalar particles in the oscillaton,
 \begin{equation}
 N \approx {M_{*}\over m_{*}} = {c^5 M \over \hbar G m}
 = {c^5 \mu \over \hbar G m^2}
 = {m_{\rm Pl}^2 \over m^2} \mu,
 \label{eq:138}
 \end{equation}
is large in comparison with unity.

That is, the quantum mass loss rate dominates and is given to good
accuracy by Eq. (\ref{eq:129}) if
 \begin{equation}
 {m^2 \over m_{\rm Pl}^2} \ll \mu < \mu_c \approx {\alpha\over x} \sim
 {\alpha\over \ln{(\gamma m_{\rm Pl}^2/ m^2)}}
 \approx{19.716\,897\,598\,580\,082
 \over 61.769+\ln{({\rm eV}/m_{*}c^2)}} \ll 1.
 \label{eq:139}
 \end{equation}
The right hand side of this requirement is actually a bit stronger
than what is needed, which is that both $\mu < \mu_c$ and that $\mu
\ll 1$, but it is not really necessary that $\mu_c \ll 1$.  We may
note that for this inequality to have any range of validity for $\mu$,
we need $m \ll m_{\rm Pl}$, which we have been assuming thoughout this
paper and shall continue to assume.

When an oscillaton is decaying, it will eventually get down to having
a small number $N$ of scalar particles, and Eq. (\ref{eq:129}) will
cease to be accurate.  In principle one could solve the $N$-body
Schr\"{o}dinger equation with Newtonian attractive potentials between
the $N$ scalar particles for the ground-state wavefunction (ignoring
for the moment the annihilations into gravitons) and then calculate
the overlap between two particles to get the two-particle annihilation
rate into two gravitons.  However, I did not do this calculation for
$N > 2$ and am not familiar with the literature where it might have
been done.

Just as the annihilation rate for large $N$ goes as the 5th power of
$\mu$ and of $N$, one would also expect that the annihilation rate for
a small number $N$ would also decrease rapidly as $N$ is reduced,
reaching a minimum for $N=2$ (if one can reach this number, though if
one has an odd number when $N$ is somewhat larger, and if the 2-body
annihilations dominate so that the scalar particles predominately
annihilate in pairs, then one would most likely end up with a
3-particle state before the final decay to two gravitons and one free
scalar particle).

So if the oscillaton decays down to two scalar particles before
annihilating completely, the decay of the final two particles is
likely to take more time than the entire decay down to that point.

The final annihilation rate is easily calculable from using the
ground-state solution of the two-particle Schr\"{o}dinger equation.
One readily gets that the probability density for one of the particles
to be at the location of the other, the quantity that takes the role
of the number density $n$ in Eq. (\ref{eq:123}), is
 \begin{equation}
 n = {1\over 4\pi}{m^9\over m_{\rm Pl}^6 c^3},
 \label{eq:140}
 \end{equation}
continuing to use $m$ with units of inverse time.
Then by Eq. (\ref{eq:123}), the annihilation rate per time (for the two
particles to annihilate) is
 \begin{equation}
 R = {m^{11}\over 2 m_{\rm Pl}^{10}}.
 \label{eq:140b}
 \end{equation}

If one uses the first Eq. (\ref{eq:128}) to convert this to the quantum
expectation value of a mass-loss rate and uses the fact that for this
2-particle state, $\mu = 2(m/m_{\rm Pl})^2$, one gets
 \begin{eqnarray}
 P_q &\equiv& -\dot{M}_{\mathrm{quantum}} = -{dM\over dt}=-{d\mu\over
 d(mt)}
 \nonumber \\
     &=& \left({m\over m_{\rm Pl}}\right)^{12}
     = {\hbar^6 G^6\over c^{30}} m^{12}
     = {1\over 32} {m^2 \over m_{\rm Pl}^2} \mu^5
 \nonumber \\
     &\approx& 3.670\,759\,777\,914\,995\,479\,155\,135\,882\,831\,790\,22\,
       Q t_{\rm Pl}^2 m^2 \mu^5.
 \label{eq:141}
 \end{eqnarray}
That is, the actual rate at which the 2-particle state annihilates is
a factor of about 3.67 times what one would get by blindly
extrapolating down to $N=2$ particles the rate given by
Eq. (\ref{eq:129}), which actually applies only for very many
particles, $N \gg 1$ (as well as $\mu \ll 1$).  Eq. (\ref{eq:141})
contains the largest positive power of $\hbar$ (6), the largest
positive power of $G$ (6), and the largest negative power of the speed
of $c$ (30), that I can ever recall seeing in a formula, though I am
not used to using formulas in which I have not just set $\hbar=G=c=1$.

If instead we take the reciprocal of the annihilation rate $R$ given
by Eq. (\ref{eq:140b}) as the expectation value of the decay time from
the 2-scalar-particle state to the 2-graviton state, and also as an
estimate for the total decay time for an oscillaton (since it
presumably dominates over the time to get down to 2 particles,
assuming that the number of particles is even when one gets close
enough to the 2-particle state that one can ignore the probability
that an odd number of scalar particles will annihilate), then we get a
total decay time of
 \begin{equation}
 t_{\rm decay} \approx 1/R = {2 m_{\rm Pl}^{10} \over m^{11}}
   = {2c^{25}\over \hbar^5 G^5 m^{11}}
   = {2\hbar^6 c^3 \over G^5 m_{*}^{11}}
   = 2 t_{\rm Pl} \left({m_{*\rm Pl} \over m{*}}\right)^{11}.
 \label{eq:142}
 \end{equation}

For example, if we take a typical axion mass, $m_{*}=10^{-5}{\rm
eV}/c^2$, then Eq. (\ref{eq:141}) gives $P_q \approx 9.11\times
10^{-398}$ and Eq. (\ref{eq:142}) gives $t_{\rm decay} = 1.88\times
10^{346}\ {\rm years}$.  To take the more extreme example in which $m$
has the value of the current Hubble expansion rate, $H_0 \approx
2.3\times 10^{-18} s^{-1} \approx 1.24\times 10^{-61} m_{\rm Pl}$,
then Eq. (\ref{eq:141}) gives $P_q \approx 1.3\times 10^{-731}$ and
Eq. (\ref{eq:142}) gives $t_{\rm decay} \approx 1.3\times 10^{680}\ 
{\rm years}$.

Thus the complete quantum decay of an oscillaton can take a very long
time and probably would not be a suitable subject for an experimental
Ph.D. thesis.  On the other hand, the slowness of both the classical
and quantum decay of oscillatons of light scalar fields shows that if
they form in the universe, they can last for astronomically long times.

\section{Quantum Decay of Multiple-Field Oscillatons}

The analysis above was for oscillatons having just one real massive
scalar field, minimally coupled to gravity.  Although in this paper we
shall not go beyond minimal coupling or consider other scalar-field
self couplings (other than mass terms), we started with a general
discussion of an arbitrary number of minimally coupled massive scalar
fields and their classical decay rates, so it would be of interest to
say also how the quantum decay goes when there are more than one
scalar field.

When the scalar fields all have different masses, then the separate
decay processes are incoherent, so the rates for each just add, with
the rate for each (in the nearly-Newtonian limit) going as the spatial
integral of the square of the mass density for that scalar field, with
the coefficient as given above.

This is at least so if we average over times long in comparison with
the reciprocals of the differences of the scalar field masses in
frequency units, which will hereby be assumed---if any mass
differences are short in comparison with the reciprocal of the decay
time of interest, then for that time we may consider these scalar
fields as having the same mass.  Intermediate cases in which the decay
times of interest are comparable to the reciprocal of any mass
differences will not be discussed here.

Therefore, we may consider separately all of the fields at one mass
(or one range of masses if the range is much less than the reciprocal
of the decay time being considered).

The simplest case is that in which there are two equal-mass scalar
fields that are oscillating at $90^{\circ}$ out of phase.  This is
equivalent to one complex scalar field that has a global $U(1)$
symmetry and hence a conserved particle number that presumably cannot
decay away, at least by perturbative quantum effects such as what
DeWitt \cite{DeW} used to calculate the annihilation of scalar
particles into gravitons.

Presumably there are nonperturbative gravitational effects in which a
nonzero particle number, though conserved by the global $U(1)$
symmetry perturbatively, forms or tunnels into a real or virtual black
hole that then decays into a different particle number (e.g., zero).
Thus at some level the global $U(1)$ invariance is surely broken by
gravity.  In \cite{HPP} we used a model of gravitational foam to
estimate that this rate would be disastrously high for point scalar
particles, suggesting that perhaps no such particles could exist in
our universe.  If so, this would of course rule out the whole idea of
oscillatons (unless they were made of composite scalars that are not
pointlike down to near the Planck scale).  But since our ideas were
admittedly rather speculative, here I shall assume that the
nonperturbative effects violating global $U(1)$ invariance are
suppressed to give rates much smaller than the particle-antiparticle
annihilation into gravitons that is allowed by the perturbative
analysis that preserves the global $U(1)$ invariance.

If this is indeed so, when there are two equal-mass scalar fields that
are oscillating at $90^{\circ}$ out of phase, effectively the decay of
each individual scalar field must destructively interfere so that the
total decay rate is zero.

At first this sounds impossible, since if one has a state with $N_1$
scalar particles of the first field and $N_2$ of the second, then the
final state in which two particles of the first field annihilate into
gravitons would have $N_1-2$ particles of the first field and $N_2$ of
the second, which would be orthogonal to the final state in which
instead two particles of the second field annihilate into gravitons,
leaving $N_1$ particles of the first field and $N_2-2$ particles of
the second field.  Therefore, how could there possibly be any
destructive interference to prevent the particles from annihilating?

However, this objection can be circumvented if the quantum state of
the oscillaton does not have a definite number of particles of both
kinds.  (Indeed, that would have to be the case if they are
oscillating $90^{\circ}$ out of phase, since phase is in some sense a
conjugate variable to particle number.  Note that the total particle
number could be precise, so that the total phase is undefined, so long
as the individual particle numbers are sufficiently indefinite that
the relative phase between the two fields is well defined.)  For
example, the quantum state for the particles could be a coherent state
that is an eigenstate of the annihilation operators for the two kinds
of particles.

Then in the case that two particles of the first real scalar field
decay, although the expectation value of the number of particles of
that field would have been reduced by two, the final state would not
need to be orthogonal to the state that would result if instead two
particles of the second real scalar field were to annihilate.
Therefore, the two decay processes can interfere.  When the two scalar
fields oscillate $90^{\circ}$ out of phase, their combination is
equivalent to a single complex scalar field with a $U(1)$ symmetry
that prevents perturbative quantum decay into gravitons.

The gravitational signal of this $U(1)$ symmetry would be that the
stress-energy tensor would have no oscillations, so one would be back
to the case of a boson star that seems to be completely stable (except
presumably to nonperturbative tunneling processes in which some or all
of the particles tunnel into a black hole, or a virtual black hole,
that would either transcend or violate what would otherwise be the
conservation of the global $U(1)$ charge \cite{HPP}).

Now consider the case in which there are an arbitrary number, say $n$,
scalar fields at some mass $m_I$.  We shall continue to assume the
nearly-Newtonian limit, in which the dimensionless rescaled real
massive scalar fields $\phi_{IJ}$, defined by Eq. (\ref{eq:5}), have
the form given by Eq. (\ref{eq:16}) in terms of the complex
dimensionless scalar fields $\psi_{IJ}$ that are very slowly varying
spatially, and even more slowly varying temporally, if at all, on the
length scale $c/m_I$ and on the time scale $1/m_I$.  Then the real
scalar fields $\phi_{IJ}$ are essentially oscillating nearly
periodically with frequency $m_I$.

When two scalar particles annihilate into (predominantly) two
gravitons, the graviton wavelengths are roughly $c/m_I$, which is a
much shorter length scale than the length scale of the variation of
the fields $\phi_{IJ}$, so each region of size of the order of $c/m_I$
annihilates essentially independently.  (That is, the quantum states
of the outgoing gravitons are essentially orthogonal for the
annihilation in the separate regions, if one uses graviton wavepackets
that have sizes more nearly comparable to their wavelengths than to
the much bigger size of the oscillaton.)  Thus we can just add up the
annihilation rates in each region, effectively getting an integral
over the oscillaton of the annihilation rate in each region.

Since we shall only be interested in the annihilation over many
oscillations of the oscillaton, we shall only consider the
annihilation rate averaged over many such periods.

In each region of size somewhat bigger than $c/m_I$ where we are
calculating the average annihilation rate, each of the real scalar
fields of mass $m_I$ is oscillating essentially with constant
amplitude and period, staying in phase with each other scalar field
over a time long compared with the oscillation period that is very
nearly $2\pi/m_I$.  In this region, we can perform an $O(n)$
transformation of the $n$ scalar fields so that all but two of the
fields are transformed to zero for the time of interest (to the
accuracy of the nearly-Newtonian approximation), and the two that
remain nonzero are oscillating $90^{\circ}$ out of phase, say
 \begin{equation}
 \phi_1 = \sum_J O_{1J}\phi_{IJ} \approx 2c_1\cos{[m_I(t-t_0)]},
 \label{eq:143}
 \end{equation}
 \begin{equation}
 \phi_2 = \sum_J O_{2J}\phi_{IJ} \approx -2c_2\sin{[m_I(t-t_0)]},
 \label{eq:144}
 \end{equation}
and
 \begin{equation}
 \phi_i = \sum_J O_{iJ}\phi_{IJ} \approx 0
 \label{eq:145}
 \end{equation}
for $i>2$, with real positive amplitudes $c_1$ and $c_2$ that are
nearly constant over the spacetime region where the time-averaged
annihilation rate is being calculated.  Without loss of generality we
can choose the $O(n)$ transformation to give $c_1 \geq c_2$.

I am not bothering to include the subscript $I$, which tells what the
mass $m_I$ is, on what I am calling $\phi_i$.  Indeed, the fact that I
am giving $\phi_i$ only a single subscript is used here to distinguish
it from the $O(n)$-related scalar fields $\phi_{IJ}$ without having to
put primes on $\phi_i$ as I would have if it had the same number of
indices as $\phi_{IJ}$.

We can also define the $O(n)$-transformed complex scalar fields
 \begin{equation}
 \psi_1 = \sum_J O_{1J}\psi_{IJ} \approx c_1 e^{im_I t_0},
 \label{eq:146}
 \end{equation}
 \begin{equation}
 \psi_2 = \sum_J O_{2J}\psi_{IJ} \approx -i c_2 e^{im_I t_0},
 \label{eq:147}
 \end{equation}
and
 \begin{equation}
 \psi_i = \sum_J O_{iJ}\psi_{IJ} \approx 0
 \label{eq:148a}
 \end{equation}
for $i > 2$.

Now for the quantum analysis, we can replace the two real scalar
fields with the one complex scalar field
 \begin{eqnarray}
 \Phi &=& {1\over\sqrt{2}}(\phi_1+i\phi_2)
      \approx {1\over \sqrt{2}}
       [(\psi_1+i\psi_2)e^{-i m_I t}+(\bar{\psi}_1
	 +i\bar{\psi}_2)e^{i m_I t}]
 \nonumber \\
      &\approx& {1\over \sqrt{2}}[(c_1+c_2)e^{-i
      m_I(t-t_0)}+(c_1-c_2)e^{i m_I(t-t_0)}]
      = \Phi_{+}e^{-i m_I t} + \Phi_{-}e^{i m_I t}
 \label{eq:148}
 \end{eqnarray}
(with $\Phi$ not to be confused with the original real scalar fields
$\Phi_{IJ}$ or the slowly-varying complex dimensionless scalar fields
$\psi_{IJ}$ that were used to represent each dimensionless real scalar
field $\phi_{IJ}$; those all had subscripts that will not appear on
the single rapidly-varying complex $\Phi$ that combines the two
rapidly-varying real scalar fields $\phi_1$ and $\phi_2$).

Because of the $U(1)$-invariance of the field equations and
stress-energy tensor of the two nonzero real scalar fields, $\phi_1$
and $\phi_2$, which are represented by this complex scalar field
$\Phi$, there is a conserved global $U(1)$ charge.  The part of the
classical field $\Phi$ that has the phase factor $e^{-i m_I(t-t_0)}$
and the coefficient $\Phi_{+}$ can be said to represent particles with
positive global $U(1)$ charge and with particle number density
 \begin{equation}
 n_{I+} \approx {m_I c^2 \over 4\pi G \hbar}\left|\Phi_{+}\right|^2
        \approx {m_I c^2 \over 4\pi G \hbar}\left|{c_1+c_2\over
        2\sqrt{2}}\right|^2
	\approx {m_I c^2 \over 4\pi G \hbar}\left|{\psi_1+i\psi_2\over
        \sqrt{2}}\right|^2
 \label{eq:149}
 \end{equation}
and the part of the classical field $\Phi$ that has the phase factor
$e^{+i m_I(t-t_0)}$ and the coefficient $\Phi_{+}$ can be said to
represent antiparticles with negative global $U(1)$ charge and with
antiparticle number density
 \begin{equation}
 n_{I-} \approx {m_I c^2 \over 4\pi G \hbar}\left|\Phi_{-}\right|^2
        \approx {m_I c^2 \over 4\pi G \hbar}\left|{c_1-c_2\over
        2\sqrt{2}}\right|^2
	\approx {m_I c^2 \over 4\pi G \hbar}
        \left|{\bar{\psi}_1+i\bar{\psi}_2\over\sqrt{2}}\right|^2.
 \label{eq:150}
 \end{equation}

In the classical limit that we are assuming, we can express these
number densities in terms of the mass density and mean-squared
pressure of the fields of mass $m_I$ in the following way:

By using Eq. (\ref{eq:21e}) for $G_{00} \approx 8\pi G \rho$ or
Eq. (\ref{eq:27dd}) directly for the mass density $\rho$ (which is
nearly constant in time), and splitting it up into the contributions
from the fields of the different masses $m_I$, one gets
 \begin{equation}
 \rho_I \approx {m_I^2\over 4\pi G} \sum_J |\psi_{IJ}|^2
        = {m_I^2\over 4\pi G} (|\psi_1|^2 + |\psi_2|^2)
	\approx {m_I^2\over 4\pi G} (c_1^2 + c_2^2)
        \approx {m_I \hbar\over c^2} n_I = m_{I*} n_I,
 \label{eq:151}
 \end{equation}
where
 \begin{equation}
 n_I = n_{I+}+n_{I-} = {m_I c^2 \over 4\pi G \hbar}(c_1^2 + c_2^2)
                     = {m_I c^2 \over 4\pi G \hbar}\sum_J|\psi_{IJ}|^2
 \label{eq:152}
 \end{equation}
is the total number density of all the scalar fields of mass $m_I$,
or, equivalently, of both the particles and the antiparticles of the
single complex scalar field $\Phi$ that classically represents all of
the real fields.  (I.e., we are ignoring vacuum fluctuations in the
transformed scalar fields $\phi_i$ with $i>2$ that are classically
zero.)

Similarly, by using Eq. (\ref{eq:23e}) for $G_{ij} \approx (8\pi
G/c^2)P\delta_{ij}$ for the oscillating nearly-isotropic pressure $P$,
and also splitting it up into the contributions from the fields of the
different masses $m_I$, one gets
 \begin{equation}
 P_I \approx -{m_I^2 c^2\over 8\pi G} \sum_J 
     (\psi_{IJ}^2 e^{-2i m_I t} +\bar{\psi}_{IJ}^2 e^{2i m_I t}).
 \label{eq:153}
 \end{equation}
In this case, if one takes the time-average of the square of the total
pressure of all the $n$ scalar fields of mass $m_I$ (or equivalently
of the single complex scalar field $\Phi$), one gets
 \begin{eqnarray}
 \langle P_I^2 \rangle &\approx& {c^4\over 2} \left({m_I^2\over 4\pi G}\right)^2
   |\sum_J \psi_{IJ}^2|^2
  \approx {c^4\over 2} \left({m_I^2\over 4\pi G}\right)^2 |\psi_1^2 +
       \psi_2^2|^2
 \nonumber \\
  &\approx& {c^4\over 2} \left({m_I^2\over 4\pi G}\right)^2
        (c_1^2-c_2^2)^2
  \approx 2 m_I^2 \hbar^2 n_{I+} n_{I-}.
 \label{eq:154}
 \end{eqnarray}
Because $P_I$ is oscillating sinusoidally, its maximum value, say
$P_{I\rm max}$ (as a function of time at each spatial location) is
$\langle 2 P_I^2 \rangle^{1/2}$.

Then from Eqs. (\ref{eq:151}), (\ref{eq:152}), and (\ref{eq:154}), one
can solve for the number densities of both the particles and the
antiparticles of the complex scalar field $\Phi$ of mass $m_I$:
 \begin{equation}
 n_{I+} \approx {1\over 2m_I\hbar}\left(\rho_I c^2 
  + \sqrt{\rho_I^2 c^4 - \langle 2 P_I^2 \rangle}\right),
 \label{eq:155}
 \end{equation}
 \begin{equation}
 n_{I-} \approx {1\over 2m_I\hbar}\left(\rho_I c^2 
  - \sqrt{\rho_I^2 c^4 -  \langle 2 P_I^2 \rangle}\right),
 \label{eq:156}
 \end{equation}
where $\langle 2 P_I^2 \rangle$ is given in terms of the $\psi_{IJ}$'s
by Eq. (\ref{eq:154}).

Another way to express this relationship, using $P_{I\rm max} =
\langle 2 P_I^2 \rangle^{1/2}$, is to note that
 \begin{equation}
 \rho_I c^2 + P_{I\rm max} = m_I\hbar(\sqrt{n_{I+}}+\sqrt{n_{I-}})^2,
 \label{eq:157}
 \end{equation}
 \begin{equation}
 \rho_I c^2 - P_{I\rm max} = m_I\hbar(\sqrt{n_{I+}}-\sqrt{n_{I-}})^2.
 \label{eq:158}
 \end{equation}
Thus when $P_{I\rm max} = 0$, there are just particles but no
antiparticles of the complex scalar field, and when $P_{I\rm max} =
\rho_I c^2$, there are equal numbers of particles and antiparticles.
This latter possibility is the case when there is only one real scalar
field, in which case it is a fiction to say that there is the complex
scalar field $\Phi$ at all, but the classical real scalar field does
act as if it were composed of equal numbers of particles and
antiparticles of the fictitious complex scalar field, $n_{I+} = n_{I-}
= (1/2)n_I$.  (Of course, then there are no vacuum fluctuations of the
nonexistent imaginary component of the complex $\Phi$, but here we are
taking the classical limit and are only considering the effects of
real particles and not of any vacuum fluctuations.)

By making the arbitrary requirement that the $O(n)$ transformation
lead to real $c_1 \geq c_2 \geq 0$ for the coefficients of the two
nonzero scalar fields after the transformation, we have made an
arbitrary choice of what to call particles and what to call
antiparticles (with the number density of particles never less than
the number of antiparticles by this choice) and hence of which
expression in Eqs. (\ref{eq:155}) and (\ref{eq:156}) has the minus
sign in front of the square root.

%Note that if there is only one real scalar field of mass $m_I$, or if
%all of the real scalar fields of mass $m_I$ oscillate with the same
%phase, then $P_I$ oscillates sinusoidally with an amplitude (maximum
%value) that is equal to $\rho_I c^2$, so $\langle 2 P_I^2 \rangle =
%\rho_I^2 c^4$.  Then the square roots in Eqs. (\ref{eq:155}) and
%(\ref{eq:156}) are zero, and hence $n_{I+} = n_{I-} = (1/2)n_I$ (equal
%numbers of particles and antiparticles).

Now when we have the possibility of both particles and antiparticles
of the complex scalar field $\Phi$, the $U(1)$ invariance prevents the
annihilation of a pair of particles or of a pair of antiparticles (at
least at the perturbative level) but allows the annihilation of a
particle-antiparticle pair.  On the other hand, the
particle-antiparticle annihilation cross-section is twice that given
by Eq. (\ref{eq:121}) \cite{DeW} for two real scalar field particles,
i.e.,
 \begin{equation}
 \sigma_{+-} = {4\pi G^2 m_*^2 \over c^3 v}
             = {4\pi \hbar^2 G^2 m^2\over c^7 v}.
 \label{eq:159}
 \end{equation}

This means that when we have $N_{I+}$ particles and $N_{I-}$
antiparticles, both of these numbers decrease at the rate
 \begin{eqnarray}
 -{dN_{I+}\over dt} &=& -{dN_{I-}\over dt}
  = \int d^3 x n_{I+} n_{I-} \sigma_{+-} v
  = {4\pi \hbar^2 G^2 m_I^2\over c^7} \int d^3 x n_{I+} n_{I-}
 \nonumber \\
  &=& {\pi G^2 \over c^7} \int d^3 x \langle 2 P_I^2 \rangle
  = {\pi G^2 \over c^7} \int d^3 x P_{I\rm max}^2
  = {m_I^4 \over 16\pi c^3} \int d^3 x |\sum_J \psi_{IJ}^2|^2.
 \label{eq:160}
 \end{eqnarray}
The total number decay rate is of course twice this.  When $n_{I+} =
n_{I-}$, so that $P_{I\rm max} = \rho_I c^2$, then the total rate
agrees with the integral of Eq. (\ref{eq:124}).

When we multiply the total number decay rate for each mass $m_I$
(which is in frequency units) by the mass $m_{*I} = \hbar m_I/c^2$ in
conventional units and sum over all $I$, we get the total mass loss rate
by scalar particle annihilation in conventional mass units:
 \begin{equation}
 -{dM_{*}\over dt}
 = {\hbar \over 8\pi c^5} \sum_I m_I^5 \int d^3 x |\sum_J \psi_{IJ}^2|^2.
 \label{eq:160b}
 \end{equation}
We can then multiply this by $G/c^3$ to get the dimensionless rate of decrease of
$M = G M_{*}/c^3$, the mass in units of time:
 \begin{equation}
 -{dM\over dt}
 = {\hbar G\over 8\pi c^8} \sum_I m_I^5 \int d^3 x |\sum_J \psi_{IJ}^2|^2.
 \label{eq:160c}
 \end{equation}

\section{Quantum and Classical Emission from the \\
 Simplest Multi-Field Spherical Oscillatons}

Now consider the particular case in which all of the real scalar
fields have the same mass, so we can drop the subscript $I$ that
labels the mass.  Furthermore, restrict attention to the case in which
all of the real scalar fields are oscillating with the same
quasi-stationary nodeless spherically symmetric mode (that of the
simplest spherical oscillaton, except that now there are more than one
real scalar field that may be oscillating with different phases).

By the argument above, we can perform an $O(n)$ transformation so that
only two real scalar fields are then oscillating with nonzero
amplitude and are $90^{\circ}$ out of phase.  By the assumption that all
of the scalar fields are oscillating in the same mode (up to phase),
this $O(n)$ transformation is constant over space, and the ratio of
the amplitudes of the two resulting nonzero modes are also constant.
By the procedure above, it can be replaced by a single complex field.

Let $N_{+}$ be the number of particles of the complex scalar field,
$N_{-}$ be the number of antiparticles, and $N=N_{+}+N_{-}$ be the
total number of particles and antiparticles.  Then if the $\psi$ given
by Eq. (\ref{eq:56}) (real in this case) represents the simplest
spherical oscillaton with one real scalar field described above, the
complex oscillaton with two real scalar fields and the same total mass
is represented (after a shift in the origin of time) by
 \begin{equation}
 \psi_1 = {\sqrt{N_{+}}+\sqrt{N_{-}}\over\sqrt{2N}}\psi,
 \label{eq:161}
 \end{equation}
 \begin{equation}
 \psi_2 = -i{\sqrt{N_{+}}-\sqrt{N_{-}}\over\sqrt{2N}}\psi.
 \label{eq:162}
 \end{equation}
Then the complex scalar field is given by
 \begin{eqnarray}
 \Phi &=& {1\over\sqrt{2}}(\phi_1+i\phi_2)
      = {1\over\sqrt{2}}[(\psi_1 e^{-imt} + \bar{\psi}_1 e^{imt})
                         +i(\psi_2 e^{-imt} + \bar{\psi}_2 e^{imt})]
 \nonumber \\
      &=& \left(\sqrt{N_{+}\over N}e^{-imt}
               + \sqrt{N_{-}\over N}e^{imt}\right)\psi.
 \label{eq:163}
 \end{eqnarray}

Now, by using Eq. (\ref{eq:160}) and doing an analysis analogous to
that which led to Eq. (\ref{eq:129}), one can show that the quantum
annihilation of scalaron particle-antiparticle pairs for the
spherically-symmetric nodeless complex oscillaton gives
 \begin{equation}
 -{dN_{+}\over dt} = -{dN_{-}\over dt}
                   = 2Q t_{\rm Pl}^{10} m^{11} N^3 N_{+} N_{-},
 \label{eq:164}
 \end{equation}
where the numerical constant $Q$ is given in Eq. (\ref{eq:131}).

These equations have the obvious constant of motion being the number
of particles minus the number of antiparticles, say
 \begin{equation}
 N_0 \equiv N_{+} - N_{-}.
 \label{eq:165}
 \end{equation}
Then, since the total number of particles is $N = N_{+} + N_{-}$, one
can write
 \begin{equation}
 N_{+} = {1\over 2}(N+N_0),
 \label{eq:166}
 \end{equation}
 \begin{equation}
 N_{-} = {1\over 2}(N-N_0).
 \label{eq:167}
 \end{equation}

Similarly, one can write the dimensionless mass parameter of the
classical configuration as
 \begin{equation}
 \mu \equiv Mm = t_{\rm Pl}^2 m^2 N
 \label{eq:168}
 \end{equation}
and also define a constant dimensionless mass parameter as
 \begin{equation}
 \mu_0 \equiv t_{\rm Pl}^2 m^2 N_0.
 \label{eq:168b}
 \end{equation}

As both $N_{+}$ and $N_{-}$ decay away at equal rates, the oscillaton
asymptotically approaches the configuration with $N = N_{+} = N_0$ and
$N_{-} = 0$, which is a static boson star with $\mu = \mu_0$.  Thus
$\mu_0$ is the asymptotic (minimum) value of $\mu$.

In terms of $N$ and $N_0$, Eq. (\ref{eq:164}) gives
 \begin{equation}
 -{dN\over dt} = Q t_{\rm Pl}^{10} m^{11} N^3 (N^2 - N_0^2)
 \label{eq:169}
 \end{equation}
from the annihilation of scalaron particle-antiparticle pairs into
pairs of gravitons (i.e., ignoring the classical emission into scalar
radiation).

Alternatively, we can write the evolution of the dimensionless mass
parameter as
 \begin{equation}
 -{d\mu\over dt} = Q t_{\rm Pl}^2 m^3 \mu^3 (\mu^2 - \mu_0^2)
 \label{eq:170}
 \end{equation}
from the quantum annihilation.  This differential equation has the
algebraic solution
 \begin{equation}
 {\mu^2-\mu_0^2\over \mu^2}e^{\mu_0^2/\mu^2}
 = e^{-2Q t_{\rm Pl}^2 m^3 \mu_0^4 (t-t_0)}
 \label{eq:171}
 \end{equation}
where $t_0$ is an arbitrary constant of integration.
Thus at late times, $\mu$ approaches $\mu_0$ exponentially rapidly,
and the configuration approaches that of a static boson star with the
conserved number of particles.

It may also be of interest to give the decay rates from the classical
emission of scalar radiation in this multi-field case with the
simplest spherical configuration.  By using Eq. (\ref{eq:49}), one
can deduce that the classical scalar field emission leads to
 \begin{equation}
 -{dN_{+}\over dt} = -{dN_{-}\over dt}
 \approx 2C {m_{\rm Pl}^6\over m^5} {N_{+} N_{-} \over N^4}
  e^{-{\alpha m_{\rm Pl}^2/m^2 N}},
 \label{eq:172}
 \end{equation}
 \begin{equation}
 -{dN\over dt} \approx C {m_{\rm Pl}^6\over m^5} {N^2 - N_0^2 \over N^4}
  e^{-{\alpha m_{\rm Pl}^2/m^2 N}},
 \label{eq:173}
 \end{equation}
and
 \begin{equation}
 -{d\mu\over dt} \approx C m {\mu^2-\mu_0^2 \over \mu^4} e^{-\alpha/\mu}.
 \label{eq:170b}
 \end{equation}

To get the totals for these rates, one must add the corresponding
expressions for the quantum rates from Eqs. (\ref{eq:164}),
(\ref{eq:169}), and (\ref{eq:170}) respectively.  For example, the
total rate at which the mass (in time units) decreases is
 \begin{eqnarray}
 -{dM\over dt} = -{d\mu\over m dt}
 &\approx& C {\mu^2-\mu_0^2 \over \mu^4} e^{-\alpha/\mu}
 + Q t_{\rm Pl}^2 m^2 \mu^3 (\mu^2 - \mu_0^2)
 \nonumber \\
 &=& \left({C\over m^2 M^4}e^{-{\alpha\over mM}}
         +Q t_{\rm Pl}^2 m^7 M^3 \right)(M^2 - M_0^2),
 \label{eq:171b}
 \end{eqnarray}
where $M_0$ = $\mu_0/m = t_{\rm Pl}^2 m N_0$ is the asymptotic mass of
the final boson star in time units, and where the numerical constants
$C \approx 3\,797\,438$, $\alpha \approx 39.4338$, and $Q \approx
0.008\,513\,224$ were given to 36 decimal places (though only the
first 30-35 of these are likely to be accurate) in
Eqs. (\ref{eq:114b}), (\ref{eq:114a}), and (\ref{eq:131})
respectively.

One can see that at late times, for $\mu_0 > 0$, as $\mu$ approaches
very near to $\mu_0$, $\mu - \mu_0 \ll \mu_0^2/\alpha$, $\mu$
approaches $\mu_0$ exponentially rapidly:
 \begin{equation}
 \mu \sim \mu_0
  + \exp{[-(C\mu_0^{-3}e^{-\alpha/\mu_0}
          + Q  t_{\rm Pl}^2 m^2 \mu_0^4) 2 m t]}.
 \label{eq:172b}
 \end{equation}

Finally, if we use the scalar field mass in conventional mass units,
$m_{*} = \hbar m/c^2$, and the oscillaton mass also in conventional
mass units, $M_{*} = c^3 M/G$, then the total mass loss rate from the
simplest spherical multiple-field oscillaton with minimum conventional
mass $M_{*0} = c^3 M_{0}/G = N_0 m_{*}$, for $m_{*} \ll M_{*} \ll
10^{-10} M_{\odot}[(1\ {\rm eV})/(m_{*} c^2)]$, is
 \begin{equation}
 -{dM_{*}\over dt}
 \approx \left(C {\hbar^2 c^5 \over G^3} {1\over m_{*}^2 M_{*}^4}
           e^{-\alpha\hbar c/(G m_{*} M_{*})}
    + Q {G^5 \over \hbar^6 c^3} m_{*}^7 M_{*}^3 \right)(M_{*}^2 - M_{*0}^2).
 \label{eq:173b}
 \end{equation}
Again may I remind the reader that the conventional mass $m_{*}$ of
the scalar field is a quantum quantity with $\hbar$ in it, which is
why there is an explicit $\hbar^2$ in the numerator of the first
(classical) term for the mass decay rate (from scalar field
radiation), to cancel the implicit $\hbar^2$ in the $m_{*}^2$ term in
the denominator.  Similarly, in the second (quantum) term, once the
implicit factor of $\hbar^7$ in the $m_{*}^7$ term is taken into
account, there is one positive power of $\hbar$ appearing, as one would
expect for this first-order quantum perturbative contribution (the
annihilation of pairs of scalarons into pairs of gravitons).

\section{Conclusions}

Oscillatons without a $U(1)$ invariance (those without a static
geometry) are unstable both classically (to emitting scalar waves) and
quantum mechanically (to having scalarons annihilate into gravitons).
The classical rate dominates for large $\mu = Mm$ but drops very fast
with decreasing $\mu$ and is nonanalytic at $\mu=0$: $P_c \approx
(C/\mu^2) e^{-\alpha/\mu}$, Eq. (\ref{eq:113}), with the numerical
constants $C$ given by Eq. (\ref{eq:114b}) and $\alpha$ given by
Eq. (\ref{eq:114a}).  The quantum rate also drops as $\mu$ drops, but
only as a power law in $\mu$: $P_q \approx Q(m/m_{\rm Pl})^2 \mu^5$,
Eq. (\ref{eq:129}), with the numerical constant $Q$ given by
Eq. (\ref{eq:131}).  The quantum rate dominates for $\mu
\,\stackrel{<}{\sim}\, 1/8$ for $m \,\stackrel{>}{\sim}\, H_0$ (the
current Hubble expansion rate, a lower bound on $m$ for any oscillaton
existing in our universe today).

An oscillaton that starts at $\mu_1 \approx u_{\rm max} \approx 0.607$
\cite{ABGMNUL} has a significant drop in $\mu$ (more than $10\%$) over
a lifetime comparable with the age of the universe if $m_{*} c^2
\,\stackrel{>}{\sim}\, 2\times 10^{-11}\,{\rm eV}$ or $M_{*}
\,\stackrel{<}{\sim}\, 3.6\, M_{\odot}$.  However, unless $m_{*} c^2
\,\stackrel{>}{\sim}\, 2.3\times 10^{13 }\,{\rm eV} = 2300\ {\rm GeV}$
or $M_{*} \,\stackrel{<}{\sim}\, 1.8\times 10^{-24} M_{\odot} \approx
3.5 \times 10^9 g$, in this decay time $\mu$ does not decrease by more
than a factor of 2.

These numerical approximations are using formulas derived in this
paper for nearly-Newtonian configurations, which have $\mu \ll 1$.
For accurate results for the not-too-small values of $\mu$ that would
arise from the decay, within astronomical times, of oscillatons that
started with the maximum mass possible for any reasonable value of the
scalaron mass, one would need to extend the results derived here to
the strong-gravity regime.  This is research that shall be left to the
future.

Multi-field oscillatons \cite{HC}, in which different real scalar
fields of the same mass oscillate out of phase, do not decay away
completely but instead asymptotically approach a stable $U(1)$-invariant
configuration with a static metric (a boson star), at a rate given by
Eq. (\ref{eq:171b}) or (\ref{eq:173b}).

\section*{Acknowledgments}

I am grateful for many discussions with Jeongwon Ho (who got me
interested in oscillatons, gave me some of the early literature on
them, and found for me that the quantum annihilation of scalarons into
gravitons was calculated in \cite{DeW}), and for various other
conversations with Bruce Campbell, Matthew Choptuik, Bryce DeWitt,
Valeri Frolov, Gary Horowitz, and Werner Israel (who immediately
recognized that oscillatons would be classically unstable).  The
numerical calculations were performed with the aid of Maple.  Finally,
financial support has been provided by the Natural Sciences and
Engineering Research Council of Canada.

%\newpage


\baselineskip 4pt

\begin{thebibliography}{99}

\bibitem{SS}  E. Seidel and W.-M. Suen,
Phys.\ Rev.\ Lett.\ {\bf 66}, 1659-1662 (1991).

\bibitem{SS2}  E. Seidel and W.-M. Suen,
Phys.\ Rev.\ Lett.\ {\bf 72}, 2516-2519 (1994), gr-qc/9309015.

\bibitem{I}  A. Iwazaki,
Prog.\ Theor.\ Phys.\ {\bf 101}, 1253-1260 (1999), hep-ph/9803374; 
Phys.\ Lett.\ B{\bf 451}, 123-128 (1999), hep-ph/9804369;
``Axionic Boson Stars in Magnetized Conducting Media and Monochromatic
Radiations,'' hep-ph/9807232;
Phys.\ Rev.\ D{\bf 60}, 025001 (1999), hep-ph/9901396;
Phys.\ Lett.\ B{\bf 455}, 192-196 (1999), astro-ph/9903251;
Phys.\ Lett.\ B{\bf 486}, 147-152 (2000), hep-ph/9906353;
``Emission of Radio Waves in Gamma Ray Bursts and Axionic Boson
Stars,'' hep-ph/9908468.

\bibitem{U-L}  L. A. Ure\~{n}a-L\'{o}pez,
Class.\ Quantum Grav.\ {\bf 19}, 2617-2632 (2002), gr-qc/0104093.

\bibitem{AGMNULW}  M. Alcubierre, S. F. Guzman, T. Matos,
D. Nunez, L. A. Ure\~{n}a-L\'{o}pez, and P. Wiederhold,
Class.\ Quantum Grav.\ {\bf 19}, 5017-5024 (2002), gr-qc/0110102. 

\bibitem{ULMB}  L. A. Ure\~{n}a-L\'{o}pez, T. Matos, and R. Becerril,
Class.\ Quantum Grav.\ {\bf 19}, 6259-6277 (2002).

\bibitem{HC}  S. H. Hawley and M. W. Choptuik,
Phys.\ Rev.\ D{\bf 67}, 024010 (2003), gr-qc/0208078. 

\bibitem{ABGMNUL}  M. Alcubierre, R. Becerril, S. F. Guzman, T. Matos,
D. Nunez, and L. A. Ure\~{n}a-L\'{o}pez,
Class.\ Quantum Grav.\ {\bf 20}, 2883-2904 (2003), gr-qc/0301105.

\bibitem{GUL}  F. S. Guzman and  L. A. Ure\~{n}a-L\'{o}pez,
Phys.\ Rev.\ D{\bf 68} 024023 (2003), astro-ph/0303440.

\bibitem{HBG}  W. Hu, R. Barkana, and A. Gruzinov,
Phys.\ Rev.\ Lett.\ {\bf 85}, 1158-1161 (2000), astro-ph/0003365.

\bibitem{FLP}  R. Friedberg, T. D. Lee, and Y. Pang,
Phys.\ Rev.\ D{\bf 35}, 3640-3657 (1987).

\bibitem{P}  R. Penrose,
Phil.\ Trans.\ R.\ Soc.\ {\bf 356}, 1-13 (1998).

\bibitem{MPT}  I. M. Moroz, R. Penrose, and P. Tod,
Class.\ Quantum Grav.\ {\bf 15}, 2733-2742 (1998).

\bibitem{TM}  P. Tod and I. M. Moroz,
Nonlinearity {\bf 12}, 201-216 (1999).

\bibitem{C}  J. Christian,
Phys.\ Rev.\ D{\bf 56}, 4844-4877 (1997), gr-qc/9701013.

\bibitem{RB}  R. Ruffini and S. Bonazzola,
Phys.\ Rev.\ {\bf 187}, 1767-1783 (1969).

\bibitem{Pagfrac}  D. N. Page,
Class.\ Quantum Grav.\ {\bf 1}, 417-427 (1984).

\bibitem{DeW}  B. S. DeWitt,
Phys.\ Rev.\ {\bf 162}, 1239-1256 (1967).

\bibitem{HPP}  S. W. Hawking, D. N. Page, and C. N. Pope,
Phys.\ Lett.\ B{\bf 86}, 175-178 (1979);
Nucl.\ Phys.\ B{\bf 170}, 283-306 (1980).



\end{thebibliography}
\end{document}